# A Statistical Study of the Gamma-Ray Burst and Supernova Association

Xiao-Fei Dong[1], Yong-Feng Huang[1,2], Zhi-Bin Zhang[3], Jin-Jun Geng[4], Chen Deng[1], Ze-Cheng Zou[1], Chen-Ran Hu[1], and Orkash Amat[1]

[1] School of Astronomy and Space Science, Nanjing University, Nanjing 210023, People's Republic of China; hyf@nju.edu.cn
[2] Key Laboratory of Modern Astronomy and Astrophysics (Nanjing University), Ministry of Education, Nanjing 210023, People's Republic of China
[3] School of Physics and Engineering, Qufu Normal University, Qufu 273165, People's Republic of China
[4] Purple Mountain Observatory, Chinese Academy of Sciences, Nanjing 210023, People's Republic of China

## Abstract

The association between long gamma-ray bursts (LGRBs) and core-collapse supernovae (SNe) has been well established since the discovery of SN 1998bw, which was linked to the low-luminosity LGRB 980425. However, long-term monitoring of several well-localized, low-redshift LGRBs has yielded compelling evidence for the absence of accompanying SNe. Notably, two long bursts, GRB 211211A and GRB 230307A, show signatures consistent with kilonova emission from compact binary mergers, indicating that at least some long events may originate from progenitors other than core-collapse SNe. In this study, we conduct a comparative analysis of two samples of LGRBs, i.e., LGRBs with and without SN associations, to investigate the differences that may reveal intrinsic distinctions in their progenitors. A detailed examination of their prompt emission properties, host galaxy environments, and event rates is performed. While the two samples exhibit considerable overlap in most observed properties, a significant discrepancy in their event rate is revealed. LGRBs without SN association have an event rate that aligns well with the star formation rate, whereas that of SN-associated LGRBs differs significantly. It indicates that LGRBs without an SN association may constitute a distinct subclass with intrinsically different progenitors.

## 1. Introduction

Gamma-ray bursts (GRBs) with a duration exceeding 2 s are usually defined as long gamma-ray bursts (LGRBs; E. P. Mazets et al. 1981; C. Kouveliotou et al. 1993; Z. B. Zhang & C. S. Choi 2008). They are generally believed to originate from core-collapse supernovae (SNe; B. Zhang et al. 2007; J. S. Bloom et al. 2008), the explosive deaths of massive stars whose iron cores collapse into neutron stars or black holes (S. Woosley & T. Janka 2005; S. E. Woosley & J. S. Bloom 2006). The first evidence for a link between LGRBs and SNe has been well established since the discovery of the energetic core-collapse SN 1998bw associated with the nearby low-luminosity LGRB (LLGRB) 980425 (T. J. Galama et al. 1998; A. Clocchiatti et al. 2011). This connection was further reinforced by the much brighter GRB 030329 and its associated SN 2003dh, whose high-quality spectrum confirmed that GRB–SN associations are a common phenomenon (Y. M. Lipkin et al. 2004; R. Vanderspek et al. 2004). To date, more than 60 SN-associated GRBs (SN-GRBs) have been observed (G. Finneran et al. 2025a). The majority of them are found to be explosions from highly stripped stars with broad spectral features, i.e., broad-line Type Ic supernovae (Ic-BL SNe; M. Modjaz 2011; J. Hjorth & J. S. Bloom 2012; M. Modjaz et al. 2016; D. A. Kann et al. 2024), whereas GRB 111209A is linked to SN 2011kl, whose properties resemble those of superluminous supernovae (SLSNe; J. Greiner et al. 2015; P. A. Mazzali et al. 2016).

However, long-term monitoring of several well-localized, low-redshift LGRBs has revealed no signatures of accompanying SNe. For instance, no clues pointing to a nebular-phase SN are found in the spectrum of the nearby LGRB 051109B ($z = 0.08$; D. A. Perley et al. 2006). Similarly, no SN signature is found in the nearby LGRBs 060505 ($z = 0.0894$; E. O. Ofek et al. 2007) and 060614 ($z = 0.125$; A. Gal-Yam et al. 2006) (J. P. U. Fynbo et al. 2006; D. Guetta & M. Della Valle 2007; R. L. C. Starling et al. 2011). In addition to the possibility that the accompanying SN is too faint to be detected, the viewing-angle effect may also lead to the nondetection of the associated SN (A. M. Soderberg et al. 2006b, 2010; B. Margalit et al. 2018; P. Beniamini et al. 2022; A. Corsi et al. 2023). On the other hand, the seemingly long GRB 211211A was surprisingly found to be associated with a kilonova, which should be produced by compact binary mergers (E. Troja et al. 2022; J. Yang et al. 2022). Moreover, a compact merger origin was hinted for the long-duration GRB 230307A by observations of the James Webb Space Telescope, which found some clues of a kilonova in the spectrum (S. Dichiara et al. 2023; A. J. Levan et al. 2024). These two events pose further challenges to the traditional core-collapse origin of LGRBs.

It is quite possible that LGRBs not associated with any SNe may have a different origin compared with SN-associated LGRBs. At least a subset of SN-less LGRBs, which are characterized by a short, hard initial pulse and a negligible spectral lag, have been suggested to be of compact binary origin, similar to short gamma-ray bursts (SGRBs; N. Gehrels et al. 2006; B. Zhang et al. 2007; L. Caito et al. 2009; E. Troja et al. 2022; A. J. Levan et al. 2024). Recently, H.-Y. Chen & O. Gottlieb (2025) studied the duration of GRBs produced by neutron star mergers. They argued that when the total mass of the two compact stars exceeds $1.36^{+0.08}_{-0.09}$ times of the neutron star Tolman–Oppenheimer–Volkoff mass, the resultant GRB should be a long one. Otherwise, an SGRB would be produced. Alternatively, R. Ruffini et al. (2014) argued that LGRBs could also originate from a tight binary system comprising a carbon–oxygen star and a neutron star, i.e., the so-called binary-driven hypernova model. Moreover, LLGRBs, which have a significantly higher local event rate, may constitute a separate population in SN-GRBs with a distinct origin (A. M. Soderberg et al. 2006a; E. Liang et al. 2007; H. Sun et al. 2015; X. F. Dong et al. 2023). In addition, Q. M. Li et al. (2023) found that SN-GRBs and kilonova-GRBs might have the same radiation mechanism. This is supported by the fact that several of their properties exhibit significant overlap, even though they arise from different origins. Despite extensive efforts, the origin of LGRBs not associated with any SNe still remains an open question.

To explore the different origins of LGRBs with and without an SN association, here we conduct a comprehensive analysis by considering various observational features involving the prompt emission properties and host galaxy properties. The redshift-dependent event rate will also be examined, which could provide a valuable census on the cosmic history of these events (N. M. Lloyd-Ronning et al. 2002; G. Ghirlanda & R. Salvaterra 2022; A. M. Hasan & W. J. Azzam 2024). LGRBs associated with SNe are likely originated from massive stars; thus, a correlation between the LGRB event rate and the star formation rate (SFR) is expected (E. Liang et al. 2007; B. Zhang 2007; S.-W. Wu et al. 2012; H. Yu et al. 2015; X. F. Dong et al. 2022; Q. M. Li et al. 2024a). Establishing such connections is crucial for utilizing GRBs as powerful probes of star formation and stellar death in the distant Universe.

The structure of this paper is organized as follows. Section 2 describes the data acquisition and the samples used for analysis. Section 3 presents the parameter distributions of different LGRB samples, along with the corresponding event rate results. Section 4 is a brief discussion and Section 5 summarizes our main findings. Throughout the paper, a flat ΛCDM Universe with a matter density $\Omega_{\rm m} = 0.315$ and a Hubble constant $H_0 = 67.3\ {\rm km\ s^{-1}\ Mpc^{-1}}$ (Planck Collaboration et al. 2020) are assumed. A solar oxygen abundance of $12 + \log({\rm O/H})_\odot = 8.69$ (M. Asplund et al. 2009) is adopted.

## 2. Data

To conduct our analysis, we collected observational GRB data from the literature and composed two samples, i.e., a sample of GRBs with SN association and a GRB sample without SN association. For comparison, we also constructed two SN samples: one consists of Ic-BL SNe not associated with any GRBs, and the other includes SLSNe also not connected to known GRBs.

### 2.1. GRBs Associated with SNe

GRB 980425 is the first GRB confirmed with an SN association, which is associated with SN 1998bw (T. J. Galama et al. 1998). G. Finneran et al. (2025a) compiled the most complete catalog of GRB and SNe associations to date and made it available as a public application, i.e., the GRBSN webtool.[5] It contains 61 GRBs, including 29 events with spectroscopic evidence supporting an SN association. However, some important parameters such as the peak luminosity of the associated SNe and those related to the host galaxy properties are not directly available in the catalog. Therefore, we have further collected these parameters from the literature to carry out our study.

In our study, we require that the redshift should be available. Fifty-five GRBs in the GRBSN webtool catalog meet this requirement. Especially, although the redshift of GRB 980326 is not directly measured, it was reasonably estimated as $z = 1$ (J. S. Bloom et al. 1999). Two GRBs, i.e., 150518A and 221009A, are excluded from our analysis due to the lack of necessary parameters needed in our study. For example, we do not have duration, isotropic energy, and peak energy for GRB 150518A. Similarly, due to the extreme brightness of GRB 221009A that caused detectors saturated during the burst (R. Pillera et al. 2022; D. Frederiks et al. 2023; W.-L. Zhang et al. 2025), we do not have accurate duration, isotropic energy, and peak energy for it.

The collected data of all the SN-GRBs are presented in Table 1. In the table, columns (1) and (2) list the GRB names and the accompanied SN names; columns (3), (4), (5), and (6) are relevant parameters of the GRBs, most of which are taken from the GRBSN webtool catalog; columns (7), (8), and (9) are parameters of their host galaxies; column (10) is the peak luminosity of the associated SN, and column (11) gives the corresponding references. For simplicity, we denote the 53 SN-GRBs in Table 1 as an $S_{\rm a}$ sample.

Note that the calculation of SN-GRB event rates requires information of redshift and isotropic gamma-ray energy ($E_{\gamma,{\rm iso}}$). GRBs 070419A and 071112C lack the parameter of $E_{\gamma,{\rm iso}}$ and are therefore excluded from the rate estimation. Nonetheless, as other parameters such as $T_{90}$ and the projected offset are available, they are retained in Table 1. The remaining 51 GRBs have complete parameter sets and are included in the event rate calculation.

### 2.2. GRBs Not Associated with SNe: Events Detected before 2004

Before the launch of the Swift satellite, it was difficult to quickly localize the position of a GRB. As a result, the number of GRBs with afterglow being detected is very small, which also means the redshift is not available for most bursts before 2004. In our study, we have collected all the GRBs earlier than 2004 with measured redshift from the literature. They are mainly detected by HETE-2, Ulysses, BeppoSAX, and BATSE satellites. The data are listed in Table 2. Here, the first column is the GRB name; the second column is the power-law (PL) index of the spectrum; columns (3), (4), and (5) are the spectrum parameters derived by adopting the Band function (D. Band et al. 1993); columns (6) and (7) are the gamma-ray fluence and the corresponding energy band; columns (8) and (9) are the gamma-ray peak flux and the corresponding energy band; column (10) is the redshift; and column (11) is $T_{90}$.

The isotropic energy $E_{\gamma,{\rm iso}}$ and peak luminosity $L_{\rm p}$ are calculated by using

$$E_{\gamma,{\rm iso}} = \frac{4\pi D_{\rm L}^2(z) S_\gamma K}{(1+z)}, \tag{1}$$

[5] https://grbsn.watchertelescope.ie/

**Table 1**
Key Parameters of the SN-GRBs

| GRB (1) | SNe (2) | Redshift (3) | $T_{90}$ (s) (4) | $E_p$ (keV) (5) | $E_{\gamma,\mathrm{iso}}$ (erg) (6) | $R_{\mathrm{off}}$ (kpc) (7) | SFR ($M_\odot$ yr$^{-1}$) (8) | $\log(Z/Z_\odot)$[c] (9) | $L_{\mathrm{p,SN}}$ (erg s$^{-1}$) (10) | References (11) |
|---|---|---|---|---|---|---|---|---|---|---|
| 970228 | ⋯ | 0.695 | 56 | $115.04 \pm 37.76$ | $(1.6 \pm 0.12) \times 10^{52}$ | ⋯ | 0.53 | −0.22 | ⋯ | (1, 2, 3, 47, 56) |
| 980326 | ⋯ | 1[a] | ⋯ | $33.8 \pm 18$ | $(4.8 \pm 0.9) \times 10^{51}$[a] | ⋯ | ⋯ | ⋯ | ⋯ | (3, 37, 51, 56) |
| 980425 | 1998bw | 0.00867 | 18 | $55 \pm 21$ | $(8.6 \pm 0.2) \times 10^{47}$ | ⋯ | 0.26 | −0.5 | $1.45 \times 10^{43}$ | (3, 47, 48) |
| 990712 | ⋯ | 0.4331 | 19 | $64.89 \pm 10.47$ | $(6.7 \pm 1.3) \times 10^{51}$ | ⋯ | 2.39 | −0.59 | ⋯ | (3, 47, 56) |
| 991208 | ⋯ | 0.7063 | 60 | $183.44 \pm 18.17$ | $(2.23 \pm 0.18) \times 10^{53}$ | ⋯ | 4.52 | −0.67 | ⋯ | (3, 47, 56) |
| 000911 | ⋯ | 1.0585 | 500 | $901.63 \pm 180.23$ | $(6.7 \pm 1.4) \times 10^{53}$ | 0.66 | 1.57 | −0.59 | ⋯ | (3, 47, 56) |
| 011121 | 2001ke | 0.362 | 47 | $778.27 \pm 194.57$ | $(7.8 \pm 2.1) \times 10^{52}$ | 4.563 | 2.24 | −0.05[b] | $5.9 \times 10^{42}$ | (3, 4, 47, 48, 56) |
| 020405 | ⋯ | 0.68986 | 40 | $209.48 \pm 5.92$ | $(1 \pm 0.09) \times 10^{53}$ | ⋯ | 3.74 | −0.2 | ⋯ | (3, 47, 56) |
| 020903 | ⋯ | 0.2506 | 3.3 | $2.69 \pm 1.43$ | $(2.4 \pm 0.6) \times 10^{49}$ | ⋯ | 2.65 | −0.44 | ⋯ | (3, 47, 53, 56) |
| 021211 | 2002lt | 1.004 | 2.8 | $63.37 \pm 25.95$ | $(1.12 \pm 0.13) \times 10^{52}$ | ⋯ | 3.01 | ⋯ | $7.2 \times 10^{42}$ | (3, 47, 56) |
| 030329 | 2003dh | 0.1685 | 23 | $85.57 \pm 19.68$ | $(1.5 \pm 0.3) \times 10^{52}$ | ⋯ | 0.11 | −0.69 | $1.01 \times 10^{43}$ | (3, 47, 56) |
| 031203 | 2003lw | 0.10536 | 37 | $<200$ | $(8.6 \pm 4) \times 10^{49}$ | 0.132 | 12.68 | −0.64 | $1.26 \times 10^{43}$ | (3, 47, 48) |
| 040924 | ⋯ | 0.858 | 2.39 | $54.9 \pm 18.84$ | $(9.5 \pm 0.9) \times 10^{51}$ | 2.197 | 1.88 | −0.46 | ⋯ | (3, 40, 47, 56) |
| 041006 | ⋯ | 0.716 | 18 | $57.11 \pm 11.66$ | $(3 \pm 0.9) \times 10^{52}$ | 2.57 | 0.34 | ⋯ | ⋯ | (3, 40, 47, 56) |
| 050416A | ⋯ | 0.6528 | 2.4 | $15.19 \pm 2.54$ | $(1 \pm 0.1) \times 10^{51}$ | 0.26 | 4.5 | −0.23 | ⋯ | (3, 40, 56) |
| 050525A | 2005nc | 0.606 | 8.84 | $79.08 \pm 6.23$ | $(2.5 \pm 0.43) \times 10^{52}$ | 0.19 | 0.07 | ⋯ | $4.47 \times 10^{42}$ | (3, 7, 40, 48, 56) |
| 050824 | ⋯ | 0.8281 | 25 | $21.5 \pm 10.5$ | $(2.09 \pm 1.68) \times 10^{51}$ | 3.597 | 1.2 | −0.58 | ⋯ | (3, 40, 46, 47) |
| 060218 | 2006aj | 0.03342 | 2100 | $4.74 \pm 0.29$ | $(5.3 \pm 0.3) \times 10^{49}$ | 0.115 | 0.05 | −0.53 | $6.47 \times 10^{42}$ | (3, 40, 47, 48, 56) |
| 060729 | ⋯ | 0.5428 | 115 | $>50$ | $(1.6 \pm 0.6) \times 10^{52}$ | 2.149 | 0.96 | ⋯ | ⋯ | (3, 40, 46, 47) |
| 060904B | ⋯ | 0.7029 | 192 | $163 \pm 31$ | $(2.4 \pm 0.2) \times 10^{52}$ | ⋯ | ⋯ | ⋯ | ⋯ | (3, 47) |
| 070419A | ⋯ | 0.9705 | 116 | … | ⋯ | ⋯ | ⋯ | ⋯ | ⋯ | (3) |
| 071112C | ⋯ | 0.823 | 15 | … | ⋯ | 1.533 | ⋯ | ⋯ | ⋯ | (8, 40) |
| 080319B | ⋯ | 0.9371 | 124.86 | $650.97 \pm 33.56$ | $(1.14 \pm 0.09) \times 10^{54}$ | 1.697 | ⋯ | ⋯ | ⋯ | (3, 40, 47, 56) |
| 081007 | 2008hw | 0.5295 | 9.01 | $61 \pm 15$ | $(1.5 \pm 0.4) \times 10^{51}$ | 0.7 | ⋯ | ⋯ | $1.4 \times 10^{43}$ | (3, 40, 47, 48) |
| 090618 | ⋯ | 0.54 | 113.34 | $211 \pm 22$ | $(2.57 \pm 0.5) \times 10^{53}$ | 4.402 | ⋯ | ⋯ | ⋯ | (3, 40, 47) |
| 091127 | 2009nz | 0.49044 | 7.42 | $35.5 \pm 1.5$ | $(1.5 \pm 0.2) \times 10^{52}$ | 1.335 | 0.37 | −0.62 | $1.2 \times 10^{43}$ | (3, 40, 46, 47, 48) |
| 100316D | 2010bh | 0.059 | 1300 | $26 \pm 16$ | $>0.0059 \times 10^{52}$ | ⋯ | 0.14 | −0.39 | $5.67 \times 10^{42}$ | (3, 47, 48, 49) |
| 100418A | ⋯ | 0.6239 | 8 | $29 \pm 2$ | $(9.9 \pm 6.3) \times 10^{50}$ | ⋯ | 4.2 | −0.17 | ⋯ | (3) |
| 111211A | ⋯ | 0.478 | 25 | … | $1.49 \times 10^{52}$ | ⋯ | 0.12 | ⋯ | ⋯ | (3, 37) |
| 101219B | 2010ma | 0.55185 | 51 | $70 \pm 8$ | $(4.2 \pm 0.5) \times 10^{51}$ | ⋯ | 16 | ⋯ | $1.5 \times 10^{43}$ | (3, 47, 48) |
| 101225A | ⋯ | 0.847 | 7000 | $38 \pm 20$ | $(1.2 \pm 0.3) \times 10^{52}$ | 0.124 | ⋯ | ⋯ | ⋯ | (3, 37, 50) |
| 111209A | 2011kl | 0.677 | 25000 | $310 \pm 89$ | $(5.82 \pm 0.73) \times 10^{53}$ | 0.079 | 0.35 | −0.74 | $2.91 \times 10^{43}$ | (3, 9, 10, 37, 48, 50) |
| 111228A | ⋯ | 0.7163 | 101.2 | $58.4 \pm 6.9$ | $(4.2 \pm 0.6) \times 10^{52}$ | ⋯ | 0.32 | ⋯ | ⋯ | (3, 8) |
| 120422A | 2012bz | 0.28253 | 5.4 | $<72$ | $(2.4 \pm 0.8) \times 10^{50}$ | 8.167 | 1.38 | −0.3 | $1.48 \times 10^{43}$ | (3, 37, 48) |
| 120714B | 2012eb | 0.3984 | 159 | $49.3 \pm 30.7$ | $(8 \pm 2) \times 10^{50}$ | ⋯ | 0.27 | −0.3 | $6.2 \times 10^{42}$ | (3, 8, 46, 48, 57) |
| 120729A | ⋯ | 0.8 | 71.5 | $310.6 \pm 31.6$ | $(2.3 \pm 1.5) \times 10^{52}$ | ⋯ | 6 | ⋯ | ⋯ | (3) |
| 130215A | 2013ez | 0.597 | 65.7 | $155 \pm 63$ | $(3.1 \pm 1.6) \times 10^{52}$ | ⋯ | ⋯ | ⋯ | $8.7 \times 10^{42}$ | (3, 15, 48) |
| 130427A | 2013cq | 0.3399 | 160 | $1028 \pm 50$ | $(8.1 \pm 1) \times 10^{53}$ | 4.013 | 0.34 | −0.12 | $9.12 \times 10^{42}$ | (3, 11, 12, 13, 40, 48) |
| 130702A | 2013dx | 0.145 | 59 | $<17.47$ | $(6.4 \pm 1.3) \times 10^{50}$ | 1.53 | 0.05 | −0.5 | $1.08 \times 10^{43}$ | (3, 14, 37, 48, 55, 61) |
| 130831A | 2013fu | 0.4791 | 32.5 | $67 \pm 4$ | $(4.6 \pm 0.2) \times 10^{51}$ | ⋯ | ⋯ | ⋯ | $6.9 \times 10^{42}$ | (3, 8, 15, 48) |
| 140606B | iPTF14bfu | 0.384 | 22.78 | $554 \pm 165$ | $(3.47 \pm 0.2) \times 10^{51}$ | ⋯ | 0.06 | ⋯ | $2.3 \times 10^{43}$ | (3, 47, 52, 62) |
| 150818A | ⋯ | 0.282 | 123.3 | $128 \pm 13$ | $(1 \pm 0.2) \times 10^{51}$ | ⋯ | ⋯ | ⋯ | ⋯ | (3) |
| 161219B | 2016jca | 0.1475 | 7 | $62.9^{+47}_{-19.9}$ | $8.5^{+8.46}_{-3.75} \times 10^{49}$ | ⋯ | ⋯ | ⋯ | $1.04 \times 10^{43}$ | (16, 48, 54) |
| 171010A | 2017htp | 0.33 | 160 | $230 \pm 0$ | $2 \times 10^{53}$ | ⋯ | ⋯ | ⋯ | $2.1 \times 10^{43}$ | (17, 48) |
| 171205A | 2017iuk | 0.037 | 189.4 | $125^{+141}_{-37}$ | $(2.18 \pm 0.63) \times 10^{49}$ | 5.81 | ⋯ | ⋯ | $6.5 \times 10^{42}$ | (18, 40, 43, 48) |
| 180728A | 2018fip | 0.117 | 8.68 | $142 \pm 20$ | $(2.33 \pm 0.1) \times 10^{51}$ | ⋯ | ⋯ | ⋯ | $(6 \times 10^{42}$ | (19, 32, 48) |

**Table 1**
(Continued)

| GRB<br>(1) | SNe<br>(2) | Redshift<br>(3) | $T_{90}$<br>(s)<br>(4) | $E_p$<br>(keV)<br>(5) | $E_{\gamma,iso}$<br>(erg)<br>(6) | $R_{off}$<br>(kpc)<br>(7) | SFR<br>($M_\odot$ yr$^{-1}$)<br>(8) | $\log(Z/Z_\odot)$[c]<br>(9) | $L_{p,SN}$<br>(erg s$^{-1}$)<br>(10) | References<br>(11) |
|---|---|---|---|---|---|---|---|---|---|---|
| 190114C | 2019jrj | 0.4245 | 362 | $998.6 \pm 11.9$ | $(3 \times 10^{53}$ | ⋯ | ⋯ | ⋯ | $5.8 \times 10^{42}$ | (20, 21, 31, 33, 48) |
| 190829A | 2019oyw | 0.0785 | 58.2 | $130 \pm 20$ | $<2 \times 10^{50}$ | ⋯ | ⋯ | ⋯ | $6.27 \times 11^{42}$ | (22, 41, 44, 48) |
| 200826A | ⋯ | 0.71 | 0.65 | 67.25 | $4.7 \times 10^{51}$ | ⋯ | ⋯ | ⋯ | ⋯ | (25, 34) |
| 201015A | ⋯ | 0.426 | 9.78 | $14 \pm 6$ | $(1.1 \pm 0.02) \times 10^{50}$ | ⋯ | ⋯ | ⋯ | ⋯ | (23, 24, 38, 42, 44) |
| 211023A | ⋯ | 0.36 | ⋯ | $96 \pm 5$ | $(7.45 \pm 0.05) \times 10^{52}$ | ⋯ | ⋯ | ⋯ | ⋯ | (26, 35) |
| 220219B | ⋯ | 0.293 | ⋯ | $39 \pm 27$ | $(2.75 \pm 0.36) \times 10^{51}$ | ⋯ | ⋯ | ⋯ | ⋯ | (27, 28, 36, 39) |
| 230812B | 2023pel | 0.36 | 3.26 | $411.35^{+8.21}_{-8.18}$ | $(8.74 \pm 1.61) \times 10^{52}$ | ⋯ | ⋯ | ⋯ | $5.75 \times 10^{42}$ | (30, 58, 59, 60) |

**Notes.** Columns (1) and (2) list the names of the GRBs associated with SNe and the corresponding SNe, respectively. Column (3) provides the redshift ($z$) of each GRB. Columns (4), (5), and (6) detail the gamma-ray properties of the associated GRBs: the duration ($T_{90}$), the peak energy ($E_{peak}$), and the isotropic energy ($E_{\gamma,iso}$). Columns (7), (8), and (9) report the host galaxy properties of the GRBs: the projected offset ($R_{off}$), the star formation rate (SFR), and the metallicity in solar units ($\log(Z/Z_\odot)$).

[a] Assume the redshift is 1.

[b] S. Savaglio et al. (2009) suggests that it could also be −1.19.

[c] Photospheric solar oxygen abundance is 8.69.

**References.** (1) D. E. Reichart (1999); (2) T. J. Galama et al. (2000); (3) Z. Cano et al. (2017b); (4) P. M. Garnavich et al. (2003); (5) J. Gorosabel et al. (2005); (6) G. Pugliese et al. (2005); (7) M. Della Valle et al. (2006); (8) S. Klose et al. (2019); (9) B. Gendre et al. (2013); (10) D. A. Kann et al. (2019); (11) M. De Pasquale et al. (2016); (12) A. Melandri et al. (2014); (13) D. Xu et al. (2013); (14) V. L. Toy et al. (2016); (15) Z. Cano et al. (2014); (16) C. Ashall et al. (2019); (17) A. Melandri et al. (2019); (18) J. Wang et al. (2018b); (19) Y. Wang et al. (2019); (20) J. D. Gropp et al. (2019); (21) A. Melandri et al. (2022); (22) Y. D. Hu et al. (2021); (23) A. Pozanenko et al. (2020); (24) L. Izzo et al. (2020); (25) T. Ahumada et al. (2021); (26) A. Rossi et al. (2022b); (27) A. Rossi et al. (2022a); (28) A. Y. Q. Ho et al. (2020); (29) M. D. Fulton et al. (2023); (30) J. F. Agui Fernandez et al. (2023); (31) H. A. Krimm et al. (2019); (32) D. Frederiks et al. (2018); (33) R. Hamburg et al. (2019); (34) D. Svinkin et al. (2020); (35) P. Minaev et al. (2021); (36) A. Tsvetkova et al. (2022); (37) Y. Li et al. (2016); (38) C. Fletcher et al. (2020); (39) S. Belkin et al. (2022); (40) J. S. Wang et al. (2016); (41) S. Lesage et al. (2019); (42) M. Patel et al. (2023); (43) V. D'Elia et al. (2018); (44) S. Dichiara et al. (2022); (45) L.-L. Zhang et al. (2023); (46) T. Krühler et al. (2015); (47) https://www.grbhosts.org; (48) Y. Aimuratov et al. (2023a); (49) C. C. Thöne et al. (2011); (50) A. J. Levan et al. (2014); (51) G. Ghirlanda et al. (2004); (52) L. P. Singer (2015); (53) L.-X. Li (2006); (54) Z. Cano et al. (2017a); (55) Y. Aimuratov et al. (2023b); (56) L. Amati et al. (2008); (57) H. Dereli et al. (2017); (58) T. Hussenot-Desenonges et al. (2024); (59) A. K. Ror et al. (2024b); (60) L. M. Roman Aguilar et al. (2025); (61) L. Amati et al. (2013); (62) Z. Cano et al. (2015).

**Table 2**
Prompt Emission Parameters of SN-less GRBs before Swift Launch

| GRB | $\Gamma$ | $\alpha$ | $\beta$ | $E_p$ (keV) | $S_\gamma$ (erg $cm^{-2}$) | Band (keV) | $F_p$ (erg $cm^{-2}$ $s^{-1}$) | Band (keV) | $z$ | $T_{90}$ (s) | $K$ | $E_{\gamma,iso}$ (erg) | $L_p$ (erg $s^{-1}$) |
|---|---|---|---|---|---|---|---|---|---|---|---|---|---|
| (1) | (2) | (3) | (4) | (5) | (6) | (7) | (8) | (9) | (10) | (11) | (12) | (13) | (14) |
| 981226A | −2.38 | ⋯ | ⋯ | ⋯ | $(5.80 \pm 0.8) \times 10^{-7}$ | 40–700 | $(3.7 \pm 0.7) \times 10^{-8}$ | 40−700 | 1.11 | 17 | 7.89 | $(1.51 \pm 0.21) \times 10^{52}$ | $(2.04 \pm 0.39) \times 10^{51}$ |
| 991208A | −1.68 | ⋯ | ⋯ | ⋯ | $4.90 \times 10^{-5}$ | 25–100 | $5.1 \times 10^{-6}$ | 25−100 | 0.7063 | 60 | 9.73 | $6.36 \times 10^{53}$ | $1.13 \times 10^{53}$ |
| 000210A | −1.44 | ⋯ | ⋯ | ⋯ | $(4.80 \pm 0.37) \times 10^{-5}$ | 40–700 | $(1.65 \pm 0.13) \times 10^{-5}$ | 40−700 | 0.846 | 9 | 3.92 | $3.62 \pm 0.28) \times 10^{53}$ | $(2.29 \pm 0.18) \times 10^{53}$ |
| 000301C | −1.79 | ⋯ | ⋯ | ⋯ | $(8.35 \pm 0.53) \times 10^{-6}$ | 40–700 | $(1.13 \pm 0.09) \times 10^{-7}$ | 40−700 | 2.0404 | 87 | 2.62 | $(2.27 \pm 0.14) \times 10^{53}$ | $(9.33 \pm 0.74) \times 10^{51}$ |
| 000926A | −0.78 | ⋯ | ⋯ | ⋯ | $(2.60 \pm 0.7) \times 10^{-7}$ | 40–700 | $(2.16 \pm 0.87) \times 10^{-7}$ | 40−700 | 2.0379 | 1.3 | 6.82 | $(1.83 \pm 0.49) \times 10^{52}$ | $(4.63 \pm 1.86) \times 10^{52}$ |
| 010222A | −1.57 | ⋯ | ⋯ | ⋯ | $(9.25 \pm 0.28) \times 10^{-5}$ | 40–700 | $(8.6 \pm 0.2) \times 10^{-6}$ | 40−700 | 1.47688 | 74 | 2.94 | $(1.56 \pm 0.05) \times 10^{54}$ | $(3.59 \pm 0.08) \times 10^{53}$ |
| 011121A | −1.22 | ⋯ | ⋯ | ⋯ | $(9.83 \pm 0.77) \times 10^{-5}$ | 40–700 | $(6.59 \pm 0.53) \times 10^{-6}$ | 40−700 | 0.36 | 47 | 7.01 | $(2.28 \pm 0.18) \times 10^{53}$ | $(2.08 \pm 0.17) \times 10^{52}$ |
| 011211A | −1 | ⋯ | ⋯ | ⋯ | $(1.35 \pm 0.22) \times 10^{-6}$ | 40–700 | $(3.5 \pm 0.8) \times 10^{-8}$ | 40−700 | 2.14 | 51 | 4.82 | $(7.35 \pm 1.2) \times 10^{52}$ | $(5.98 \pm 1.37) \times 10^{51}$ |
| 020405A | −2 | ⋯ | ⋯ | ⋯ | $(4.22 \pm 0.24) \times 10^{-5}$ | 40–700 | $(3.21 \pm 0.19) \times 10^{-6}$ | 40−700 | 0.6908 | 40 | 3.22 | $(1.73 \pm 0.1) \times 10^{53}$ | $(2.23 \pm 0.13) \times 10^{52}$ |
| 030323A | −1.62 | ⋯ | ⋯ | ⋯ | $(1.23 \pm 0.368) \times 10^{-6}$ | 2–400 | $(1.36 \pm 0.74) \times 10^{-7}$ | 2−400 | 3.3736 | 26 | 2.17 | $(6.52 \pm 1.95) \times 10^{52}$ | $(3.16 \pm 1.73) \times 10^{52}$ |
| 970508A | ⋯ | −1.71 | −2.2 | 79 | $(1.80 \pm 0.3) \times 10^{-6}$ | 40–700 | $(3.4 \pm 0.1) \times 10^{-7}$ | 40−700 | 0.835 | 14 | 0.9 | $(3.05 \pm 0.51) \times 10^{51}$ | $(1.06 \pm 0.03) \times 10^{51}$ |
| 970828A | ⋯ | −0.45 | −2.06 | 379 | $9.60 \times 10^{-5}$ | 20–2000 | $(5.93 \pm 0.34) \times 10^{-6}$ | 30−10,000 | 0.9578 | 146.59 | 1.62 | $3.85 \times 10^{53}$ | $(4.66 \pm 0.27) \times 10^{52}$ |
| 971214A | ⋯ | −0.76 | −2.7 | 154 | $(8.80 \pm 0.9) \times 10^{-6}$ | 40–700 | $(6.8 \pm 0.7) \times 10^{-7}$ | 40−700 | 3.42 | 6 | 0.18 | $(3.88 \pm 0.4) \times 10^{52}$ | $(1.33 \pm 0.14) \times 10^{52}$ |
| 980329A | ⋯ | −0.64 | −2.2 | 236 | $(6.50 \pm 0.5) \times 10^{-5}$ | 40–700 | $(3.1 \pm 0.1) \times 10^{-6}$ | 40−700 | 3.5 | 19 | 0.09 | $(1.46 \pm 0.11) \times 10^{53}$ | $(3.12 \pm 0.1) \times 10^{52}$ |
| 980613A | ⋯ | −1.43 | −2.7 | 92.6 | $(1.00 \pm 0.2) \times 10^{-6}$ | 40–700 | $(1.6 \pm 0.4) \times 10^{-7}$ | 40−700 | 1.0969 | 42 | 0.79 | $(2.54 \pm 0.51) \times 10^{51}$ | $(8.51 \pm 2.13) \times 10^{50}$ |
| 980703A | ⋯ | −1.2 | −1.93 | 281 | $(3.98 \pm 0.09) \times 10^{-5}$ | 20–2000 | $(2.64 \pm 0.51) \times 10^{-6}$ | 30−10,000 | 0.966 | 76 | 0.28 | $(2.78 \pm 0.06) \times 10^{52}$ | $(3.63 \pm 0.7) \times 10^{51}$ |
| 990123A | ⋯ | −0.89 | −2.45 | 780 | $(3.00 \pm 0.4) \times 10^{-4}$ | 40–700 | $(1.7 \pm 0.5) \times 10^{-5}$ | 40−700 | 1.6004 | 61 | 0.1 | $(1.91 \pm 0.25) \times 10^{53}$ | $(2.82 \pm 0.83) \times 10^{52}$ |
| 990506A | ⋯ | −1.17 | −2.62 | 341 | $(1.69 \pm 0.01) \times 10^{-4}$ | 20–2000 | $(9.36 \pm 0.2) \times 10^{-6}$ | 30−10,000 | 1.30658 | 129 | 0.2 | $(1.57 \pm 0.01) \times 10^{53}$ | $(2.01 \pm 0.04) \times 10^{52}$ |
| 990510A | ⋯ | −1.23 | −2.7 | 161 | $(1.90 \pm 0.2) \times 10^{-5}$ | 40–700 | $(2.47 \pm 0.21) \times 10^{-6}$ | 40−700 | 1.619 | 57 | 0.41 | $(5.33 \pm 0.56) \times 10^{52}$ | $(1.81 \pm 0.15) \times 10^{52}$ |
| 990705A | ⋯ | −1.05 | −2.2 | 188 | $(7.50 \pm 0.8) \times 10^{-5}$ | 40–700 | $(3.7 \pm 0.1) \times 10^{-6}$ | 40−700 | 0.8424 | 32 | 0.41 | $(5.82 \pm 0.62) \times 10^{52}$ | $(5.29 \pm 0.14) \times 10^{51}$ |
| 991216A | ⋯ | −1.2 | −2.22 | 382 | $(1.75 \pm 0.005) \times 10^{-4}$ | 20–2000 | $(6.14 \pm 0.12) \times 10^{-5}$ | 30−10,000 | 1.02 | 45 | 0.21 | $1.04 \times 10^{53}$ | $(7.37 \pm 0.14) \times 10^{52}$ |
| 000131A | ⋯ | −0.91 | −2.02 | 168 | $4.18 \times 10^{-5}$ | 20–2000 | $(2.67 \pm 0.41) \times 10^{-6}$ | 30−10,000 | 4.5 | 50 | 0.14 | $2.29 \times 10^{53}$ | $(8.03 \pm 1.23) \times 10^{52}$ |
| 000418*A* | ⋯ | −1 | −2.3 | 134 | $(2.00 \pm 0.7) \times 10^{-7}$ | 40–700 | $(1 \pm 0.36) \times 10^{-7}$ | 40−700 | 1.1181 | 2 | 0.49 | $1.7 \times 10^{52}$ | $(3.46 \pm 1.25) \times 10^{50}$ |
| 010921A | ⋯ | −1.55 | −2.3 | 88.63 | $(1.34 \pm 0.17) \times 10^{-5}$ | 40–700 | $(1.04 \pm 0.16) \times 10^{-6}$ | 40−700 | 0.4509 | 22 | 0.91 | $(6.42 \pm 0.81) \times 10^{51}$ | $(7.23 \pm 1.11) \times 10^{50}$ |
| 020124A | ⋯ | −0.79 | −2.6 | 86.93 | $(8.10 \pm 0.886) \times 10^{-6}$ | 2–400 | $(4.57 \pm 0.86) \times 10^{-7}$ | 2−400 | 3.198 | 51.17 | 0.28 | $(5.08 \pm 0.56) \times 10^{52}$ | $(1.2 \pm 0.23) \times 10^{52}$ |
| 020127A | ⋯ | −1.03 | −2.3 | 104 | $(2.72 \pm 0.443) \times 10^{-6}$ | 2–400 | $(3.73 \pm 0.69) \times 10^{-7}$ | 2−400 | 1.9 | 6.99 | 0.36 | $(8.96 \pm 1.46) \times 10^{51}$ | $(3.56 \pm 0.66) \times 10^{51}$ |
| 020305A | ⋯ | −0.86 | −2.3 | 143 | $1.04 \times 10^{-5}$ | 30–400 | $4.7 \times 10^{-7}$ | 25−100 | 2.8 | 39.06 | 0.56 | $1.04 \times 10^{53}$ | $1.78 \times 10^{52}$ |
| 020813A | ⋯ | −0.94 | −1.57 | 142.1 | $(9.79 \pm 0.127) \times 10^{-5}$ | 2–400 | $(1.96 \pm 0.13) \times 10^{-6}$ | 2−400 | 1.255 | 87.34 | 0.35 | $(1.42 \pm 0.02) \times 10^{53}$ | $(6.44 \pm 0.41) \times 10^{51}$ |
| 020819B | ⋯ | −0.9 | −2 | 50 | $7.80 \times 10^{-6}$ | 25–100 | $(7 \pm 0) \times 10^{-7}$ | 25−100 | 0.41 | 20 | 1.45 | $4.89 \times 10^{51}$ | $6.19 \times 10^{50}$ |
| 021004A | ⋯ | −1.01 | −2.3 | 79.79 | $(2.55 \pm 0.685) \times 10^{-6}$ | 2–400 | $(1.06 \pm 0.2) \times 10^{-7}$ | 2−400 | 2.3351 | 48.94 | 0.39 | $(1.3 \pm 0.35) \times 10^{52}$ | $(1.8 \pm 0.33) \times 10^{51}$ |
| 030115A | ⋯ | −1.28 | −2.3 | 82.79 | $(2.31 \pm 0.398) \times 10^{-6}$ | 2–400 | $(2.64 \pm 0.45) \times 10^{-7}$ | 2−400 | 2 | 20.33 | 0.42 | $(9.79 \pm 1.69) \times 10^{51}$ | $(3.37 \pm 0.57) \times 10^{51}$ |
| 030226A | ⋯ | −0.89 | −2.3 | 97.12 | $(5.61 \pm 0.693) \times 10^{-6}$ | 2–400 | $(1.32 \pm 0.28) \times 10^{-7}$ | 2−400 | 1.986 | 76.23 | 0.36 | $(1.97 \pm 0.24) \times 10^{52}$ | $(1.39 \pm 0.29) \times 10^{51}$ |
| 030328A | ⋯ | −1.14 | −2.09 | 126.3 | $(3.70 \pm 0.14) \times 10^{-5}$ | 2–400 | $(5.4 \pm 0.39) \times 10^{-7}$ | 2−400 | 1.5216 | 138.27 | 0.37 | $(8.21 \pm 0.31) \times 10^{52}$ | $(3.03 \pm 0.22) \times 10^{51}$ |
| 030429A | ⋯ | −1.12 | −2.3 | 35.04 | $(8.54 \pm 1.48) \times 10^{-7}$ | 2–400 | $(8.19 \pm 1.71) \times 10^{-8}$ | 2−400 | 2.658 | 12.95 | 0.6 | $(8.47 \pm 1.47) \times 10^{51}$ | $(2.97 \pm 0.62) \times 10^{51}$ |
| 030528A | ⋯ | −1.33 | −2.65 | 31.84 | $(1.19 \pm 0.076) \times 10^{-5}$ | 2–400 | $(3.39 \pm 0.3) \times 10^{-7}$ | 2−400 | 0.782 | 62.8 | 0.67 | $(1.31 \pm 0.08) \times 10^{52}$ | $(6.66 \pm 0.58) \times 10^{50}$ |
| 041219A | ⋯ | −1.48 | −2.01 | 156 | $(8.67 \pm 0.054) \times 10^{-5}$ | 20–200 | $3.6 \times 10^{-6}$ | 20−200 | 0.31 | 520 | 0.85 | $(1.78 \pm 0.01) \times 10^{52}$ | $(9.69 \pm 0.39) \times 10^{50}$ |

**Notes.** Column (1) lists the names of the GRBs. Column (2) provides the spectral index obtained from a PL fit to the prompt emission spectrum. Columns (3), (4), and (5) present the spectral parameters of the Band function: the low-energy index ($\alpha$), high-energy index ($\beta$), and peak energy ($E_p$), respectively. Column (6) gives the gamma-ray fluence ($S_\gamma$) in units of erg $cm^{-2}$, while column (7) indicates the corresponding observational energy band in keV. Column (8) reports the 1 s resolution gamma-ray peak flux ($F_p$) in units of erg $cm^{-2}$ $s^{-1}$, and column (9) provides the corresponding observational energy band. Column (10) lists the redshift of each GRB. Column (11) shows the duration of the burst. Column (12) gives the $K$-correction factor. Finally, columns (13) and (14) present the isotropic-equivalent gamma-ray energy ($E_{\gamma,iso}$) and the peak luminosity ($L_p$), respectively, both calculated in the rest-frame 15–150 keV band.
All relevant data in the first 11 columns are taken from Y. Li et al. (2016); D. Turpin et al. (2016); J. S. Wang et al. (2016).

and

$$L_{\rm p} = 4\pi D_{\rm L}^2(z) F_{\rm p} K, \tag{2}$$

where $S_\gamma$ is the gamma-ray fluence (usually in units of erg cm$^{-2}$) and $F_{\rm p}$ is the peak flux (usually in units of erg cm$^{-2}$ s$^{-1}$). $D_{\rm L}(z)$ is the luminosity distance given by

$$D_{\rm L}(z) = \frac{c}{H_0}(1+z)\int_0^z \frac{dz}{\sqrt{1-\Omega_m+\Omega_m(1+z)^3}}. \tag{3}$$

$K$ is the $K$-correction factor accounting for the bolometric energy distortion induced in transforming from the lab frame to the GRB rest frame, which is

$$K = \frac{\int_{15/(1+z)}^{150/(1+z)} EN(E)dE}{\int_{E_{\rm min}}^{E_{\rm max}} EN(E)dE}. \tag{4}$$

Here $E_{\rm max}$ and $E_{\rm min}$ are the observational energy range of photons as listed in column (7) of Table 2, and $N(E)$ denotes the photon spectrum of GRBs.

As for the spectrum of the prompt gamma-ray emission, there are mainly two cases. First, the spectrum of some GRBs is best fitted with a PL function, i.e.,

$$N(E) = AE^{-\Gamma}, \tag{5}$$

where $\Gamma$ is the photon index. However, other GRBs are best fitted with the so-called "Band function," which is

$$N(E) = \begin{cases} A\left(\frac{E}{100\ {\rm keV}}\right)^{\alpha}\exp\left(-\frac{E}{E_0}\right), E < (\alpha-\beta)E_0, \\ A\left(\frac{E}{100\ {\rm keV}}\right)^{\beta}\left[\frac{E_0(\alpha-\beta)}{100\ {\rm keV}}\right]^{\alpha-\beta}\exp(\beta-\alpha), E \geqslant (\alpha-\beta)E_0, \end{cases} \tag{6}$$

where $\alpha$ is the low-energy photon index, $\beta$ is the high-energy photon index, and $E_0$ is the break energy. The peak energy ($E_{\rm p}$) in the $E^2N(E)$ energy spectrum can be calculated as $E_{\rm p} = (2+\alpha)E_0$. The best-fit spectrum parameters for these two cases are given in Table 2 correspondingly. Note that a very small portion of GRBs are best fitted with a cutoff power-law (CPL) function, which is essentially the first half of Equation (6). For these events, we adopt $\beta = -2.3$ in calculating their luminosity and energy, which is the peak value of the $\beta$ distribution derived from multiple GRB samples (R. D. Preece et al. 2000; D. Gruber et al. 2014; Y. Li et al. 2016).

Using the above equations, we have calculated the $K$-correction factor for each burst. The 15–150 keV isotropic energy ($E_{\gamma,\rm iso}$) and peak luminosity ($L_{\rm p}$) are also calculated in the cosmological rest frame. The results are presented in columns (12), (13), and (14) of Table 2.

### 2.3. GRBs Not Associated with SNe: Swift Bursts

The launch of the Swift satellite (N. Gehrels et al. 2004) leads to a significant increase in the number of well-localized GRBs with redshift being measured. Table 3 presents 337 such LGRBs observed by the Burst Alert Telescope (BAT) on board Swift[6] between 2005 January and 2024 October. Note that no evidence of SN association is observed for all these events in Table 3. However, note that due to the narrowness of the energy band of Swift-BAT (15–150 keV; T. Sakamoto et al. 2011), the spectra of most of these events can only be fitted by a simple PL model, which gives the spectral parameters of $\Gamma$, $S_\gamma$, and $F_{\rm p}$. A small fraction of them ($\sim$15%) can be fitted by using the CPL model, which further gives $\alpha$, $E_{\rm p}$, $S_\gamma$, and $F_{\rm p}$.

Table 3 presents the data of 337 Swift GRBs for which no evidence of SN association was observed. Columns (2)–(6) give the values of spectral index, $E_{\rm p}$, $S_\gamma$, and $F_{\rm p}$ obtained from a PL or CPL fitting. Columns (7) and (8) are the redshift and $T_{90}$, respectively. Columns (9), (10), and (11) are the $K$-correction factor and the rest-frame 15–150 keV band, $E_{\gamma,\rm iso}$, and $L_{\rm p}$ parameters. For simplicity, we denote all the GRBs in Tables 2 and 3, which are not associated with any SNe, as the $S_{\rm w}$ sample. Note that the host galaxy is identified only for a small portion of GRBs. We list those GRBs (still without an SN association) with an identified host galaxy in Table 4, which presents some key parameters such as the projected offset ($R_{\rm off}$) and SFR and metallicity ($\log(Z/Z_\odot)$) of the host galaxies.

### 2.4. Supernovae

Thanks to the successful operation of the Swift satellite, the number of SN-GRBs has tripled, compared with that of the pre-Swift era. It is interesting to note that all the currently known SNe associated with LGRBs are Ic-BL SNe (M. Modjaz et al. 2016). We thus have also collected a sample of Ic-BL SNe, i.e., the SN$_{\rm IcBL}$ sample, as listed in Table 5. To be more specific, those Ic-BL SNe that are confirmed to be associated with a GRB are removed from our SN sample. Redshift and $r$-band peak luminosity ($L_{\rm r,p}$) of each SN are included in the table.

Interestingly, the ultralong gamma-ray burst (ULGRB), GRB 111209A, is thought to be associated with the luminous SN, SN 2011kl (J. Lin et al. 2020). This peculiarity points to possible connection between long bursts and SLSNe. Therefore, we have also collected 265 SLSNe to form a sample denoted as SN$_{\rm SL}$. The relevant parameters of SN$_{\rm SL}$, such as redshift and $L_{\rm r,p}$, are taken from S. Gomez et al. (2024). Comparison between Ic-BL SNe, SLSNe, and SN-GRBs could shed new light on the nature of LGRBs.

## 3. Results

We have compared GRBs with and without associated SNe in terms of both their prompt emission features and host galaxy properties. Their difference in the Amati relation, redshift–$T_{90}$ distribution, projected offset in their host galaxy, as well as the SFR and metallicity of their host galaxy are investigated. The evolution of the comoving event rate density between the SN-GRBs and SNe-less GRBs is also studied.

### 3.1. Amati Relation

Figure 1 shows the distribution of the GRBs on the $E_{\rm p}$–$E_{\gamma,\rm iso}$ plane. These two parameters are closely correlated, which is the so-called Amati relation (L. Amati et al. 2008), i.e.,

$$E_{\rm p,i} = kE_{\gamma,\rm iso}^{\rm m}, \tag{7}$$

where $E_{\rm p,i} = E_{\rm p}(1+z)$ is the rest-frame peak energy of the prompt emission and $E_{\gamma,\rm iso}$ is the rest-frame isotropic gamma-ray energy. The original Amati relation obtained by L. Amati et al. (2008) is also shown in Figure 1, which has a PL index of $m = 0.57 \pm 0.01$. We see that most GRBs without SN

[6] Swift GRB table: https://swift.gsfc.nasa.gov/archive/grb_table/.

**Table 3**
Prompt Emission Parameters of SN-less Swift GRBs

| GRB | $\Gamma$ | $\alpha$ | $E_{\rm p}$ (keV) | $S_\gamma$ (erg cm$^{-2}$) | $F_{\rm p}$ (erg cm$^{-2}$ s$^{-1}$) | $z$ | $T_{90}$ (s) | $K$ | $E_{\gamma,\rm iso}$ (erg) | $L_{\rm p}$ (erg s$^{-1}$) |
|---|---|---|---|---|---|---|---|---|---|---|
| (1) | (2) | (3) | (4) | (5) | (6) | (7) | (8) | (9) | (10) | (11) |
| 050126 | −1.34 | ⋯ | ⋯ | $(8.38 \pm 0.8) \times 10^{-7}$ | $(5.78 \pm 1.38) \times 10^{-8}$ | 1.29 | 24.8 | 0.58 | $(2.15 \pm 0.2) \times 10^{51}$ | $(3.39 \pm 0.81) \times 10^{50}$ |
| 050223 | −1.85 | ⋯ | ⋯ | $(6.36 \pm 0.65) \times 10^{-7}$ | $(4.51 \pm 1.05) \times 10^{-8}$ | 0.5915 | 22.5 | 0.93 | $(5.5 \pm 0.56) \times 10^{50}$ | $(6.21 \pm 1.44) \times 10^{49}$ |
| 050315 | −2.11 | ⋯ | ⋯ | $(3.22 \pm 0.15) \times 10^{-6}$ | $(1.14 \pm 0.13) \times 10^{-7}$ | 1.949 | 95.6 | 1.13 | $(3.46 \pm 0.16) \times 10^{52}$ | $(3.6 \pm 0.41) \times 10^{51}$ |
| 050318 | −1.9 | ⋯ | ⋯ | $(1.08 \pm 0.08) \times 10^{-6}$ | $(2.02 \pm 0.13) \times 10^{-7}$ | 1.44 | 32 | 0.91 | $(5.39 \pm 0.38) \times 10^{51}$ | $(2.47 \pm 0.16) \times 10^{51}$ |
| 050319 | −2.02 | ⋯ | ⋯ | $(1.31 \pm 0.15) \times 10^{-6}$ | $(9.27 \pm 1.28) \times 10^{-8}$ | 3.24 | 152.5 | 1.03 | $(3.08 \pm 0.35) \times 10^{52}$ | $(9.25 \pm 1.28) \times 10^{51}$ |
| 050401 | −1.4 | ⋯ | ⋯ | $(8.22 \pm 0.31) \times 10^{-6}$ | $(8.49 \pm 0.73) \times 10^{-7}$ | 2.9 | 33.3 | 0.44 | $(6.91 \pm 0.26) \times 10^{52}$ | $(2.79 \pm 0.24) \times 10^{52}$ |
| 050406 | −2.43 | ⋯ | ⋯ | $(6.8 \pm 1.4) \times 10^{-8}$ | $(1.88 \pm 0.52) \times 10^{-8}$ | 2.44 | 5.4 | 1.7 | $(1.64 \pm 0.34) \times 10^{51}$ | $(1.56 \pm 0.43) \times 10^{51}$ |
| 050505 | −1.41 | ⋯ | ⋯ | $(2.49 \pm 0.18) \times 10^{-6}$ | $(1.46 \pm 0.24) \times 10^{-7}$ | 4.27 | 58.9 | 0.38 | $(3.32 \pm 0.24) \times 10^{52}$ | $(1.03 \pm 0.17) \times 10^{52}$ |
| 050724 | −1.89 | ⋯ | ⋯ | $(9.98 \pm 1.2) \times 10^{-7}$ | $(2.1 \pm 0.19) \times 10^{-7}$ | 0.257 | 96 | 0.98 | $(1.6 \pm 0.19) \times 10^{50}$ | $(4.21 \pm 0.39) \times 10^{49}$ |
| 050730 | −1.53 | ⋯ | ⋯ | $(2.38 \pm 0.15) \times 10^{-6}$ | $(4.13 \pm 1.05) \times 10^{-8}$ | 3.96855 | 156.5 | 0.47 | $(3.55 \pm 0.23) \times 10^{52}$ | $(3.06 \pm 0.78) \times 10^{51}$ |
| 050802 | −1.54 | ⋯ | ⋯ | $(2 \pm 0.16) \times 10^{-6}$ | $(2.05 \pm 0.33) \times 10^{-7}$ | 1.71 | 19 | 0.63 | $(9.51 \pm 0.75) \times 10^{51}$ | $(2.65 \pm 0.42) \times 10^{51}$ |
| 050803 | −1.38 | ⋯ | ⋯ | $(2.15 \pm 0.14) \times 10^{-6}$ | $(7.68 \pm 0.88) \times 10^{-8}$ | 0.422 | 87.9 | 0.8 | $(7.96 \pm 0.5) \times 10^{50}$ | $(4.04 \pm 0.46) \times 10^{49}$ |
| 050814 | −1.8 | ⋯ | ⋯ | $(2.01 \pm 0.22) \times 10^{-6}$ | $(4.74 \pm 1.67) \times 10^{-8}$ | 5.3 | 150.9 | 0.69 | $(6.89 \pm 0.75) \times 10^{52}$ | $(1.02 \pm 0.36) \times 10^{52}$ |
| 050819 | −2.71 | ⋯ | ⋯ | $(3.5 \pm 0.55) \times 10^{-7}$ | $(1.81 \pm 0.57) \times 10^{-8}$ | 2.5043 | 37.7 | 2.44 | $(1.26 \pm 0.2) \times 10^{52}$ | $(2.29 \pm 0.72) \times 10^{51}$ |
| 050820A | −1.25 | ⋯ | ⋯ | $(3.44 \pm 0.24) \times 10^{-6}$ | $(2.07 \pm 0.19) \times 10^{-7}$ | 2.612 | 26 | 0.38 | $(2.09 \pm 0.15) \times 10^{52}$ | $(4.56 \pm 0.43) \times 10^{51}$ |
| 050826 | −1.16 | ⋯ | ⋯ | $(4.13 \pm 0.72) \times 10^{-7}$ | $(3.34 \pm 1.14) \times 10^{-8}$ | 0.297 | 35.5 | 0.8 | $(7.35 \pm 1.28) \times 10^{49}$ | $(7.71 \pm 2.64) \times 10^{48}$ |
| 050908 | −1.88 | ⋯ | ⋯ | $(4.83 \pm 0.51) \times 10^{-7}$ | $(4.52 \pm 0.9) \times 10^{-8}$ | 3.35 | 19.4 | 0.84 | $(9.78 \pm 1.03) \times 10^{51}$ | $(3.98 \pm 0.8) \times 10^{51}$ |
| 050915A | −1.39 | ⋯ | ⋯ | $(8.5 \pm 0.88) \times 10^{-7}$ | $(6.14 \pm 1.12) \times 10^{-8}$ | 2.5273 | 52 | 0.46 | $(5.94 \pm 0.61) \times 10^{51}$ | $(1.51 \pm 0.27) \times 10^{51}$ |
| 050922C | −1.37 | ⋯ | ⋯ | $(1.62 \pm 0.05) \times 10^{-6}$ | $(5.84 \pm 0.26) \times 10^{-7}$ | 2.198 | 4.5 | 0.48 | $(9.21 \pm 0.31) \times 10^{51}$ | $(1.06 \pm 0.05) \times 10^{52}$ |
| 051001 | −2.05 | ⋯ | ⋯ | $(1.74 \pm 0.15) \times 10^{-6}$ | $(2.95 \pm 0.66) \times 10^{-8}$ | 2.4296 | 189.1 | 1.06 | $(2.61 \pm 0.22) \times 10^{52}$ | $(1.52 \pm 0.34) \times 10^{51}$ |
| 051006 | −1.51 | ⋯ | ⋯ | $(1.34 \pm 0.14) \times 10^{-6}$ | $(1.23 \pm 0.23) \times 10^{-7}$ | 1.059 | 34.8 | 0.7 | $(2.83 \pm 0.3) \times 10^{51}$ | $(5.34 \pm 0.99) \times 10^{50}$ |
| 051016B | −2.4 | ⋯ | ⋯ | $(1.7 \pm 0.22) \times 10^{-7}$ | $(6.85 \pm 0.84) \times 10^{-8}$ | 0.9364 | 4 | 1.3 | $(5.23 \pm 0.68) \times 10^{50}$ | $(4.08 \pm 0.5) \times 10^{50}$ |
| 051109A | −1.51 | ⋯ | ⋯ | $(2.2 \pm 0.27) \times 10^{-6}$ | $(2.98 \pm 0.52) \times 10^{-7}$ | 2.346 | 37.2 | 0.55 | $(1.61 \pm 0.2) \times 10^{52}$ | $(7.31 \pm 1.28) \times 10^{51}$ |
| 051109B | −1.97 | ⋯ | ⋯ | $(2.56 \pm 0.41) \times 10^{-7}$ | $(3.42 \pm 0.81) \times 10^{-8}$ | 0.08 | 14.3 | 1 | $(3.79 \pm 0.61) \times 10^{48}$ | $(5.47 \pm 1.29) \times 10^{47}$ |
| 051111 | −1.32 | ⋯ | ⋯ | $(4.08 \pm 0.13) \times 10^{-6}$ | $(2.18 \pm 0.17) \times 10^{-7}$ | 1.549 | 46.1 | 0.53 | $(1.35 \pm 0.04) \times 10^{52}$ | $(1.85 \pm 0.15) \times 10^{51}$ |
| 051117B | −1.53 | ⋯ | ⋯ | $(1.77 \pm 0.37) \times 10^{-7}$ | $(3.68 \pm 1.05) \times 10^{-8}$ | 0.481 | 9 | 0.83 | $(8.89 \pm 1.86) \times 10^{49}$ | $(2.74 \pm 0.78) \times 10^{49}$ |
| 060108 | −2.03 | ⋯ | ⋯ | $(3.69 \pm 0.37) \times 10^{-7}$ | $(4.68 \pm 0.73) \times 10^{-8}$ | 2.03 | 14.3 | 1.03 | $(3.92 \pm 0.39) \times 10^{51}$ | $(1.5 \pm 0.23) \times 10^{51}$ |
| 060123 | −1.9 | ⋯ | ⋯ | $(3 \times 10^{-7}$ | $2.56 \times 10^{-9}$ | 1.099 | 900 | 0.93 | $9.03 \times 10^{50}$ | $1.62 \times 10^{49}$ |
| 060210 | −1.53 | ⋯ | ⋯ | $(7.66 \pm 0.41) \times 10^{-6}$ | $(2.04 \pm 0.21) \times 10^{-7}$ | 3.91 | 255 | 0.47 | $(1.12 \pm 0.06) \times 10^{53}$ | $(1.47 \pm 0.15) \times 10^{52}$ |
| 060223A | −1.74 | ⋯ | ⋯ | $(6.73 \pm 0.48) \times 10^{-7}$ | $(9.25 \pm 1.23) \times 10^{-8}$ | 4.41 | 11.3 | 0.64 | $(1.62 \pm 0.12) \times 10^{52}$ | $(1.21 \pm 0.16) \times 10^{52}$ |
| 060418 | −1.7 | ⋯ | ⋯ | $(8.33 \pm 0.25) \times 10^{-6}$ | $(4.55 \pm 0.24) \times 10^{-7}$ | 1.49 | 103.1 | 0.76 | $(3.69 \pm 0.11) \times 10^{52}$ | $(5.01 \pm 0.27) \times 10^{51}$ |
| 060510B | −1.78 | ⋯ | ⋯ | $(4.07 \pm 0.18) \times 10^{-6}$ | $(3.84 \pm 0.74) \times 10^{-8}$ | 4.9 | 275.2 | 0.68 | $(1.21 \pm 0.05) \times 10^{53}$ | $(6.75 \pm 1.3) \times 10^{51}$ |
| 060512 | −2.48 | ⋯ | ⋯ | $(2.32 \pm 0.4) \times 10^{-7}$ | $(4.51 \pm 1.03) \times 10^{-8}$ | 0.4428 | 8.5 | 1.19 | $(1.41 \pm 0.24) \times 10^{50}$ | $(3.95 \pm 0.9) \times 10^{49}$ |
| 060522 | −1.56 | ⋯ | ⋯ | $(1.14 \pm 0.11) \times 10^{-6}$ | $(4.07 \pm 1.11) \times 10^{-8}$ | 5.11 | 71.1 | 0.45 | $(2.41 \pm 0.23) \times 10^{52}$ | $(5.26 \pm 1.44) \times 10^{51}$ |
| 060526 | −2.01 | ⋯ | ⋯ | $(1.26 \pm 0.17) \times 10^{-6}$ | $(1.02 \pm 0.11) \times 10^{-7}$ | 3.21 | 298.2 | 1.01 | $(2.88 \pm 0.38) \times 10^{52}$ | $(9.84 \pm 1.06) \times 10^{51}$ |
| 060602A | −1.25 | ⋯ | ⋯ | $(1.61 \pm 0.16) \times 10^{-6}$ | $(4.74 \pm 1.69) \times 10^{-8}$ | 0.787 | 75 | 0.65 | $(1.73 \pm 0.17) \times 10^{51}$ | $(9.12 \pm 3.26) \times 10^{49}$ |
| 060604 | −2.01 | ⋯ | ⋯ | $(4.02 \pm 1.06) \times 10^{-7}$ | $(2.08 \pm 0.8) \times 10^{-8}$ | 2.1357 | 95 | 1.01 | $(4.57 \pm 1.2) \times 10^{51}$ | $(7.42 \pm 2.84) \times 10^{50}$ |
| 060605 | −1.55 | ⋯ | ⋯ | $(6.97 \pm 0.9) \times 10^{-7}$ | $(3.42 \pm 0.89) \times 10^{-8}$ | 3.8 | 79.1 | 0.49 | $(1.02 \pm 0.13) \times 10^{52}$ | $(2.4 \pm 0.63) \times 10^{51}$ |
| 060607A | −1.47 | ⋯ | ⋯ | $(2.55 \pm 0.11) \times 10^{-6}$ | $(1.08 \pm 0.1) \times 10^{-7}$ | 3.082 | 102.2 | 0.47 | $(2.55 \pm 0.11) \times 10^{52}$ | $(4.4 \pm 0.41) \times 10^{51}$ |
| 060614 | −2.02 | ⋯ | ⋯ | $(2.04 \pm 0.04) \times 10^{-5}$ | $(7.01 \pm 0.45) \times 10^{-7}$ | 0.13 | 108.7 | 1 | $(8.19 \pm 0.15) \times 10^{50}$ | $(3.18 \pm 0.2) \times 10^{49}$ |
| 060714 | −1.93 | ⋯ | ⋯ | $(2.83 \pm 0.17) \times 10^{-6}$ | $(8.1 \pm 0.82) \times 10^{-8}$ | 2.71 | 115 | 0.91 | $(4.38 \pm 0.26) \times 10^{52}$ | $(4.65 \pm 0.47) \times 10^{51}$ |
| 060719 | −1.91 | ⋯ | ⋯ | $(1.5 \pm 0.09) \times 10^{-6}$ | $(1.38 \pm 0.13) \times 10^{-7}$ | 1.532 | 66.9 | 0.92 | $(8.46 \pm 0.51) \times 10^{51}$ | $(1.97 \pm 0.18) \times 10^{51}$ |
| 060814 | −1.53 | ⋯ | ⋯ | $(1.46 \pm 0.02) \times 10^{-5}$ | $(5.45 \pm 0.22) \times 10^{-7}$ | 0.84 | 145.3 | 0.75 | $(2.08 \pm 0.03) \times 10^{52}$ | $(1.43 \pm 0.06) \times 10^{51}$ |
| 060906 | −2.03 | ⋯ | ⋯ | $(2.21 \pm 0.14) \times 10^{-6}$ | $(1.2 \pm 0.17) \times 10^{-7}$ | 3.685 | 43.5 | 1.05 | $(6.52 \pm 0.4) \times 10^{52}$ | $(1.65 \pm 0.24) \times 10^{52}$ |
| 060912A | −1.74 | ⋯ | ⋯ | $(1.35 \pm 0.06) \times 10^{-6}$ | $(5.88 \pm 0.3) \times 10^{-7}$ | 0.937 | 5 | 0.84 | $(2.69 \pm 0.12) \times 10^{51}$ | $(2.27 \pm 0.12) \times 10^{51}$ |
| 060926 | −2.54 | ⋯ | ⋯ | $(2.19 \pm 0.25) \times 10^{-7}$ | $(5.47 \pm 0.7) \times 10^{-8}$ | 3.208 | 8 | 2.17 | $(1.07 \pm 0.12) \times 10^{52}$ | $(1.13 \pm 0.14) \times 10^{52}$ |

**Table 3**
(Continued)

| GRB | $\Gamma$ | $\alpha$ | $E_p$ (keV) | $S_\gamma$ (erg cm$^{-2}$) | $F_p$ (erg cm$^{-2}$ s$^{-1}$) | $z$ | $T_{90}$ (s) | $K$ | $E_{\gamma,iso}$ (erg) | $L_p$ (erg s$^{-1}$) |
|---|---|---|---|---|---|---|---|---|---|---|
| (1) | (2) | (3) | (4) | (5) | (6) | (7) | (8) | (9) | (10) | (11) |
| 061007 | −1.03 | ⋯ | ⋯ | $(4.44 \pm 0.06) \times 10^{-5}$ | $(1.35 \pm 0.03) \times 10^{-6}$ | 1.261 | 75.3 | 0.45 | $(8.52 \pm 0.11) \times 10^{52}$ | $(5.88 \pm 0.15) \times 10^{51}$ |
| 061021 | −1.3 | ⋯ | ⋯ | $(2.96 \pm 0.1) \times 10^{-6}$ | $(5.06 \pm 0.22) \times 10^{-7}$ | 0.3463 | 46.2 | 0.81 | $(7.33 \pm 0.25) \times 10^{50}$ | $(1.69 \pm 0.07) \times 10^{50}$ |
| 061110A | −1.67 | ⋯ | ⋯ | $(1.06 \pm 0.08) \times 10^{-6}$ | $(3.74 \pm 0.85) \times 10^{-8}$ | 0.758 | 40.7 | 0.83 | $(1.36 \pm 0.1) \times 10^{51}$ | $(8.42 \pm 1.91) \times 10^{49}$ |
| 061110B | −1.03 | ⋯ | ⋯ | $(1.33 \pm 0.12) \times 10^{-6}$ | $(4.18 \pm 1.02) \times 10^{-8}$ | 3.44 | 134 | 0.24 | $(7.9 \pm 0.72) \times 10^{51}$ | $(1.1 \pm 0.27) \times 10^{51}$ |
| 061121 | −1.41 | ⋯ | ⋯ | $(1.37 \pm 0.02) \times 10^{-5}$ | $(1.67 \pm 0.04) \times 10^{-6}$ | 1.314 | 81.3 | 0.61 | $(3.83 \pm 0.06) \times 10^{52}$ | $(1.08 \pm 0.02) \times 10^{52}$ |
| 061210 | −1.56 | ⋯ | ⋯ | $(1.11 \pm 0.18) \times 10^{-6}$ | $(3.93 \pm 0.35) \times 10^{-7}$ | 0.41 | 85.3 | 0.86 | $(4.14 \pm 0.66) \times 10^{50}$ | $(2.07 \pm 0.18) \times 10^{50}$ |
| 061222A | −1.35 | ⋯ | ⋯ | $(7.99 \pm 0.16) \times 10^{-6}$ | $(6.92 \pm 0.21) \times 10^{-7}$ | 2.088 | 71.4 | 0.48 | $(4.14 \pm 0.08) \times 10^{52}$ | $(1.11 \pm 0.03) \times 10^{52}$ |
| 061222B | −1.97 | ⋯ | ⋯ | $(2.24 \pm 0.18) \times 10^{-6}$ | $(9.9 \pm 2.24) \times 10^{-8}$ | 3.355 | 40 | 0.96 | $(5.19 \pm 0.42) \times 10^{52}$ | $(9.99 \pm 2.26) \times 10^{51}$ |
| 070103 | −1.95 | ⋯ | ⋯ | $(3.38 \pm 0.46) \times 10^{-7}$ | $(6.53 \pm 0.94) \times 10^{-8}$ | 2.6208 | 18.6 | 0.94 | $(5.08 \pm 0.69) \times 10^{51}$ | $(3.55 \pm 0.51) \times 10^{51}$ |
| 070110 | −1.58 | ⋯ | ⋯ | $(1.62 \pm 0.11) \times 10^{-6}$ | $(4.4 \pm 0.88) \times 10^{-8}$ | 2.352 | 88.4 | 0.6 | $(1.3 \pm 0.09) \times 10^{52}$ | $(1.18 \pm 0.24) \times 10^{51}$ |
| 070129 | −2.01 | ⋯ | ⋯ | $(2.98 \pm 0.27) \times 10^{-6}$ | $(3.37 \pm 0.73) \times 10^{-8}$ | 2.3384 | 460.6 | 1.01 | $(3.97 \pm 0.36) \times 10^{52}$ | $(1.5 \pm 0.33) \times 10^{51}$ |
| 070208 | −1.94 | ⋯ | ⋯ | $(4.45 \pm 1.01) \times 10^{-7}$ | $(5.67 \pm 1.39) \times 10^{-8}$ | 1.165 | 47.7 | 0.95 | $(1.54 \pm 0.35) \times 10^{51}$ | $(4.26 \pm 1.04) \times 10^{50}$ |
| 070306 | −1.66 | ⋯ | ⋯ | $(5.38 \pm 0.29) \times 10^{-6}$ | $(2.89 \pm 0.15) \times 10^{-7}$ | 1.497 | 209.5 | 0.73 | $(2.31 \pm 0.12) \times 10^{52}$ | $(3.1 \pm 0.16) \times 10^{51}$ |
| 070318 | −1.42 | ⋯ | ⋯ | $(2.48 \pm 0.11) \times 10^{-6}$ | $(1.38 \pm 0.12) \times 10^{-7}$ | 0.836 | 74.6 | 0.7 | $(3.28 \pm 0.15) \times 10^{51}$ | $(3.36 \pm 0.29) \times 10^{50}$ |
| 070411 | −1.72 | ⋯ | ⋯ | $(2.7 \pm 0.16) \times 10^{-6}$ | $(6.29 \pm 0.9) \times 10^{-8}$ | 2.954 | 121.5 | 0.68 | $(3.61 \pm 0.21) \times 10^{52}$ | $(3.32 \pm 0.47) \times 10^{51}$ |
| 070506 | −1.73 | ⋯ | ⋯ | $(2.1 \pm 0.23) \times 10^{-7}$ | $(6.61 \pm 0.89) \times 10^{-8}$ | 2.31 | 4.3 | 0.72 | $(1.96 \pm 0.21) \times 10^{51}$ | $(2.04 \pm 0.28) \times 10^{51}$ |
| 070529 | −1.34 | ⋯ | ⋯ | $(2.57 \pm 0.25) \times 10^{-6}$ | $(1.16 \pm 0.29) \times 10^{-7}$ | 2.4996 | 109.2 | 0.44 | $(1.66 \pm 0.16) \times 10^{52}$ | $(2.64 \pm 0.66) \times 10^{51}$ |
| 070611 | −1.66 | ⋯ | ⋯ | $(3.91 \pm 0.57) \times 10^{-7}$ | $(5.82 \pm 1.49) \times 10^{-8}$ | 2.04 | 12.2 | 0.69 | $(2.77 \pm 0.4) \times 10^{51}$ | $(1.25 \pm 0.32) \times 10^{51}$ |
| 070612A | −1.69 | ⋯ | ⋯ | $(1.06 \pm 0.06) \times 10^{-5}$ | $(1.06 \pm 0.27) \times 10^{-7}$ | 0.617 | 368.8 | 0.86 | $(9.24 \pm 0.52) \times 10^{51}$ | $(1.49 \pm 0.37) \times 10^{50}$ |
| 070714B | −1.36 | ⋯ | ⋯ | $(7.2 \pm 0.9) \times 10^{-7}$ | $(2.18 \pm 0.16) \times 10^{-7}$ | 0.92 | 64 | 0.66 | $(1.08 \pm 0.14) \times 10^{51}$ | $(6.28 \pm 0.47) \times 10^{50}$ |
| 070721B | −1.34 | ⋯ | ⋯ | $(3.6 \pm 0.2) \times 10^{-6}$ | $(1.22 \pm 0.24) \times 10^{-7}$ | 3.626 | 340 | 0.36 | $(3.6 \pm 0.2) \times 10^{52}$ | $(5.65 \pm 1.13) \times 10^{51}$ |
| 070802 | −1.79 | ⋯ | ⋯ | $(2.8 \pm 0.5) \times 10^{-7}$ | $(2.68 \pm 0.67) \times 10^{-8}$ | 2.45 | 16.4 | 0.77 | $(3.08 \pm 0.55) \times 10^{51}$ | $(1.02 \pm 0.25) \times 10^{51}$ |
| 070810A | −2.04 | ⋯ | ⋯ | $(6.9 \pm 0.6) \times 10^{-7}$ | $(1.15 \pm 0.12) \times 10^{-7}$ | 2.17 | 11 | 1.05 | $(8.35 \pm 0.73) \times 10^{51}$ | $(4.41 \pm 0.46) \times 10^{51}$ |
| 071003 | −1.36 | ⋯ | ⋯ | $(8.3 \pm 0.3) \times 10^{-6}$ | $(5.09 \pm 0.32) \times 10^{-7}$ | 1.1 | 150 | 0.62 | $(1.68 \pm 0.06) \times 10^{52}$ | $(2.16 \pm 0.14) \times 10^{51}$ |
| 071010A | −2.33 | ⋯ | ⋯ | $(2 \pm 0.4) \times 10^{-7}$ | $(4.33 \pm 1.62) \times 10^{-8}$ | 0.98 | 6 | 1.25 | $(6.48 \pm 1.3) \times 10^{50}$ | $(2.77 \pm 1.04) \times 10^{50}$ |
| 071021 | −1.7 | ⋯ | ⋯ | $(1.3 \pm 0.2) \times 10^{-6}$ | $(4.88 \pm 0.7) \times 10^{-8}$ | 2.452 | 225 | 0.69 | $(1.28 \pm 0.2) \times 10^{52}$ | $(1.66 \pm 0.24) \times 10^{51}$ |
| 071028B | −1.45 | ⋯ | ⋯ | $(2.5 \pm 0.8) \times 10^{-7}$ | $(1.09 \pm 0.39) \times 10^{-7}$ | 0.94 | 55 | 0.69 | $(4.13 \pm 1.32) \times 10^{50}$ | $(3.49 \pm 1.24) \times 10^{50}$ |
| 071031 | −2.42 | ⋯ | ⋯ | $(9 \pm 1.3) \times 10^{-7}$ | $(2.62 \pm 0.52) \times 10^{-8}$ | 2.692 | 180 | 1.73 | $(2.62 \pm 0.38) \times 10^{52}$ | $(2.81 \pm 0.56) \times 10^{51}$ |
| 071117 | −1.57 | ⋯ | ⋯ | $(2.4 \pm 0.1) \times 10^{-6}$ | $(8.33 \pm 0.29) \times 10^{-7}$ | 1.331 | 6.6 | 0.69 | $(7.84 \pm 0.33) \times 10^{51}$ | $(6.34 \pm 0.22) \times 10^{51}$ |
| 071122 | −1.77 | ⋯ | ⋯ | $(5.8 \pm 1.1) \times 10^{-7}$ | $(2.71 \pm 1.35) \times 10^{-8}$ | 1.14 | 68.7 | 0.84 | $(1.7 \pm 0.32) \times 10^{51}$ | $(1.69 \pm 0.85) \times 10^{50}$ |
| 080210 | −1.77 | ⋯ | ⋯ | $(1.8 \pm 0.1) \times 10^{-6}$ | $(1.08 \pm 0.14) \times 10^{-7}$ | 2.641 | 45 | 0.74 | $(2.17 \pm 0.12) \times 10^{52}$ | $(4.76 \pm 0.59) \times 10^{51}$ |
| 080310 | −2.32 | ⋯ | ⋯ | $(2.3 \pm 0.2) \times 10^{-6}$ | $(7.06 \pm 1.09) \times 10^{-8}$ | 2.4266 | 365 | 1.48 | $(4.79 \pm 0.42) \times 10^{52}$ | $(5.04 \pm 0.78) \times 10^{51}$ |
| 080319C | −1.37 | ⋯ | ⋯ | $(3.6 \pm 0.1) \times 10^{-6}$ | $(4.18 \pm 0.24) \times 10^{-7}$ | 1.95 | 34 | 0.51 | $(1.74 \pm 0.05) \times 10^{52}$ | $(5.96 \pm 0.34) \times 10^{51}$ |
| 080330 | −2.53 | ⋯ | ⋯ | $(3.4 \pm 0.8) \times 10^{-7}$ | $(4.54 \pm 1.01) \times 10^{-8}$ | 1.51 | 61 | 1.63 | $(3.31 \pm 0.78) \times 10^{51}$ | $(1.11 \pm 0.25) \times 10^{51}$ |
| 080411 | −1.75 | ⋯ | ⋯ | $(2.64 \pm 0.01) \times 10^{-5}$ | $(2.95 \pm 0.06) \times 10^{-6}$ | 1.03 | 56 | 0.84 | $(6.31 \pm 0.02) \times 10^{52}$ | $(1.43 \pm 0.03) \times 10^{52}$ |
| 080413A | −1.57 | ⋯ | ⋯ | $(3.5 \pm 0.1) \times 10^{-6}$ | $(4.13 \pm 0.15) \times 10^{-7}$ | 2.433 | 46 | 0.59 | $(2.91 \pm 0.08) \times 10^{52}$ | $(1.18 \pm 0.04) \times 10^{52}$ |
| 080516 | −1.82 | ⋯ | ⋯ | $(2.6 \pm 0.4) \times 10^{-7}$ | $(1.19 \pm 0.2) \times 10^{-7}$ | 3.2 | 5.8 | 0.77 | $(4.5 \pm 0.69) \times 10^{51}$ | $(8.67 \pm 1.44) \times 10^{51}$ |
| 080520 | −2.9 | ⋯ | ⋯ | $(5.5 \pm 1.7) \times 10^{-8}$ | $(2.25 \pm 0.45) \times 10^{-8}$ | 1.545 | 2.8 | 2.32 | $(7.94 \pm 2.46) \times 10^{50}$ | $(8.25 \pm 1.65) \times 10^{50}$ |
| 080604 | −1.78 | ⋯ | ⋯ | $(8 \pm 0.9) \times 10^{-7}$ | $(2.7 \pm 0.67) \times 10^{-8}$ | 1.416 | 82 | 0.82 | $(3.48 \pm 0.39) \times 10^{51}$ | $(2.84 \pm 0.71) \times 10^{50}$ |
| 080607 | −1.31 | ⋯ | ⋯ | $2.4 \times 10^{-5}$ | $(1.91 \pm 0.09) \times 10^{-6}$ | 3.036 | 79 | 0.38 | $1.88 \times 10^{53}$ | $(6.03 \pm 0.29) \times 10^{52}$ |
| 080707 | −1.77 | ⋯ | ⋯ | $(5.2 \pm 0.6) \times 10^{-7}$ | $(6.77 \pm 0.68) \times 10^{-8}$ | 1.23 | 27.1 | 0.83 | $(1.75 \pm 0.2) \times 10^{51}$ | $(5.07 \pm 0.51) \times 10^{50}$ |
| 080710 | −1.47 | ⋯ | ⋯ | $(1.4 \pm 0.2) \times 10^{-6}$ | $(7.7 \pm 1.54) \times 10^{-8}$ | 0.845 | 120 | 0.72 | $(1.94 \pm 0.28) \times 10^{51}$ | $(1.97 \pm 0.39) \times 10^{50}$ |
| 080721 | −1.11 | ⋯ | ⋯ | $(1.2 \pm 0.1) \times 10^{-5}$ | $(1.88 \pm 0.16) \times 10^{-6}$ | 2.602 | 16.2 | 0.32 | $(6.08 \pm 0.51) \times 10^{52}$ | $(3.42 \pm 0.29) \times 10^{52}$ |
| 080804 | −1.19 | ⋯ | ⋯ | $(3.6 \pm 0.2) \times 10^{-6}$ | $(2.69 \pm 0.35) \times 10^{-7}$ | 2.2045 | 34 | 0.39 | $(1.67 \pm 0.09) \times 10^{52}$ | $(3.99 \pm 0.51) \times 10^{51}$ |
| 080805 | −1.53 | ⋯ | ⋯ | $(2.5 \pm 0.1) \times 10^{-6}$ | $(8.25 \pm 0.75) \times 10^{-8}$ | 1.505 | 78 | 0.65 | $(9.63 \pm 0.39) \times 10^{51}$ | $(7.96 \pm 0.72) \times 10^{50}$ |
| 080810 | −1.34 | ⋯ | ⋯ | $(4.6 \pm 0.2) \times 10^{-6}$ | $(1.63 \pm 0.16) \times 10^{-7}$ | 3.35 | 106 | 0.38 | $(4.21 \pm 0.18) \times 10^{52}$ | $(6.49 \pm 0.65) \times 10^{51}$ |

**Table 3**
(Continued)

| GRB | Γ | α | $E_p$ (keV) | $S_\gamma$ (erg cm$^{-2}$) | $F_p$ (erg cm$^{-2}$ s$^{-1}$) | $z$ | $T_{90}$ (s) | $K$ | $E_{\gamma,iso}$ (erg) | $L_p$ (erg s$^{-1}$) |
|---|---|---|---|---|---|---|---|---|---|---|
| (1) | (2) | (3) | (4) | (5) | (6) | (7) | (8) | (9) | (10) | (11) |
| 080905B | −1.78 | ⋯ | ⋯ | $(1.8 \pm 0.2) \times 10^{-6}$ | $(3.37 \pm 0.67) \times 10^{-8}$ | 2.374 | 128 | 0.77 | $(1.86 \pm 0.21) \times 10^{52}$ | $(1.18 \pm 0.24) \times 10^{51}$ |
| 080906 | −1.59 | ⋯ | ⋯ | $(3.5 \pm 0.2) \times 10^{-6}$ | $(7.31 \pm 1.46) \times 10^{-8}$ | 2 | 147 | 0.64 | $(2.23 \pm 0.13) \times 10^{52}$ | $(1.4 \pm 0.28) \times 10^{51}$ |
| 080928 | −1.77 | ⋯ | ⋯ | $(2.5 \pm 0.2) \times 10^{-6}$ | $(1.42 \pm 0.07) \times 10^{-7}$ | 1.692 | 280 | 0.8 | $(1.47 \pm 0.12) \times 10^{52}$ | $(2.25 \pm 0.11) \times 10^{51}$ |
| 081008 | −1.69 | ⋯ | ⋯ | $(4.3 \pm 0.2) \times 10^{-6}$ | $(9.1 \pm 0.7) \times 10^{-8}$ | 1.9685 | 185.5 | 0.71 | $(2.98 \pm 0.14) \times 10^{52}$ | $(1.87 \pm 0.14) \times 10^{51}$ |
| 081028A | −1.25 | ⋯ | ⋯ | $(3.7 \pm 0.2) \times 10^{-6}$ | $(4.23 \pm 0.85) \times 10^{-8}$ | 3.038 | 260 | 0.35 | $(2.67 \pm 0.14) \times 10^{52}$ | $(1.23 \pm 0.25) \times 10^{51}$ |
| 081029 | −1.43 | ⋯ | ⋯ | $(2.1 \pm 0.2) \times 10^{-6}$ | $(3.92 \pm 1.57) \times 10^{-8}$ | 3.8479 | 270 | 0.41 | $(2.58 \pm 0.25) \times 10^{52}$ | $(2.33 \pm 0.93) \times 10^{51}$ |
| 081118 | −2.1 | ⋯ | ⋯ | $(1.2 \pm 0.1) \times 10^{-6}$ | $(3.54 \pm 1.18) \times 10^{-8}$ | 2.58 | 67 | 1.14 | $(2.13 \pm 0.18) \times 10^{52}$ | $(2.25 \pm 0.75) \times 10^{51}$ |
| 081203A | −1.54 | ⋯ | ⋯ | $(7.7 \pm 0.3) \times 10^{-6}$ | $(2.17 \pm 0.15) \times 10^{-7}$ | 2.1 | 294 | 0.59 | $(4.99 \pm 0.19) \times 10^{52}$ | $(4.35 \pm 0.3) \times 10^{51}$ |
| 090102 | −1.36 | ⋯ | ⋯ | $(6.8 \pm 0.3) \times 10^{-8}$ | $(4.44 \pm 0.65) \times 10^{-7}$ | 1.547 | 27 | 0.55 | $(2.33 \pm 0.1) \times 10^{50}$ | $(3.88 \pm 0.56) \times 10^{51}$ |
| 090113 | −1.6 | ⋯ | ⋯ | $(7.6 \pm 0.4) \times 10^{-7}$ | $(1.82 \pm 0.15) \times 10^{-7}$ | 1.7493 | 9.1 | 0.67 | $(3.98 \pm 0.21) \times 10^{51}$ | $(2.62 \pm 0.21) \times 10^{51}$ |
| 090205 | −2.15 | ⋯ | ⋯ | $(1.9 \pm 0.3) \times 10^{-7}$ | $(2.9 \pm 0.58) \times 10^{-8}$ | 4.7 | 8.8 | 1.3 | $(1.02 \pm 0.16) \times 10^{52}$ | $(8.84 \pm 1.77) \times 10^{51}$ |
| 090313 | −1.91 | ⋯ | ⋯ | $(1.4 \pm 0.2) \times 10^{-6}$ | $(5.1 \pm 1.91) \times 10^{-8}$ | 3.375 | 79 | 0.88 | $(3 \pm 0.43) \times 10^{52}$ | $(4.78 \pm 1.79) \times 10^{51}$ |
| 090407 | −1.73 | ⋯ | ⋯ | $(1.1 \pm 0.2) \times 10^{-6}$ | $(4.13 \pm 0.69) \times 10^{-8}$ | 1.4485 | 310 | 0.79 | $(4.77 \pm 0.87) \times 10^{51}$ | $(4.38 \pm 0.73) \times 10^{50}$ |
| 090418A | −1.48 | ⋯ | ⋯ | $(4.6 \pm 0.2) \times 10^{-6}$ | $(1.46 \pm 0.23) \times 10^{-7}$ | 1.608 | 56 | 0.61 | $(1.88 \pm 0.08) \times 10^{52}$ | $(1.55 \pm 0.24) \times 10^{51}$ |
| 090516A | −1.84 | ⋯ | ⋯ | $(9 \pm 0.6) \times 10^{-6}$ | $(1.05 \pm 0.13) \times 10^{-7}$ | 4.109 | 210 | 0.77 | $(2.32 \pm 0.15) \times 10^{53}$ | $(1.39 \pm 0.17) \times 10^{52}$ |
| 090519 | −1.02 | ⋯ | ⋯ | $(1.2 \pm 0.1) \times 10^{-6}$ | $(5.59 \pm 1.86) \times 10^{-8}$ | 3.9 | 64 | 0.21 | $(7.8 \pm 0.65) \times 10^{51}$ | $(1.78 \pm 0.59) \times 10^{51}$ |
| 090715B | −1.57 | ⋯ | ⋯ | $(5.7 \pm 0.2) \times 10^{-6}$ | $(2.8 \pm 0.15) \times 10^{-7}$ | 3 | 266 | 0.55 | $(6.32 \pm 0.22) \times 10^{52}$ | $(1.24 \pm 0.07) \times 10^{52}$ |
| 090726 | −2.25 | ⋯ | ⋯ | $(8.6 \pm 1) \times 10^{-7}$ | $(3.9 \pm 1.11) \times 10^{-8}$ | 2.71 | 67 | 1.39 | $(2.03 \pm 0.24) \times 10^{52}$ | $(3.41 \pm 0.97) \times 10^{51}$ |
| 090809 | −1.34 | ⋯ | ⋯ | $(3.4 \pm 0.5) \times 10^{-7}$ | $(8.96 \pm 1.63) \times 10^{-8}$ | 2.737 | 5.4 | 0.42 | $(2.46 \pm 0.36) \times 10^{51}$ | $(2.42 \pm 0.44) \times 10^{51}$ |
| 090812 | −1.24 | ⋯ | ⋯ | $(5.8 \pm 0.2) \times 10^{-6}$ | $(3.06 \pm 0.17) \times 10^{-7}$ | 2.452 | 66.7 | 0.39 | $(3.24 \pm 0.11) \times 10^{52}$ | $(5.89 \pm 0.33) \times 10^{51}$ |
| 091020 | −1.53 | ⋯ | ⋯ | $(3.7 \pm 0.1) \times 10^{-6}$ | $(3.15 \pm 0.23) \times 10^{-7}$ | 1.71 | 34.6 | 0.63 | $(1.74 \pm 0.05) \times 10^{52}$ | $(4.02 \pm 0.29) \times 10^{51}$ |
| 091024 | −1.2 | ⋯ | ⋯ | $(6.1 \pm 0.3) \times 10^{-6}$ | $(1.73 \pm 0.26) \times 10^{-7}$ | 1.092 | 109.8 | 0.55 | $(1.08 \pm 0.05) \times 10^{52}$ | $(6.41 \pm 0.96) \times 10^{50}$ |
| 091109A | −1.31 | ⋯ | ⋯ | $(1.6 \pm 0.2) \times 10^{-6}$ | $(1.07 \pm 0.33) \times 10^{-7}$ | 3.5 | 48 | 0.35 | $(1.47 \pm 0.18) \times 10^{52}$ | $(4.43 \pm 1.36) \times 10^{51}$ |
| 091208B | −1.74 | ⋯ | ⋯ | $(3.3 \pm 0.2) \times 10^{-6}$ | $(1.04 \pm 0.07) \times 10^{-6}$ | 1.063 | 14.9 | 0.83 | $(8.3 \pm 0.5) \times 10^{51}$ | $(5.41 \pm 0.36) \times 10^{51}$ |
| 100219A | −1.34 | ⋯ | ⋯ | $(3.7 \pm 0.6) \times 10^{-7}$ | $(3.26 \pm 0.81) \times 10^{-8}$ | 4.5 | 18.8 | 0.32 | $(4.64 \pm 0.75) \times 10^{51}$ | $(2.24 \pm 0.56) \times 10^{51}$ |
| 100302A | −1.72 | ⋯ | ⋯ | $(3.1 \pm 0.4) \times 10^{-7}$ | $(3.46 \pm 0.69) \times 10^{-8}$ | 4.813 | 17.9 | 0.61 | $(8.11 \pm 1.05) \times 10^{51}$ | $(5.25 \pm 1.05) \times 10^{51}$ |
| 100316B | −2.23 | ⋯ | ⋯ | $(2 \pm 0.2) \times 10^{-7}$ | $(7.3 \pm 0.56) \times 10^{-8}$ | 1.18 | 3.8 | 1.2 | $(8.91 \pm 0.89) \times 10^{50}$ | $(7.09 \pm 0.55) \times 10^{50}$ |
| 100413A | −1.25 | ⋯ | ⋯ | $(6.2 \pm 0.2) \times 10^{-6}$ | $(5.92 \pm 0.85) \times 10^{-8}$ | 3.9 | 191 | 0.3 | $(5.81 \pm 0.19) \times 10^{52}$ | $(2.72 \pm 0.39) \times 10^{51}$ |
| 100424A | −1.83 | ⋯ | ⋯ | $(1.5 \pm 0.1) \times 10^{-6}$ | $(2.64 \pm 0.66) \times 10^{-8}$ | 2.465 | 104 | 0.81 | $(1.75 \pm 0.12) \times 10^{52}$ | $(1.07 \pm 0.27) \times 10^{51}$ |
| 100425A | −2.42 | ⋯ | ⋯ | $(4.7 \pm 0.9) \times 10^{-7}$ | $(7.33 \pm 1.05) \times 10^{-8}$ | 1.755 | 37 | 1.53 | $(5.68 \pm 1.09) \times 10^{51}$ | $(2.44 \pm 0.35) \times 10^{51}$ |
| 100513A | −1.62 | ⋯ | ⋯ | $(1.4 \pm 0.1) \times 10^{-6}$ | $(4.33 \pm 0.72) \times 10^{-8}$ | 4.772 | 84 | 0.51 | $(3.04 \pm 0.22) \times 10^{52}$ | $(5.42 \pm 0.9) \times 10^{51}$ |
| 100615A | −1.87 | ⋯ | ⋯ | $(5 \pm 0.1) \times 10^{-6}$ | $(3.5 \pm 0.13) \times 10^{-7}$ | 1.398 | 39 | 0.89 | $(2.3 \pm 0.05) \times 10^{52}$ | $(3.87 \pm 0.14) \times 10^{51}$ |
| 100621A | −1.9 | ⋯ | ⋯ | $2.1 \pm 0 \times 10^{-5}$ | $(8.2 \pm 0.19) \times 10^{-7}$ | 0.542 | 63.6 | 0.96 | $1.56 \pm 0 \times 10^{52}$ | $(9.37 \pm 0.22) \times 10^{50}$ |
| 100704A | −1.73 | ⋯ | ⋯ | $(6 \pm 0.2) \times 10^{-6}$ | $(2.96 \pm 0.14) \times 10^{-7}$ | 3.6 | 197.5 | 0.66 | $(1.08 \pm 0.04) \times 10^{53}$ | $(2.45 \pm 0.11) \times 10^{52}$ |
| 100728A | −1.18 | ⋯ | ⋯ | $3.8 \times 10^{-5}$ | $(4.45 \pm 0.17) \times 10^{-7}$ | 1.567 | 198.5 | 0.46 | $1.12 \times 10^{53}$ | $(3.37 \pm 0.13) \times 10^{51}$ |
| 100814A | −1.47 | ⋯ | ⋯ | $(9 \pm 0.2) \times 10^{-6}$ | $(1.92 \pm 0.15) \times 10^{-7}$ | 1.44 | 174.5 | 0.62 | $(3.06 \pm 0.07) \times 10^{52}$ | $(1.6 \pm 0.13) \times 10^{51}$ |
| 100901A | −1.52 | ⋯ | ⋯ | $(2.1 \pm 0.3) \times 10^{-6}$ | $(6.03 \pm 1.51) \times 10^{-8}$ | 1.408 | 439 | 0.66 | $(7.2 \pm 1.03) \times 10^{51}$ | $(4.98 \pm 1.24) \times 10^{50}$ |
| 100902A | −1.98 | ⋯ | ⋯ | $(3.2 \pm 0.2) \times 10^{-6}$ | $(6.2 \pm 0.62) \times 10^{-8}$ | 4.5 | 428.8 | 0.97 | $(1.19 \pm 0.07) \times 10^{53}$ | $(1.27 \pm 0.13) \times 10^{52}$ |
| 100906A | −1.78 | ⋯ | ⋯ | $1.2 \times 10^{-5}$ | $(6.81 \pm 0.27) \times 10^{-7}$ | 1.727 | 114.4 | 0.8 | $7.37 \times 10^{52}$ | $(1.14 \pm 0.05) \times 10^{52}$ |
| 110128A | −1.31 | ⋯ | ⋯ | $(7.2 \pm 1.4) \times 10^{-7}$ | $(6.6 \pm 1.65) \times 10^{-8}$ | 2.339 | 30.7 | 0.44 | $(4.13 \pm 0.8) \times 10^{51}$ | $(1.26 \pm 0.32) \times 10^{51}$ |
| 110205A | −1.8 | ⋯ | ⋯ | $1.7 \times 10^{-5}$ | $(2.41 \pm 0.13) \times 10^{-7}$ | 1.98 | 257 | 0.8 | $1.34 \times 10^{53}$ | $(5.66 \pm 0.31) \times 10^{51}$ |
| 110213A | −1.83 | ⋯ | ⋯ | $(5.9 \pm 0.4) \times 10^{-6}$ | $(1.06 \pm 0.4) \times 10^{-7}$ | 1.46 | 48 | 0.86 | $(2.84 \pm 0.19) \times 10^{52}$ | $(1.25 \pm 0.47) \times 10^{51}$ |
| 110731A | −1.15 | ⋯ | ⋯ | $(6 \pm 0.1) \times 10^{-6}$ | $(9.71 \pm 0.26) \times 10^{-7}$ | 2.83 | 38.8 | 0.32 | $(3.5 \pm 0.06) \times 10^{52}$ | $(2.17 \pm 0.06) \times 10^{52}$ |
| 110801A | −1.84 | ⋯ | ⋯ | $(4.7 \pm 0.3) \times 10^{-6}$ | $(7.23 \pm 1.31) \times 10^{-8}$ | 1.858 | 385 | 0.85 | $(3.48 \pm 0.22) \times 10^{52}$ | $(1.53 \pm 0.28) \times 10^{51}$ |
| 110808A | −2.32 | ⋯ | ⋯ | $(3.3 \pm 0.8) \times 10^{-7}$ | $(2.17 \pm 1.09) \times 10^{-8}$ | 1.348 | 48 | 1.31 | $(2.09 \pm 0.51) \times 10^{51}$ | $(3.23 \pm 1.61) \times 10^{50}$ |
| 110818A | −1.58 | ⋯ | ⋯ | $(4 \pm 0.2) \times 10^{-6}$ | $(1.17 \pm 0.22) \times 10^{-7}$ | 3.36 | 103 | 0.54 | $(5.23 \pm 0.26) \times 10^{52}$ | $(6.7 \pm 1.26) \times 10^{51}$ |

**Table 3**
(Continued)

| GRB (1) | $\Gamma$ (2) | $\alpha$ (3) | $E_{\rm p}$ (keV) (4) | $S_\gamma$ (erg cm$^{-2}$) (5) | $F_{\rm p}$ (erg cm$^{-2}$ s$^{-1}$) (6) | $z$ (7) | $T_{90}$ (s) (8) | $K$ (9) | $E_{\gamma,{\rm iso}}$ (erg) (10) | $L_{\rm p}$ (erg s$^{-1}$) (11) |
|---|---|---|---|---|---|---|---|---|---|---|
| 111008A | −1.86 | ⋯ | ⋯ | $(5.3 \pm 0.3) \times 10^{-6}$ | $(4.17 \pm 0.46) \times 10^{-7}$ | 5 | 63.46 | 0.78 | $(1.87 \pm 0.11) \times 10^{53}$ | $(8.83 \pm 0.97) \times 10^{52}$ |
| 111107A | −1.49 | ⋯ | ⋯ | $(8.8 \pm 0.8) \times 10^{-7}$ | $(9.16 \pm 1.53) \times 10^{-8}$ | 2.893 | 26.6 | 0.5 | $(8.34 \pm 0.76) \times 10^{51}$ | $(3.38 \pm 0.56) \times 10^{51}$ |
| 111123A | −1.68 | ⋯ | ⋯ | $(7.3 \pm 0.3) \times 10^{-6}$ | $(6.33 \pm 0.7) \times 10^{-8}$ | 3.1516 | 290 | 0.63 | $(1.01 \pm 0.04) \times 10^{53}$ | $(3.64 \pm 0.4) \times 10^{51}$ |
| 111225A | −1.7 | ⋯ | ⋯ | $(1.3 \pm 0.12) \times 10^{-6}$ | $(4.88 \pm 0.7) \times 10^{-8}$ | 0.297 | 106.8 | 0.92 | $(2.66 \pm 0.25) \times 10^{50}$ | $(1.3 \pm 0.19) \times 10^{49}$ |
| 111229A | −1.85 | ⋯ | ⋯ | $(3.4 \pm 0.7) \times 10^{-7}$ | $(6.54 \pm 1.31) \times 10^{-8}$ | 1.3805 | 25.4 | 0.88 | $(1.5 \pm 0.31) \times 10^{51}$ | $(6.89 \pm 1.38) \times 10^{50}$ |
| 120118B | −2.08 | ⋯ | ⋯ | $(1.8 \pm 0.1) \times 10^{-6}$ | $(1.31 \pm 0.18) \times 10^{-7}$ | 2.943 | 23.26 | 1.12 | $(3.92 \pm 0.22) \times 10^{52}$ | $(1.12 \pm 0.15) \times 10^{52}$ |
| 120119A | −1.38 | ⋯ | ⋯ | $1.7 \times 10^{-5}$ | $(8.24 \pm 0.24) \times 10^{-7}$ | 1.728 | 253.8 | 0.54 | $7 \times 10^{52}$ | $(9.26 \pm 0.27) \times 10^{51}$ |
| 120327A | −1.52 | ⋯ | ⋯ | $(3.6 \pm 0.1) \times 10^{-6}$ | $(2.94 \pm 0.15) \times 10^{-7}$ | 2.81 | 62.9 | 0.53 | $(3.42 \pm 0.09) \times 10^{52}$ | $(1.06 \pm 0.05) \times 10^{52}$ |
| 120404A | −1.85 | ⋯ | ⋯ | $(1.6 \pm 0.1) \times 10^{-6}$ | $(7.85 \pm 1.31) \times 10^{-8}$ | 2.876 | 38.7 | 0.82 | $(2.45 \pm 0.15) \times 10^{52}$ | $(4.66 \pm 0.78) \times 10^{51}$ |
| 120712A | −1.36 | ⋯ | ⋯ | $(1.8 \pm 0.1) \times 10^{-6}$ | $(1.94 \pm 0.16) \times 10^{-7}$ | 4 | 14.7 | 0.36 | $(2.06 \pm 0.11) \times 10^{52}$ | $(1.11 \pm 0.09) \times 10^{52}$ |
| 120722A | −1.9 | ⋯ | ⋯ | $(1.2 \pm 0.2) \times 10^{-6}$ | $(6.41 \pm 1.92) \times 10^{-8}$ | 0.9586 | 42.4 | 0.93 | $(2.78 \pm 0.46) \times 10^{51}$ | $(2.9 \pm 0.87) \times 10^{50}$ |
| 120805A | −1.2 | ⋯ | ⋯ | $(8.2 \pm 1.4) \times 10^{-7}$ | $(3.2 \pm 1.73) \times 10^{-8}$ | 3.1 | 48 | 0.32 | $(5.64 \pm 0.96) \times 10^{51}$ | $(9.02 \pm 4.88) \times 10^{50}$ |
| 120815A | −2.29 | ⋯ | ⋯ | $(4.9 \pm 0.7) \times 10^{-7}$ | $(1.21 \pm 0.16) \times 10^{-7}$ | 2.358 | 9.7 | 1.42 | $(9.31 \pm 1.33) \times 10^{51}$ | $(7.7 \pm 1.05) \times 10^{51}$ |
| 120907A | −1.73 | ⋯ | ⋯ | $(6.7 \pm 1.1) \times 10^{-7}$ | $(2 \pm 0.28) \times 10^{-7}$ | 0.97 | 16.9 | 0.83 | $(1.41 \pm 0.23) \times 10^{51}$ | $(8.29 \pm 1.14) \times 10^{50}$ |
| 121024A | −1.41 | ⋯ | ⋯ | $(1.1 \pm 0.1) \times 10^{-6}$ | $(1.03 \pm 0.16) \times 10^{-7}$ | 2.298 | 69 | 0.49 | $(6.95 \pm 0.63) \times 10^{51}$ | $(2.14 \pm 0.33) \times 10^{51}$ |
| 121027A | −1.82 | ⋯ | ⋯ | $(2 \pm 0.1) \times 10^{-6}$ | $(8.61 \pm 1.33) \times 10^{-8}$ | 1.773 | 62.6 | 0.83 | $(1.34 \pm 0.07) \times 10^{52}$ | $(1.6 \pm 0.25) \times 10^{51}$ |
| 121128A | −1.32 | ⋯ | ⋯ | $(6.9 \pm 0.4) \times 10^{-6}$ | $(1.06 \pm 0.03) \times 10^{-6}$ | 2.2 | 23.3 | 0.45 | $(3.7 \pm 0.21) \times 10^{52}$ | $(1.82 \pm 0.06) \times 10^{52}$ |
| 121201A | −1.9 | ⋯ | ⋯ | $(7.8 \pm 1) \times 10^{-7}$ | $(5.13 \pm 0.64) \times 10^{-8}$ | 3.385 | 85 | 0.86 | $(1.65 \pm 0.21) \times 10^{52}$ | $(4.76 \pm 0.6) \times 10^{51}$ |
| 121211A | −2.36 | ⋯ | ⋯ | $(1.3 \pm 0.27) \times 10^{-6}$ | $(5.35 \pm 1.6) \times 10^{-8}$ | 1.023 | 182 | 1.29 | $(4.72 \pm 0.97) \times 10^{51}$ | $(3.93 \pm 1.18) \times 10^{50}$ |
| 130131B | −1.15 | ⋯ | ⋯ | $(3.4 \pm 0.4) \times 10^{-7}$ | $(8.83 \pm 1.77) \times 10^{-8}$ | 2.539 | 4.3 | 0.34 | $(1.76 \pm 0.21) \times 10^{51}$ | $(1.62 \pm 0.32) \times 10^{51}$ |
| 130408A | −1.28 | ⋯ | ⋯ | $(2.3 \pm 0.4) \times 10^{-6}$ | $(4.09 \pm 0.84) \times 10^{-7}$ | 3.758 | 28 | 0.33 | $(2.18 \pm 0.38) \times 10^{52}$ | $(1.84 \pm 0.38) \times 10^{52}$ |
| 130427B | −1.64 | ⋯ | ⋯ | $(1.5 \pm 0.1) \times 10^{-6}$ | $(2.15 \pm 0.29) \times 10^{-7}$ | 2.78 | 27 | 0.62 | $(1.65 \pm 0.11) \times 10^{52}$ | $(8.91 \pm 1.19) \times 10^{51}$ |
| 130505A | −1.18 | ⋯ | ⋯ | $(2.1 \pm 0.1) \times 10^{-5}$ | $(2.62 \pm 0.27) \times 10^{-6}$ | 2.27 | 88 | 0.38 | $(9.94 \pm 0.47) \times 10^{52}$ | $(4.05 \pm 0.42) \times 10^{52}$ |
| 130511A | −1.35 | ⋯ | ⋯ | $(2.2 \pm 0.4) \times 10^{-7}$ | $(1.05 \pm 0.16) \times 10^{-7}$ | 1.3033 | 5.43 | 0.58 | $(5.77 \pm 1.05) \times 10^{50}$ | $(6.37 \pm 0.98) \times 10^{50}$ |
| 130514A | −1.8 | ⋯ | ⋯ | $(9.1 \pm 0.2) \times 10^{-6}$ | $(1.87 \pm 0.2) \times 10^{-7}$ | 3.6 | 204 | 0.74 | $(1.82 \pm 0.04) \times 10^{53}$ | $(1.72 \pm 0.18) \times 10^{52}$ |
| 130604A | −1.51 | ⋯ | ⋯ | $(1.4 \pm 0.1) \times 10^{-6}$ | $(6.05 \pm 1.51) \times 10^{-8}$ | 1.06 | 37.7 | 0.7 | $(2.97 \pm 0.21) \times 10^{51}$ | $(2.64 \pm 0.66) \times 10^{50}$ |
| 130606A | −1.52 | ⋯ | ⋯ | $(2.9 \pm 0.2) \times 10^{-6}$ | $(1.96 \pm 0.15) \times 10^{-7}$ | 5.91 | 276.58 | 0.4 | $(6.68 \pm 0.46) \times 10^{52}$ | $(3.12 \pm 0.24) \times 10^{52}$ |
| 130610A | −1.27 | ⋯ | ⋯ | $(2.5 \pm 0.1) \times 10^{-6}$ | $(1.43 \pm 0.17) \times 10^{-7}$ | 2.092 | 46.4 | 0.44 | $(1.19 \pm 0.05) \times 10^{52}$ | $(2.1 \pm 0.25) \times 10^{51}$ |
| 131030A | −1.3 | ⋯ | ⋯ | $2.9 \times 10^{-5}$ | $(2.33 \pm 0.06) \times 10^{-6}$ | 1.293 | 41.1 | 0.56 | $7.21 \times 10^{52}$ | $(1.33 \pm 0.03) \times 10^{52}$ |
| 131103A | −1.97 | ⋯ | ⋯ | $(8.2 \pm 1) \times 10^{-7}$ | $(9.34 \pm 1.87) \times 10^{-8}$ | 0.599 | 17.3 | 0.99 | $(7.69 \pm 0.94) \times 10^{50}$ | $(1.4 \pm 0.28) \times 10^{50}$ |
| 131105A | −1.45 | ⋯ | ⋯ | $(7.1 \pm 0.5) \times 10^{-6}$ | $(2.72 \pm 0.47) \times 10^{-7}$ | 1.686 | 112.3 | 0.58 | $(3.02 \pm 0.21) \times 10^{52}$ | $(3.11 \pm 0.53) \times 10^{51}$ |
| 140114A | −2.06 | ⋯ | ⋯ | $(3.2 \pm 0.1) \times 10^{-6}$ | $(5.4 \pm 0.6) \times 10^{-8}$ | 3 | 139.7 | 1.09 | $(7 \pm 0.22) \times 10^{52}$ | $(4.73 \pm 0.53) \times 10^{51}$ |
| 140213A | −1.8 | ⋯ | ⋯ | $1.2 \times 10^{-5}$ | $(1.57 \pm 0.05) \times 10^{-6}$ | 1.2076 | 60 | 0.85 | $3.99 \times 10^{52}$ | $(1.15 \pm 0.04) \times 10^{52}$ |
| 140301A | −1.96 | ⋯ | ⋯ | $(4.4 \pm 0.8) \times 10^{-7}$ | $(4.38 \pm 1.25) \times 10^{-8}$ | 1.416 | 31 | 0.97 | $(2.25 \pm 0.41) \times 10^{51}$ | $(5.39 \pm 1.54) \times 10^{50}$ |
| 140304A | −1.29 | ⋯ | ⋯ | $(1.2 \pm 0.1) \times 10^{-6}$ | $(1.41 \pm 0.17) \times 10^{-7}$ | 5.283 | 15.6 | 0.27 | $(1.6 \pm 0.13) \times 10^{52}$ | $(1.19 \pm 0.14) \times 10^{52}$ |
| 140311A | −1.67 | ⋯ | ⋯ | $(2.3 \pm 0.3) \times 10^{-6}$ | $(9.18 \pm 3.53) \times 10^{-8}$ | 4.95 | 71.4 | 0.56 | $(5.7 \pm 0.74) \times 10^{52}$ | $(1.35 \pm 0.52) \times 10^{52}$ |
| 140318A | −1.35 | ⋯ | ⋯ | $(2.9 \pm 0.5) \times 10^{-7}$ | $(4.05 \pm 1.62) \times 10^{-8}$ | 1.02 | 8.43 | 0.63 | $(5.14 \pm 0.89) \times 10^{50}$ | $(1.45 \pm 0.58) \times 10^{50}$ |
| 140419A | −1.21 | ⋯ | ⋯ | $1.6 \times 10^{-5}$ | $(4.22 \pm 0.17) \times 10^{-7}$ | 3.956 | 94.7 | 0.28 | $1.43 \times 10^{53}$ | $(1.86 \pm 0.08) \times 10^{52}$ |
| 140423A | −1.33 | ⋯ | ⋯ | $(9.4 \pm 0.3) \times 10^{-6}$ | $(1.72 \pm 0.16) \times 10^{-7}$ | 3.26 | 134 | 0.38 | $(8.22 \pm 0.26) \times 10^{52}$ | $(6.4 \pm 0.61) \times 10^{51}$ |
| 140430A | −2 | ⋯ | ⋯ | $(1.1 \pm 0.2) \times 10^{-6}$ | $(1.54 \pm 0.12) \times 10^{-7}$ | 1.6 | 173.6 | 1 | $(7.32 \pm 1.33) \times 10^{51}$ | $(2.66 \pm 0.21) \times 10^{51}$ |
| 140506A | −1.68 | ⋯ | ⋯ | $(2.8 \pm 0.3) \times 10^{-6}$ | $(7.67 \pm 0.63) \times 10^{-7}$ | 0.889 | 111.1 | 0.82 | $(4.86 \pm 0.52) \times 10^{51}$ | $(2.51 \pm 0.21) \times 10^{51}$ |
| 140512A | −1.45 | ⋯ | ⋯ | $(1.4 \pm 0.03) \times 10^{-5}$ | $(5.28 \pm 0.23) \times 10^{-7}$ | 0.725 | 154.8 | 0.74 | $(1.46 \pm 0.03) \times 10^{52}$ | $(9.5 \pm 0.42) \times 10^{50}$ |
| 140629A | −1.86 | ⋯ | ⋯ | $(2.4 \pm 0.2) \times 10^{-6}$ | $(2.74 \pm 0.26) \times 10^{-7}$ | 2.275 | 42 | 0.85 | $(2.55 \pm 0.21) \times 10^{52}$ | $(9.53 \pm 0.91) \times 10^{51}$ |
| 140703A | −1.74 | ⋯ | ⋯ | $(3.9 \pm 0.3) \times 10^{-6}$ | $(1.92 \pm 0.41) \times 10^{-7}$ | 3.14 | 67.1 | 0.69 | $(5.86 \pm 0.45) \times 10^{52}$ | $(1.19 \pm 0.26) \times 10^{52}$ |
| 140710A | −2 | ⋯ | ⋯ | $(2.3 \pm 0.3) \times 10^{-7}$ | $(1.17 \pm 0.18) \times 10^{-7}$ | 0.558 | 3.52 | 1 | $(1.89 \pm 0.25) \times 10^{50}$ | $(1.5 \pm 0.24) \times 10^{50}$ |
| 140907A | −1.72 | ⋯ | ⋯ | $(4.3 \pm 0.2) \times 10^{-6}$ | $(1.73 \pm 0.14) \times 10^{-7}$ | 1.21 | 79.2 | 0.8 | $(1.35 \pm 0.06) \times 10^{52}$ | $(1.2 \pm 0.1) \times 10^{51}$ |

**Table 3**
(Continued)

| GRB | $\Gamma$ | $\alpha$ | $E_{\rm p}$ (keV) | $S_\gamma$ (erg cm$^{-2}$) | $F_{\rm p}$ (erg cm$^{-2}$ s$^{-1}$) | $z$ | $T_{90}$ (s) | $K$ | $E_{\gamma,\rm iso}$ (erg) | $L_{\rm p}$ (erg s$^{-1}$) |
|---|---|---|---|---|---|---|---|---|---|---|
| (1) | (2) | (3) | (4) | (5) | (6) | (7) | (8) | (9) | (10) | (11) |
| 141004A | −1.86 | ⋯ | ⋯ | $(6.7 \pm 0.3) \times 10^{-7}$ | $(3.97 \pm 0.2) \times 10^{-7}$ | 0.57 | 3.92 | 0.94 | $(5.4 \pm 0.24) \times 10^{50}$ | $(5.03 \pm 0.25) \times 10^{50}$ |
| 141026A | −2.34 | ⋯ | ⋯ | $(1.3 \pm 0.1) \times 10^{-6}$ | $(2.16 \pm 1.08) \times 10^{-8}$ | 3.35 | 146 | 1.65 | $(5.18 \pm 0.4) \times 10^{52}$ | $(3.73 \pm 1.87) \times 10^{51}$ |
| 141109A | −1.52 | ⋯ | ⋯ | $(6.8 \pm 0.3) \times 10^{-6}$ | $(1.88 \pm 0.15) \times 10^{-7}$ | 2.993 | 200 | 0.51 | $(7.02 \pm 0.31) \times 10^{52}$ | $(7.76 \pm 0.62) \times 10^{51}$ |
| 141121A | −1.73 | ⋯ | ⋯ | $(5.3 \pm 0.4) \times 10^{-6}$ | $(6.2 \pm 2.07) \times 10^{-8}$ | 1.47 | 549.9 | 0.78 | $(2.36 \pm 0.18) \times 10^{52}$ | $(6.8 \pm 2.27) \times 10^{50}$ |
| 141221A | −1.74 | ⋯ | ⋯ | $(2.1 \pm 0.1) \times 10^{-6}$ | $(2.12 \pm 0.14) \times 10^{-7}$ | 1.452 | 36.9 | 0.79 | $(9.22 \pm 0.44) \times 10^{51}$ | $(2.29 \pm 0.15) \times 10^{51}$ |
| 141225A | −1.32 | ⋯ | ⋯ | $(2.5 \pm 0.2) \times 10^{-6}$ | $(1.07 \pm 0.33) \times 10^{-7}$ | 0.915 | 40.24 | 0.64 | $(3.62 \pm 0.29) \times 10^{51}$ | $(2.96 \pm 0.91) \times 10^{50}$ |
| 150206A | −1.33 | ⋯ | ⋯ | $(1.39 \pm 0.03) \times 10^{-5}$ | $(8.26 \pm 0.33) \times 10^{-7}$ | 2.087 | 83.2 | 0.47 | $(7.05 \pm 0.15) \times 10^{52}$ | $(1.29 \pm 0.05) \times 10^{52}$ |
| 150301B | −1.46 | ⋯ | ⋯ | $(1.8 \pm 0.1) \times 10^{-6}$ | $(2.32 \pm 0.15) \times 10^{-7}$ | 1.5169 | 12.44 | 0.61 | $(6.58 \pm 0.37) \times 10^{51}$ | $(2.14 \pm 0.14) \times 10^{51}$ |
| 150314A | −1.08 | ⋯ | ⋯ | $(2.2 \pm 0.03) \times 10^{-5}$ | $(3.5 \pm 0.08) \times 10^{-6}$ | 1.758 | 14.79 | 0.39 | $(6.85 \pm 0.09) \times 10^{52}$ | $(3 \pm 0.07) \times 10^{52}$ |
| 150323A | −1.85 | ⋯ | ⋯ | $(6.1 \pm 0.2) \times 10^{-6}$ | $(3.53 \pm 0.2) \times 10^{-7}$ | 0.593 | 149.6 | 0.93 | $(5.3 \pm 0.17) \times 10^{51}$ | $(4.89 \pm 0.27) \times 10^{50}$ |
| 150403A | −1.23 | ⋯ | ⋯ | $(1.7 \pm 0.03) \times 10^{-5}$ | $(1.5 \pm 0.05) \times 10^{-6}$ | 2.06 | 40.9 | 0.42 | $(7.57 \pm 0.13) \times 10^{52}$ | $(2.05 \pm 0.07) \times 10^{52}$ |
| 150413A | −1.75 | ⋯ | ⋯ | $(4.3 \pm 0.4) \times 10^{-6}$ | $(1.09 \pm 0.2) \times 10^{-7}$ | 3.2 | 263.6 | 0.7 | $(6.73 \pm 0.63) \times 10^{52}$ | $(7.18 \pm 1.35) \times 10^{51}$ |
| 150727A | −1.14 | ⋯ | ⋯ | $(3.7 \pm 0.2) \times 10^{-6}$ | $(8.87 \pm 1.77) \times 10^{-8}$ | 0.313 | 88 | 0.79 | $(7.23 \pm 0.39) \times 10^{50}$ | $(2.28 \pm 0.46) \times 10^{49}$ |
| 150910A | −1.42 | ⋯ | ⋯ | $(4.8 \pm 0.4) \times 10^{-6}$ | $(8.65 \pm 3.15) \times 10^{-8}$ | 1.359 | 112.2 | 0.61 | $(1.43 \pm 0.12) \times 10^{52}$ | $(6.07 \pm 2.21) \times 10^{50}$ |
| 150915A | −2.51 | ⋯ | ⋯ | $(8 \pm 1.8) \times 10^{-7}$ | $(2.54 \pm 1.01) \times 10^{-8}$ | 1.968 | 164.7 | 1.74 | $(1.35 \pm 0.3) \times 10^{52}$ | $(1.27 \pm 0.51) \times 10^{51}$ |
| 151021A | −1.57 | ⋯ | ⋯ | $(2.8 \pm 0.1) \times 10^{-5}$ | $(7.15 \pm 0.59) \times 10^{-7}$ | 2.33 | 110.2 | 0.6 | $(2.19 \pm 0.08) \times 10^{53}$ | $(1.86 \pm 0.15) \times 10^{52}$ |
| 151027A | −1.72 | ⋯ | ⋯ | $(7.8 \pm 0.2) \times 10^{-6}$ | $(4.7 \pm 0.41) \times 10^{-7}$ | 0.38 | 129.69 | 0.91 | $(2.64 \pm 0.07) \times 10^{51}$ | $(2.19 \pm 0.19) \times 10^{50}$ |
| 151031A | −2.41 | ⋯ | ⋯ | $(3.2 \pm 0.3) \times 10^{-7}$ | $(8.93 \pm 1.05) \times 10^{-8}$ | 1.167 | 5 | 1.37 | $(1.6 \pm 0.15) \times 10^{51}$ | $(9.69 \pm 1.14) \times 10^{50}$ |
| 151112A | −1.77 | ⋯ | ⋯ | $(9.4 \pm 1.2) \times 10^{-7}$ | $(1.29 \pm 0.14) \times 10^{-7}$ | 4.1 | 19.32 | 0.69 | $(2.16 \pm 0.28) \times 10^{52}$ | $(1.5 \pm 0.16) \times 10^{52}$ |
| 151215A | −1.99 | ⋯ | ⋯ | $(3.1 \pm 0.7) \times 10^{-7}$ | $(9.88 \pm 1.23) \times 10^{-8}$ | 2.59 | 17.8 | 0.99 | $(4.81 \pm 1.09) \times 10^{51}$ | $(5.5 \pm 0.69) \times 10^{51}$ |
| 160121A | −1.77 | ⋯ | ⋯ | $(6.1 \pm 0.5) \times 10^{-7}$ | $(8.12 \pm 1.35) \times 10^{-8}$ | 1.96 | 12 | 0.78 | $(4.58 \pm 0.38) \times 10^{51}$ | $(1.81 \pm 0.3) \times 10^{51}$ |
| 160131A | −1.24 | ⋯ | ⋯ | $(2.04 \pm 0.05) \times 10^{-5}$ | $(5.44 \pm 0.25) \times 10^{-7}$ | 0.97 | 325 | 0.6 | $(3.09 \pm 0.08) \times 10^{52}$ | $(1.62 \pm 0.08) \times 10^{51}$ |
| 160203A | −1.93 | ⋯ | ⋯ | $(1.2 \pm 0.1) \times 10^{-6}$ | $(8.23 \pm 2.53) \times 10^{-8}$ | 3.52 | 20.2 | 0.9 | $(2.83 \pm 0.24) \times 10^{52}$ | $(8.76 \pm 2.69) \times 10^{51}$ |
| 160314A | −1.53 | ⋯ | ⋯ | $(2.8 \pm 0.4) \times 10^{-7}$ | $(6.75 \pm 1.5) \times 10^{-8}$ | 0.726 | 8.73 | 0.77 | $(3.06 \pm 0.44) \times 10^{50}$ | $(1.27 \pm 0.28) \times 10^{50}$ |
| 160327A | −1.84 | ⋯ | ⋯ | $(1.4 \pm 0.1) \times 10^{-6}$ | $(1.18 \pm 0.13) \times 10^{-7}$ | 4.99 | 28 | 0.75 | $(4.75 \pm 0.34) \times 10^{52}$ | $(2.41 \pm 0.27) \times 10^{52}$ |
| 160410A | −0.93 | ⋯ | ⋯ | $(7.8 \pm 0.8) \times 10^{-7}$ | $(3.38 \pm 0.29) \times 10^{-7}$ | 1.717 | 8.2 | 0.34 | $(2.03 \pm 0.21) \times 10^{51}$ | $(2.39 \pm 0.2) \times 10^{51}$ |
| 160425A | −2.2 | ⋯ | ⋯ | $(2.1 \pm 0.2) \times 10^{-6}$ | $(1.59 \pm 0.11) \times 10^{-7}$ | 0.555 | 304.58 | 1.09 | $(1.86 \pm 0.18) \times 10^{51}$ | $(2.2 \pm 0.16) \times 10^{50}$ |
| 161014A | −1.47 | ⋯ | ⋯ | $(1.9 \pm 0.2) \times 10^{-6}$ | $(2.23 \pm 0.46) \times 10^{-7}$ | 2.823 | 18.3 | 0.49 | $(1.7 \pm 0.18) \times 10^{52}$ | $(7.63 \pm 1.58) \times 10^{51}$ |
| 161017A | −1.55 | ⋯ | ⋯ | $(5.3 \pm 0.2) \times 10^{-6}$ | $(2.08 \pm 0.15) \times 10^{-7}$ | 2.0127 | 216.3 | 0.61 | $(3.26 \pm 0.12) \times 10^{52}$ | $(3.86 \pm 0.28) \times 10^{51}$ |
| 161108A | −1.85 | ⋯ | ⋯ | $(1.1 \pm 0.1) \times 10^{-6}$ | $(3.93 \pm 0.65) \times 10^{-8}$ | 1.159 | 105.1 | 0.89 | $(3.53 \pm 0.32) \times 10^{51}$ | $(2.72 \pm 0.45) \times 10^{50}$ |
| 161129A | −1.57 | ⋯ | ⋯ | $(3.6 \pm 0.1) \times 10^{-6}$ | $(2.51 \pm 0.15) \times 10^{-7}$ | 0.645 | 35.53 | 0.81 | $(3.22 \pm 0.09) \times 10^{51}$ | $(3.69 \pm 0.22) \times 10^{50}$ |
| 170202A | −1.68 | ⋯ | ⋯ | $(3.3 \pm 0.1) \times 10^{-6}$ | $(3.31 \pm 0.21) \times 10^{-7}$ | 3.645 | 46.2 | 0.61 | $(5.59 \pm 0.17) \times 10^{52}$ | $(2.6 \pm 0.17) \times 10^{52}$ |
| 170519A | −1.94 | ⋯ | ⋯ | $(1.1 \pm 0.2) \times 10^{-6}$ | $(4.41 \pm 0.63) \times 10^{-8}$ | 0.818 | 216.4 | 0.96 | $(1.91 \pm 0.35) \times 10^{51}$ | $(1.39 \pm 0.2) \times 10^{50}$ |
| 170531B | −1.95 | ⋯ | ⋯ | $(2 \pm 0.2) \times 10^{-6}$ | $(5.02 \pm 1.26) \times 10^{-8}$ | 2.366 | 164.13 | 0.94 | $(2.53 \pm 0.25) \times 10^{52}$ | $(2.14 \pm 0.53) \times 10^{51}$ |
| 170604A | −1.31 | ⋯ | ⋯ | $(5.1 \pm 0.4) \times 10^{-6}$ | $(3.46 \pm 0.74) \times 10^{-7}$ | 1.329 | 26.7 | 0.56 | $(1.33 \pm 0.1) \times 10^{52}$ | $(2.11 \pm 0.45) \times 10^{51}$ |
| 170705A | −1.65 | ⋯ | ⋯ | $(9.5 \pm 0.3) \times 10^{-6}$ | $(9.9 \pm 0.28) \times 10^{-7}$ | 2.01 | 217.3 | 0.68 | $(6.52 \pm 0.21) \times 10^{52}$ | $(2.04 \pm 0.06) \times 10^{52}$ |
| 170903A | −1.94 | ⋯ | ⋯ | $(2.4 \pm 0.2) \times 10^{-6}$ | $(2.46 \pm 0.32) \times 10^{-7}$ | 0.886 | 29.2 | 0.96 | $(4.88 \pm 0.41) \times 10^{51}$ | $(9.43 \pm 1.21) \times 10^{50}$ |
| 171020A | −1.04 | ⋯ | ⋯ | $(1.2 \pm 0.1) \times 10^{-6}$ | $(6.47 \pm 1.85) \times 10^{-8}$ | 1.87 | 41.9 | 0.36 | $(3.86 \pm 0.32) \times 10^{51}$ | $(5.98 \pm 1.71) \times 10^{50}$ |
| 171222A | −2.07 | ⋯ | ⋯ | $(1.9 \pm 0.2) \times 10^{-6}$ | $(4.18 \pm 1.2) \times 10^{-8}$ | 2.409 | 174.8 | 1.09 | $(2.87 \pm 0.3) \times 10^{52}$ | $(2.16 \pm 0.62) \times 10^{51}$ |
| 180115A | −1.66 | ⋯ | ⋯ | $(7.6 \pm 1.1) \times 10^{-7}$ | $(4.26 \pm 1.42) \times 10^{-8}$ | 2.487 | 40.9 | 0.65 | $(7.29 \pm 1.05) \times 10^{51}$ | $(1.42 \pm 0.47) \times 10^{51}$ |
| 180205A | −2.04 | ⋯ | ⋯ | $(1 \pm 0.1) \times 10^{-6}$ | $(2.06 \pm 0.18) \times 10^{-7}$ | 1.409 | 15.5 | 1.04 | $(5.43 \pm 0.54) \times 10^{51}$ | $(2.69 \pm 0.24) \times 10^{51}$ |
| 180325A | −1.18 | ⋯ | ⋯ | $(6.5 \pm 0.2) \times 10^{-6}$ | $(8.19 \pm 0.26) \times 10^{-7}$ | 2.25 | 94.1 | 0.38 | $(3.05 \pm 0.09) \times 10^{52}$ | $(1.25 \pm 0.04) \times 10^{52}$ |
| 180404A | −1.95 | ⋯ | ⋯ | $(1.3 \pm 0.1) \times 10^{-6}$ | $(8.16 \pm 1.26) \times 10^{-8}$ | 1 | 35.2 | 0.97 | $(3.38 \pm 0.26) \times 10^{51}$ | $(4.24 \pm 0.65) \times 10^{50}$ |
| 180510B | −2 | ⋯ | ⋯ | $(2.1 \pm 0.2) \times 10^{-6}$ | $(7.38 \pm 1.23) \times 10^{-8}$ | 1.305 | 134.3 | 1 | $(9.5 \pm 0.9) \times 10^{51}$ | $(7.7 \pm 1.28) \times 10^{50}$ |
| 180624A | −1.91 | ⋯ | ⋯ | $(5.8 \pm 0.3) \times 10^{-6}$ | $(8.93 \pm 1.28) \times 10^{-8}$ | 2.855 | 486.4 | 0.89 | $(9.52 \pm 0.49) \times 10^{52}$ | $(5.65 \pm 0.81) \times 10^{51}$ |
| 181010A | −1.52 | ⋯ | ⋯ | $(6.9 \pm 0.7) \times 10^{-7}$ | $(1.05 \pm 0.15) \times 10^{-7}$ | 1.39 | 16.4 | 0.66 | $(2.32 \pm 0.24) \times 10^{51}$ | $(8.47 \pm 1.21) \times 10^{50}$ |

**Table 3**
(Continued)

| GRB (1) | $\Gamma$ (2) | $\alpha$ (3) | $E_p$ (keV) (4) | $S_\gamma$ (erg cm$^{-2}$) (5) | $F_p$ (erg cm$^{-2}$ s$^{-1}$) (6) | $z$ (7) | $T_{90}$ (s) (8) | $K$ (9) | $E_{\gamma,\mathrm{iso}}$ (erg) (10) | $L_p$ (erg s$^{-1}$) (11) |
|---|---|---|---|---|---|---|---|---|---|---|
| 181020A | −1.25 | ··· | ··· | $(8.1 \pm 0.3) \times 10^{-6}$ | $(6.52 \pm 0.25) \times 10^{-7}$ | 2.938 | 238 | 0.36 | $(5.64 \pm 0.21) \times 10^{52}$ | $(1.79 \pm 0.07) \times 10^{52}$ |
| 181110A | −2.02 | ··· | ··· | $(9.9 \pm 0.3) \times 10^{-6}$ | $(2.26 \pm 0.18) \times 10^{-7}$ | 1.505 | 138.4 | 1.02 | $(5.98 \pm 0.18) \times 10^{52}$ | $(3.42 \pm 0.28) \times 10^{51}$ |
| 190114A | −2.06 | ··· | ··· | $(8 \pm 1.2) \times 10^{-7}$ | $(3 \pm 1.2) \times 10^{-8}$ | 3.3765 | 66.6 | 1.09 | $(2.14 \pm 0.32) \times 10^{52}$ | $(3.51 \pm 1.4) \times 10^{51}$ |
| 190324A | −1.46 | ··· | ··· | $(7.2 \pm 0.2) \times 10^{-6}$ | $(9.2 \pm 0.39) \times 10^{-7}$ | 1.1715 | 28.4 | 0.66 | $(1.74 \pm 0.05) \times 10^{52}$ | $(4.83 \pm 0.2) \times 10^{51}$ |
| 190719C | −1.64 | ··· | ··· | $(5.1 \pm 0.3) \times 10^{-6}$ | $(3.93 \pm 0.21) \times 10^{-7}$ | 2.469 | 185.7 | 0.64 | $(4.72 \pm 0.28) \times 10^{52}$ | $(1.26 \pm 0.07) \times 10^{52}$ |
| 191004B | −1.1 | ··· | ··· | $(2.5 \pm 0.1) \times 10^{-6}$ | $(4.51 \pm 0.18) \times 10^{-7}$ | 3.503 | 37.7 | 0.26 | $(1.68 \pm 0.07) \times 10^{52}$ | $(1.36 \pm 0.05) \times 10^{52}$ |
| 191011A | −1.94 | ··· | ··· | $(3.3 \pm 0.4) \times 10^{-7}$ | $(1.13 \pm 0.13) \times 10^{-7}$ | 1.722 | 7.37 | 0.94 | $(2.37 \pm 0.29) \times 10^{51}$ | $(2.22 \pm 0.25) \times 10^{51}$ |
| 191019A | −2.25 | ··· | ··· | $(1 \pm 0.03) \times 10^{-5}$ | $(3.18 \pm 0.22) \times 10^{-7}$ | 0.248 | 64.35 | 1.06 | $(1.61 \pm 0.05) \times 10^{51}$ | $(6.38 \pm 0.45) \times 10^{49}$ |
| 191221B | −1.24 | ··· | ··· | $(1.9 \pm 0.1) \times 10^{-5}$ | $(4.08 \pm 0.59) \times 10^{-7}$ | 1.19 | 48 | 0.55 | $(3.97 \pm 0.21) \times 10^{52}$ | $(1.86 \pm 0.27) \times 10^{51}$ |
| 200829A | −0.83 | ··· | ··· | $(5.1 \pm 0.1) \times 10^{-5}$ | $(9.92 \pm 0.16) \times 10^{-6}$ | 1.25 | 13.04 | 0.39 | $(8.22 \pm 0.16) \times 10^{52}$ | $(3.6 \pm 0.06) \times 10^{52}$ |
| 201014A | −2.55 | ··· | ··· | $(3.1 \pm 0.7) \times 10^{-7}$ | $(1.5 \pm 1) \times 10^{-8}$ | 4.56 | 36.2 | 2.57 | $(3.14 \pm 0.71) \times 10^{52}$ | $(8.45 \pm 5.63) \times 10^{51}$ |
| 201020A | −2.25 | ··· | ··· | $(9.2 \pm 0.7) \times 10^{-7}$ | $(1.17 \pm 0.11) \times 10^{-7}$ | 2.903 | 14.17 | 1.41 | $(2.47 \pm 0.19) \times 10^{52}$ | $(1.22 \pm 0.12) \times 10^{52}$ |
| 201021C | −1.63 | ··· | ··· | $(1 \pm 0.1) \times 10^{-6}$ | $(7.19 \pm 1.44) \times 10^{-8}$ | 1.07 | 24.38 | 0.76 | $(2.35 \pm 0.23) \times 10^{51}$ | $(3.49 \pm 0.7) \times 10^{50}$ |
| 201024A | −2.1 | ··· | ··· | $(8.1 \pm 0.9) \times 10^{-7}$ | $(2.19 \pm 0.35) \times 10^{-7}$ | 0.999 | 5 | 1.07 | $(2.33 \pm 0.26) \times 10^{51}$ | $(1.26 \pm 0.2) \times 10^{51}$ |
| 201104B | −1.47 | ··· | ··· | $(1.8 \pm 0.1) \times 10^{-6}$ | $(3.16 \pm 0.23) \times 10^{-7}$ | 1.954 | 8.66 | 0.56 | $(9.72 \pm 0.54) \times 10^{51}$ | $(5.04 \pm 0.37) \times 10^{51}$ |
| 201216C | −1.43 | ··· | ··· | $(4.5 \pm 0.1) \times 10^{-5}$ | $(1.41 \pm 0.09) \times 10^{-6}$ | 1.1 | 48 | 0.66 | $(9.57 \pm 0.21) \times 10^{52}$ | $(6.3 \pm 0.38) \times 10^{51}$ |
| 201221A | −1.4 | ··· | ··· | $(1.8 \pm 0.2) \times 10^{-6}$ | $(7.94 \pm 2.38) \times 10^{-8}$ | 5.7 | 44.5 | 0.32 | $(3.18 \pm 0.35) \times 10^{52}$ | $(9.38 \pm 2.81) \times 10^{51}$ |
| 210222B | −2.37 | ··· | ··· | $(3.4 \pm 0.7) \times 10^{-7}$ | $(7.46 \pm 1.6) \times 10^{-8}$ | 2.198 | 12.82 | 1.54 | $(6.18 \pm 1.27) \times 10^{51}$ | $(4.34 \pm 0.93) \times 10^{51}$ |
| 210321A | −2.15 | ··· | ··· | $(6.7 \pm 0.8) \times 10^{-7}$ | $(9.84 \pm 2.32) \times 10^{-8}$ | 1.487 | 8.21 | 1.15 | $(4.45 \pm 0.53) \times 10^{51}$ | $(1.63 \pm 0.38) \times 10^{51}$ |
| 210420B | −1.49 | ··· | ··· | $(1.5 \pm 0.2) \times 10^{-6}$ | $(3.82 \pm 1.53) \times 10^{-8}$ | 1.4 | 158.8 | 0.64 | $(4.97 \pm 0.66) \times 10^{51}$ | $(3.03 \pm 1.21) \times 10^{50}$ |
| 210504A | −1.64 | ··· | ··· | $(2.7 \pm 0.2) \times 10^{-6}$ | $(6.44 \pm 1.43) \times 10^{-8}$ | 2.077 | 135.06 | 0.67 | $(1.93 \pm 0.14) \times 10^{52}$ | $(1.41 \pm 0.31) \times 10^{51}$ |
| 210517A | −1.85 | ··· | ··· | $(1.5 \pm 0.3) \times 10^{-7}$ | $(9.81 \pm 1.31) \times 10^{-8}$ | 2.486 | 3.06 | 0.83 | $(1.82 \pm 0.36) \times 10^{51}$ | $(4.16 \pm 0.55) \times 10^{51}$ |
| 210610A | −1.41 | ··· | ··· | $(1 \pm 0.1) \times 10^{-6}$ | $(1.9 \pm 0.32) \times 10^{-7}$ | 3.54 | 13.62 | 0.41 | $(1.08 \pm 0.11) \times 10^{52}$ | $(9.32 \pm 1.55) \times 10^{51}$ |
| 210619B | −1.41 | ··· | ··· | $(9.5 \pm 0.1) \times 10^{-5}$ | $(9.09 \pm 0.17) \times 10^{-6}$ | 1.937 | 60.9 | 0.53 | $(4.75 \pm 0.05) \times 10^{53}$ | $(1.33 \pm 0.03) \times 10^{53}$ |
| 210722A | −1.62 | ··· | ··· | $(2.5 \pm 0.2) \times 10^{-6}$ | $(1.59 \pm 0.36) \times 10^{-7}$ | 1.145 | 50.2 | 0.75 | $(6.57 \pm 0.53) \times 10^{51}$ | $(8.95 \pm 2.03) \times 10^{50}$ |
| 210822A | −1.3 | ··· | ··· | $(2 \pm 0.04) \times 10^{-5}$ | $(2.29 \pm 0.06) \times 10^{-6}$ | 1.736 | 180.8 | 0.49 | $(7.65 \pm 0.15) \times 10^{52}$ | $(2.4 \pm 0.06) \times 10^{52}$ |
| 211024B | −1.63 | ··· | ··· | $(6.8 \pm 0.3) \times 10^{-6}$ | $(6.47 \pm 1.44) \times 10^{-8}$ | 1.1137 | 603.5 | 0.76 | $(1.72 \pm 0.08) \times 10^{52}$ | $(3.45 \pm 0.77) \times 10^{50}$ |
| 211207A | −1.72 | ··· | ··· | $(1.9 \pm 0.4) \times 10^{-7}$ | $(6.22 \pm 1.38) \times 10^{-8}$ | 2.272 | 3.73 | 0.72 | $(1.71 \pm 0.36) \times 10^{51}$ | $(1.83 \pm 0.41) \times 10^{51}$ |
| 220117A | −1.84 | ··· | ··· | $(1.6 \pm 0.2) \times 10^{-6}$ | $(1.25 \pm 0.26) \times 10^{-7}$ | 4.961 | 49.81 | 0.75 | $(5.39 \pm 0.67) \times 10^{52}$ | $(2.51 \pm 0.53) \times 10^{52}$ |
| 220521A | −1.97 | ··· | ··· | $(8.1 \pm 0.8) \times 10^{-7}$ | $(2.93 \pm 0.25) \times 10^{-7}$ | 5.6 | 13.55 | 0.94 | $(4.12 \pm 0.41) \times 10^{52}$ | $(9.82 \pm 0.84) \times 10^{52}$ |
| 221110A | −1.41 | ··· | ··· | $(5.1 \pm 0.5) \times 10^{-7}$ | $(8.69 \pm 1.58) \times 10^{-8}$ | 4.06 | 8.98 | 0.38 | $(6.44 \pm 0.63) \times 10^{51}$ | $(5.55 \pm 1.01) \times 10^{51}$ |
| 221226B | −1.16 | ··· | ··· | $(5.4 \pm 0.5) \times 10^{-7}$ | $(2.29 \pm 0.26) \times 10^{-7}$ | 2.694 | 3.44 | 0.33 | $(3.03 \pm 0.28) \times 10^{51}$ | $(4.74 \pm 0.55) \times 10^{51}$ |
| 230116D | −1.38 | ··· | ··· | $(8.12 \pm 1.2) \times 10^{-7}$ | $(4.8 \pm 1.6) \times 10^{-8}$ | 3.81 | 41 | 0.38 | $(9.11 \pm 1.35) \times 10^{51}$ | $(2.59 \pm 0.86) \times 10^{51}$ |
| 230325A | −1.82 | ··· | ··· | $(1.3 \pm 0.2) \times 10^{-6}$ | $(2.19 \pm 0.27) \times 10^{-7}$ | 1.664 | 38.05 | 0.84 | $(7.8 \pm 1.2) \times 10^{51}$ | $(3.49 \pm 0.42) \times 10^{51}$ |
| 230414B | −1.6 | ··· | ··· | $(4.7 \pm 1) \times 10^{-7}$ | $(4.37 \pm 1.46) \times 10^{-8}$ | 3.568 | 25.98 | 0.54 | $(6.85 \pm 1.46) \times 10^{51}$ | $(2.91 \pm 0.97) \times 10^{51}$ |
| 230506C | −1.89 | ··· | ··· | $(1.7 \pm 0.2) \times 10^{-6}$ | $(1.54 \pm 0.32) \times 10^{-7}$ | 3.7 | 31 | 0.84 | $(4.07 \pm 0.48) \times 10^{52}$ | $(1.74 \pm 0.36) \times 10^{52}$ |
| 230818A | −1.36 | ··· | ··· | $(1.9 \pm 0.1) \times 10^{-6}$ | $(5.41 \pm 0.4) \times 10^{-7}$ | 2.42 | 9.82 | 0.46 | $(1.21 \pm 0.06) \times 10^{52}$ | $(1.18 \pm 0.09) \times 10^{52}$ |
| 231118A | −1.4 | ··· | ··· | $(3.7 \pm 0.2) \times 10^{-6}$ | $(7.7 \pm 0.4) \times 10^{-7}$ | 0.8304 | 37.63 | 0.7 | $(4.77 \pm 0.26) \times 10^{51}$ | $(1.82 \pm 0.09) \times 10^{51}$ |
| 231210B | −1.08 | ··· | ··· | $(1.4 \pm 0.2) \times 10^{-6}$ | $(3.55 \pm 0.64) \times 10^{-7}$ | 3.13 | 7.47 | 0.27 | $(8.2 \pm 1.17) \times 10^{51}$ | $(8.58 \pm 1.54) \times 10^{51}$ |
| 231215A | −1.01 | ··· | ··· | $(1 \pm 0.1) \times 10^{-5}$ | $(8.98 \pm 1.31) \times 10^{-7}$ | 2.305 | 20.75 | 0.31 | $(3.93 \pm 0.39) \times 10^{52}$ | $(1.17 \pm 0.17) \times 10^{52}$ |
| 240205B | −2.23 | ··· | ··· | $2.6 \times 10^{-5}$ | $(2.68 \pm 0.06) \times 10^{-6}$ | 0.824 | 47.29 | 1.15 | $5.45 \times 10^{52}$ | $(1.03 \pm 0.02) \times 10^{52}$ |
| 240419A | −1.71 | ··· | ··· | $(8.4 \pm 2.1) \times 10^{-8}$ | $(4.17 \pm 0.69) \times 10^{-8}$ | 5.178 | 3 | 0.59 | $(2.37 \pm 0.59) \times 10^{51}$ | $(7.26 \pm 1.21) \times 10^{51}$ |
| 240529A | −1.68 | ··· | ··· | $2.1 \times 10^{-5}$ | $(5.41 \pm 0.42) \times 10^{-7}$ | 2.695 | 160.67 | 0.66 | $2.32 \times 10^{53}$ | $(2.21 \pm 0.17) \times 10^{52}$ |
| 240809A | −1.17 | ··· | ··· | $(9.7 \pm 0.2) \times 10^{-6}$ | $(1.12 \pm 0.04) \times 10^{-6}$ | 1.495 | 72.86 | 0.47 | $(2.66 \pm 0.05) \times 10^{52}$ | $(7.67 \pm 0.24) \times 10^{51}$ |
| 240825A | −1.2 | ··· | ··· | $(2.5 \pm 0.04) \times 10^{-5}$ | $(8.64 \pm 0.15) \times 10^{-6}$ | 0.659 | 57.2 | 0.67 | $(1.93 \pm 0.03) \times 10^{52}$ | $(1.11 \pm 0.02) \times 10^{52}$ |
| 241010A | −1.88 | ··· | ··· | $(2 \pm 0.1) \times 10^{-6}$ | $(2.78 \pm 0.13) \times 10^{-7}$ | 0.977 | 30.86 | 0.92 | $(4.73 \pm 0.24) \times 10^{51}$ | $(1.3 \pm 0.06) \times 10^{51}$ |

**Table 3**
(Continued)

| GRB (1) | $\Gamma$ (2) | $\alpha$ (3) | $E_p$ (keV) (4) | $S_\gamma$ (erg cm$^{-2}$) (5) | $F_p$ (erg cm$^{-2}$ s$^{-1}$) (6) | $z$ (7) | $T_{90}$ (s) (8) | $K$ (9) | $E_{\gamma,iso}$ (erg) (10) | $L_p$ (erg s$^{-1}$) (11) |
|---|---|---|---|---|---|---|---|---|---|---|
| 241026A | −1.56 | ··· | ··· | $(2.5 \pm 0.2) \times 10^{-6}$ | $(2.96 \pm 0.37) \times 10^{-7}$ | 2.79 | 25.2 | 0.56 | $(2.48 \pm 0.2) \times 10^{52}$ | $(1.11 \pm 0.14) \times 10^{52}$ |
| 241030A | −1.64 | ··· | ··· | $(2.73 \pm 0.04) \times 10^{-5}$ | $(8.44 \pm 0.21) \times 10^{-7}$ | 1.411 | 173.3 | 0.73 | $(1.04 \pm 0.02) \times 10^{53}$ | $(7.79 \pm 0.2) \times 10^{51}$ |
| 241030B | −1.4 | ··· | ··· | $(1.2 \pm 0.1) \times 10^{-6}$ | $(2.54 \pm 0.48) \times 10^{-7}$ | 2.82 | 6.7 | 0.45 | $(9.75 \pm 0.81) \times 10^{51}$ | $(7.88 \pm 1.48) \times 10^{51}$ |
| 060115 | ··· | −1.01 | 66.11 | $(1.72 \pm 0.2) \times 10^{-5}$ | $(5.88 \pm 0.81) \times 10^{-8}$ | 3.53 | 139.6 | 0.5 | $(8.7 \pm 2.25) \times 10^{52}$ | $(3.5 \pm 0.48) \times 10^{51}$ |
| 060206 | ··· | −1.12 | 80.74 | $(3.73 \pm 0.23) \times 10^{-6}$ | $(1.96 \pm 0.12) \times 10^{-7}$ | 4.045 | 7.6 | 0.44 | $(1.17 \pm 0.12) \times 10^{53}$ | $(1.43 \pm 0.09) \times 10^{52}$ |
| 060707 | ··· | −0.42 | 61.95 | $(8.92 \pm 4.4) \times 10^{-6}$ | $(6.97 \pm 1.59) \times 10^{-8}$ | 3.43 | 66.2 | 0.4 | $(6.23 \pm 1.61) \times 10^{52}$ | $(3.11 \pm 0.71) \times 10^{51}$ |
| 060908 | ··· | −0.93 | 147.69 | $(1.94 \pm 1.4) \times 10^{-6}$ | $(2.48 \pm 0.2) \times 10^{-7}$ | 1.8836 | 19.3 | 0.46 | $(6.13 \pm 1.16) \times 10^{52}$ | $(2.96 \pm 0.24) \times 10^{51}$ |
| 060927 | ··· | −0.81 | 70.67 | $(4.65 \pm 0.22) \times 10^{-6}$ | $(1.9 \pm 0.12) \times 10^{-7}$ | 5.6 | 22.5 | 0.3 | $(1.13 \pm 0.18) \times 10^{53}$ | $(2.01 \pm 0.13) \times 10^{52}$ |
| 070508 | ··· | −1.22 | 443.85 | $(1.99 \pm 0.98) \times 10^{-6}$ | $(1.98 \pm 0.05) \times 10^{-6}$ | 0.82 | 20.9 | 0.67 | $(5.34 \pm 2.36) \times 10^{52}$ | $(4.34 \pm 0.11) \times 10^{51}$ |
| 070521 | ··· | −1.11 | 237.68 | $(1.16 \pm 0.26) \times 10^{-5}$ | $(5.37 \pm 0.22) \times 10^{-7}$ | 0.553 | 37.9 | 0.75 | $(1.43 \pm 0.48) \times 10^{52}$ | $(5.03 \pm 0.21) \times 10^{50}$ |
| 080207 | ··· | −1.25 | 124.61 | $(8.07 \pm 0.83) \times 10^{-6}$ | $(7.38 \pm 2.21) \times 10^{-8}$ | 2.0858 | 340 | 0.57 | $(7.67 \pm 3.75) \times 10^{52}$ | $(1.4 \pm 0.42) \times 10^{51}$ |
| 080413B | ··· | −1.23 | 77.65 | $(1.11 \pm 0.06) \times 10^{-5}$ | $(1.28 \pm 0.05) \times 10^{-6}$ | 1.1 | 8 | 0.8 | $(2.01 \pm 0.83) \times 10^{52}$ | $(6.97 \pm 0.3) \times 10^{51}$ |
| 080603B | ··· | −1.21 | 74.94 | $(7.76 \pm 3.2) \times 10^{-6}$ | $(2.39 \pm 0.14) \times 10^{-7}$ | 2.69 | 60 | 0.6 | $(7.54 \pm 2.41) \times 10^{52}$ | $(8.86 \pm 0.51) \times 10^{51}$ |
| 080605 | ··· | −1.15 | 290.7 | $(3.4 \pm 0.39) \times 10^{-6}$ | $(1.64 \pm 0.05) \times 10^{-6}$ | 1.6398 | 20 | 0.5 | $(1.01 \pm 0.46) \times 10^{53}$ | $(1.51 \pm 0.05) \times 10^{52}$ |
| 080913 | ··· | −0.39 | 111.5 | $(2.29 \pm 1) \times 10^{-5}$ | $(1.21 \pm 0.17) \times 10^{-7}$ | 6.44 | 8 | 0.11 | $(8.25 \pm 0.41) \times 10^{52}$ | $(6.64 \pm 0.95) \times 10^{51}$ |
| 080916A | ··· | −1.19 | 120.96 | $(5.32 \pm 0.59) \times 10^{-6}$ | $(2.01 \pm 0.15) \times 10^{-7}$ | 0.689 | 60 | 0.79 | $(1.21 \pm 0.4) \times 10^{52}$ | $(3.41 \pm 0.25) \times 10^{50}$ |
| 081121 | ··· | −0.69 | 208.39 | $(4.44 \pm 0.25) \times 10^{-6}$ | $(4.04 \pm 0.92) \times 10^{-7}$ | 2.512 | 14 | 0.28 | $(8.75 \pm 1.72) \times 10^{52}$ | $(5.95 \pm 1.35) \times 10^{51}$ |
| 081221 | ··· | −1.3 | 90.39 | $(1.44 \pm 1.2) \times 10^{-6}$ | $(1.27 \pm 0.03) \times 10^{-6}$ | 2.26 | 34 | 0.63 | $(7.06 \pm 14.14) \times 10^{52}$ | $(3.23 \pm 0.09) \times 10^{52}$ |
| 081222 | ··· | −1.2 | 217.88 | $(1.2 \pm 1.18) \times 10^{-5}$ | $(6.11 \pm 0.16) \times 10^{-7}$ | 2.77 | 24 | 0.42 | $(1.62 \pm 0.36) \times 10^{53}$ | $(1.71 \pm 0.04) \times 10^{52}$ |
| 090423 | ··· | −0.76 | 53.23 | $(2.08 \pm 0.41) \times 10^{-5}$ | $(1.08 \pm 0.13) \times 10^{-7}$ | 8 | 10.3 | 0.28 | $(1.33 \pm 0.15) \times 10^{53}$ | $(2.43 \pm 0.29) \times 10^{52}$ |
| 090424 | ··· | −1.24 | 152.79 | $(1.03 \pm 0.43) \times 10^{-5}$ | $(5.38 \pm 0.15) \times 10^{-6}$ | 0.544 | 48 | 0.81 | $(9.7 \pm 13.33) \times 10^{51}$ | $(5.27 \pm 0.15) \times 10^{51}$ |
| 090926B | ··· | −0.52 | 80.55 | $(1.48 \pm 0.28) \times 10^{-5}$ | $(2.43 \pm 0.23) \times 10^{-7}$ | 1.24 | 109.7 | 0.64 | $(2.12 \pm 1.92) \times 10^{52}$ | $(1.43 \pm 0.13) \times 10^{51}$ |
| 091018 | ··· | −1.76 | 19.43 | $(2.91 \pm 1.33) \times 10^{-5}$ | $(5.56 \pm 0.22) \times 10^{-7}$ | 0.971 | 4.4 | 1.17 | $(5.76 \pm 4.15) \times 10^{51}$ | $(3.25 \pm 0.13) \times 10^{51}$ |
| 091029 | ··· | −1.46 | 61.32 | $(7.18 \pm 0.67) \times 10^{-6}$ | $(1.15 \pm 0.06) \times 10^{-7}$ | 2.752 | 39.2 | 0.74 | $(7.89 \pm 3.09) \times 10^{52}$ | $(5.56 \pm 0.31) \times 10^{51}$ |
| 100816A | ··· | −0.74 | 172.19 | $(4.44 \pm 1.96) \times 10^{-5}$ | $(9.6 \pm 0.35) \times 10^{-7}$ | 0.8034 | 2.9 | 0.62 | $(1.84 \pm 0.21) \times 10^{52}$ | $(1.85 \pm 0.07) \times 10^{51}$ |
| 110422A | ··· | −0.83 | 148 | $(7.07 \pm 1.13) \times 10^{-6}$ | $(2.57 \pm 0.08) \times 10^{-6}$ | 1.77 | 25.9 | 0.45 | $(5.35 \pm 14.81) \times 10^{52}$ | $(2.57 \pm 0.08) \times 10^{52}$ |
| 110503A | ··· | −1.02 | 229.44 | $(1.07 \pm 0.22) \times 10^{-5}$ | $(1.14 \pm 0.05) \times 10^{-7}$ | 1.613 | 10 | 0.48 | $(7.42 \pm 3.24) \times 10^{52}$ | $(9.61 \pm 0.43) \times 10^{50}$ |
| 110715A | ··· | −1.25 | 119.8 | $(5.64 \pm 0.59) \times 10^{-6}$ | $(3.95 \pm 0.08) \times 10^{-6}$ | 0.82 | 13 | 0.78 | $(1.68 \pm 1.66) \times 10^{52}$ | $(1.01 \pm 0.02) \times 10^{52}$ |
| 110726A | ··· | −0.64 | 46.53 | $(2.18 \pm 0.48) \times 10^{-5}$ | $(6.05 \pm 1.21) \times 10^{-8}$ | 1.036 | 5.2 | 0.93 | $(1.26 \pm 0.06) \times 10^{52}$ | $(3.32 \pm 0.66) \times 10^{50}$ |
| 120326A | ··· | −1.4 | 46.84 | $(1.48 \pm 4.1) \times 10^{-5}$ | $(2.8 \pm 0.12) \times 10^{-7}$ | 1.798 | 69.6 | 0.89 | $(3.43 \pm 1.91) \times 10^{52}$ | $(5.74 \pm 0.25) \times 10^{51}$ |
| 120724A | ··· | −0.75 | 26.73 | $(4.73 \pm 3) \times 10^{-6}$ | $(2.88 \pm 0.96) \times 10^{-8}$ | 1.48 | 72.8 | 1.24 | $(1.91 \pm 0.48) \times 10^{52}$ | $(5.1 \pm 1.7) \times 10^{50}$ |
| 120802A | ··· | −1.22 | 57.26 | $(9.04 \pm 18.1) \times 10^{-6}$ | $(1.92 \pm 0.13) \times 10^{-7}$ | 3.796 | 50 | 0.6 | $(1.01 \pm 0.33) \times 10^{53}$ | $(1.62 \pm 0.11) \times 10^{52}$ |
| 120811C | ··· | −1.33 | 47.31 | $(2.38 \pm 0.8) \times 10^{-5}$ | $(2.5 \pm 0.12) \times 10^{-7}$ | 2.671 | 26.8 | 0.79 | $(6.21 \pm 3.94) \times 10^{52}$ | $(1.2 \pm 0.06) \times 10^{52}$ |
| 120922A | ··· | −1.58 | 46.09 | $(6.13 \pm 2.4) \times 10^{-6}$ | $(1.22 \pm 0.12) \times 10^{-7}$ | 3.1 | 173 | 0.84 | $(8.23 \pm 11.07) \times 10^{52}$ | $(8.91 \pm 0.89) \times 10^{51}$ |
| 130420A | ··· | −1.52 | 33.36 | $(1.53 \pm 2.1) \times 10^{-5}$ | $(1.93 \pm 0.11) \times 10^{-7}$ | 1.297 | 123.5 | 1.06 | $(1.58 \pm 3.36) \times 10^{52}$ | $(2.1 \pm 0.12) \times 10^{51}$ |
| 130612A | ··· | 0.84 | 37.26 | $(5.73 \pm 1.9) \times 10^{-6}$ | $(9.01 \pm 1.59) \times 10^{-8}$ | 2.006 | 4 | 0.87 | $(3.24 \pm 0.2) \times 10^{52}$ | $(2.36 \pm 0.42) \times 10^{51}$ |
| 130701A | ··· | −0.9 | 89.16 | $(9.42 \pm 11) \times 10^{-6}$ | $(1.27 \pm 0.05) \times 10^{-6}$ | 1.155 | 4.38 | 0.69 | $(2.2 \pm 1.08) \times 10^{52}$ | $(6.73 \pm 0.28) \times 10^{51}$ |
| 131117A | ··· | 0.17 | 44.35 | $(1.21 \pm 0.4) \times 10^{-5}$ | $(4.14 \pm 0.59) \times 10^{-8}$ | 4.18 | 11 | 0.42 | $(6.34 \pm 0.36) \times 10^{52}$ | $(3.06 \pm 0.44) \times 10^{51}$ |
| 140206A | ··· | −1.14 | 133.61 | $(7.49 \pm 2.4) \times 10^{-6}$ | $(1.48 \pm 0.04) \times 10^{-6}$ | 2.73 | 93.6 | 0.46 | $(1.05 \pm 1.25) \times 10^{53}$ | $(4.33 \pm 0.11) \times 10^{52}$ |
| 140515A | ··· | −0.98 | 56.37 | $(4.79 \pm 1) \times 10^{-6}$ | $(5.81 \pm 0.65) \times 10^{-8}$ | 6.32 | 23.4 | 0.38 | $(1.38 \pm 0.14) \times 10^{53}$ | $(1.04 \pm 0.12) \times 10^{52}$ |
| 140518A | ··· | −0.98 | 47.92 | $(6.19 \pm 1.6) \times 10^{-6}$ | $(6.11 \pm 0.61) \times 10^{-8}$ | 4.707 | 60.5 | 0.53 | $(1.05 \pm 0.22) \times 10^{53}$ | $(7.65 \pm 0.77) \times 10^{51}$ |
| 141220A | ··· | −0.61 | 116.1 | $(1.42 \pm 0.52) \times 10^{-5}$ | $(7.43 \pm 0.58) \times 10^{-7}$ | 1.3195 | 7.21 | 0.53 | $(2.85 \pm 0.64) \times 10^{52}$ | $(4.22 \pm 0.33) \times 10^{51}$ |
| 151029A | ··· | −0.28 | 33.95 | $(4.28 \pm 3.6) \times 10^{-5}$ | $(9.31 \pm 1.55) \times 10^{-8}$ | 1.423 | 8.95 | 1.07 | $(1.93 \pm 0.22) \times 10^{52}$ | $(1.28 \pm 0.21) \times 10^{51}$ |
| 160227A | ··· | −0.75 | 75.53 | $(4.68 \pm 2.6) \times 10^{-6}$ | $(4.33 \pm 0.72) \times 10^{-8}$ | 2.38 | 316.5 | 0.51 | $(5.21 \pm 2.14) \times 10^{52}$ | $(1.01 \pm 0.17) \times 10^{51}$ |
| 160804A | ··· | −1.41 | 64.23 | $(2.67 \pm 0.68) \times 10^{-6}$ | $(1.88 \pm 0.19) \times 10^{-7}$ | 0.736 | 144.2 | 0.92 | $(8.59 \pm 15.25) \times 10^{51}$ | $(4.36 \pm 0.45) \times 10^{50}$ |

**Table 3**
(Continued)

| GRB (1) | $\Gamma$ (2) | $\alpha$ (3) | $E_p$ (keV) (4) | $S_\gamma$ (erg cm$^{-2}$) (5) | $F_p$ (erg cm$^{-2}$ s$^{-1}$) (6) | $z$ (7) | $T_{90}$ (s) (8) | $K$ (9) | $E_{\gamma,iso}$ (erg) (10) | $L_p$ (erg s$^{-1}$) (11) |
|---|---|---|---|---|---|---|---|---|---|---|
| 161117A | ··· | −1.2 | 73.05 | $(1.55 \pm 0.6) \times 10^{-5}$ | $(4.62 \pm 0.2) \times 10^{-7}$ | 1.549 | 125.7 | 0.74 | $(3.37 \pm 9.23) \times 10^{52}$ | $(5.43 \pm 0.24) \times 10^{51}$ |
| 170113A | ··· | −0.74 | 71.81 | $(1.26 \pm 0.35) \times 10^{-5}$ | $(7.82 \pm 0.71) \times 10^{-8}$ | 1.968 | 20.66 | 0.58 | $(4.05 \pm 0.38) \times 10^{52}$ | $(1.31 \pm 0.12) \times 10^{51}$ |
| 180314A | ··· | −0.84 | 94.19 | $(1.34 \pm 1.6) \times 10^{-5}$ | $(6 \pm 0.46) \times 10^{-7}$ | 1.445 | 51.2 | 0.6 | $(3.13 \pm 3.65) \times 10^{52}$ | $(4.87 \pm 0.37) \times 10^{51}$ |
| 180329B | ··· | −0.9 | 50.14 | $(8.05 \pm 7.3) \times 10^{-6}$ | $(8.7 \pm 2.49) \times 10^{-8}$ | 1.998 | 210 | 0.76 | $(3.81 \pm 2.51) \times 10^{52}$ | $(1.98 \pm 0.57) \times 10^{51}$ |
| 180620B | ··· | −1.31 | 137.14 | $(1.94 \pm 4.5) \times 10^{-5}$ | $(2.65 \pm 0.15) \times 10^{-7}$ | 1.1175 | 198.8 | 0.72 | $(3.29 \pm 2.4) \times 10^{52}$ | $(1.34 \pm 0.07) \times 10^{51}$ |
| 190106A | ··· | −1.37 | 154.63 | $(3.34 \pm 7.1) \times 10^{-6}$ | $(4.03 \pm 0.22) \times 10^{-7}$ | 1.86 | 76.8 | 0.62 | $(8.41 \pm 3.26) \times 10^{52}$ | $(6.26 \pm 0.34) \times 10^{51}$ |
| 200205B | ··· | −1.35 | 61.08 | $(7.31 \pm 20) \times 10^{-6}$ | $(1.29 \pm 0.13) \times 10^{-7}$ | 1.465 | 458 | 0.83 | $(2.86 \pm 2.53) \times 10^{52}$ | $(1.49 \pm 0.15) \times 10^{51}$ |
| 210210A | ··· | −1.53 | 19.91 | $(6.61 \pm 1.71) \times 10^{-6}$ | $(3.49 \pm 0.25) \times 10^{-7}$ | 0.715 | 6.6 | 1.23 | $(3.36 \pm 1.65) \times 10^{51}$ | $(1.01 \pm 0.07) \times 10^{51}$ |
| 210411C | ··· | −1.62 | 14.36 | $(6.42 \pm 11.4) \times 10^{-6}$ | $(2.27 \pm 0.14) \times 10^{-7}$ | 2.826 | 12.8 | 1.53 | $(4 \pm 3.34) \times 10^{52}$ | $(2.41 \pm 0.15) \times 10^{52}$ |
| 210610B | ··· | −0.98 | 428.31 | $(4.61 \pm 6.2) \times 10^{-6}$ | $(1.21 \pm 0.06) \times 10^{-6}$ | 1.13 | 69.38 | 0.51 | $(7.49 \pm 6.3) \times 10^{52}$ | $(4.51 \pm 0.23) \times 10^{51}$ |
| 210731A | ··· | 0.25 | 106.54 | $(4.83 \pm 2.5) \times 10^{-5}$ | $(1.51 \pm 0.28) \times 10^{-7}$ | 1.2525 | 22.51 | 0.44 | $(1.96 \pm 0.41) \times 10^{52}$ | $(6.27 \pm 1.17) \times 10^{50}$ |
| 220101A | ··· | −1.16 | 483.17 | $(1.37 \pm 1) \times 10^{-5}$ | $(6.16 \pm 0.25) \times 10^{-7}$ | 4.61 | 173.36 | 0.26 | $(5.07 \pm 2.62) \times 10^{53}$ | $(3.63 \pm 0.15) \times 10^{52}$ |
| 240218A | ··· | −1.41 | 142.14 | $(7.55 \pm 3.1) \times 10^{-6}$ | $(2.16 \pm 0.22) \times 10^{-7}$ | 6.782 | 66.93 | 0.38 | $(3.87 \pm 1.42) \times 10^{53}$ | $(4.57 \pm 0.46) \times 10^{52}$ |
| 240414A | ··· | −1.15 | 125.68 | $(1.25 \pm 0.61) \times 10^{-5}$ | $(1.13 \pm 0.15) \times 10^{-7}$ | 1.833 | 88.28 | 0.57 | $(6.09 \pm 1.7) \times 10^{52}$ | $(1.55 \pm 0.21) \times 10^{51}$ |
| 240912A | ··· | −1.45 | 193.99 | $(6.11 \pm 5.4) \times 10^{-6}$ | $(1.15 \pm 0.04) \times 10^{-6}$ | 1.234 | 113.24 | 0.72 | $(5.64 \pm 13.09) \times 10^{52}$ | $(7.47 \pm 0.24) \times 10^{51}$ |

**Notes.** Column (1) lists the names of the GRBs. Column (2) provides the spectral index derived from the PL fit to the prompt emission spectrum in the 15–150 keV energy range. Columns (3) and (4) report the spectral index and the peak energy obtained from the CPL model, also measured in the 15–150 keV range. Column (5) gives the gamma-ray fluence ($S_\gamma$) in units of erg cm$^{-2}$. Column (6) presents the 1 s resolution gamma-ray peak flux ($F_p$), also in the 15–150 keV band, in units of erg cm$^{-2}$ s$^{-1}$. Column (7) lists the redshift ($z$) of each GRB. Column (8) provides the duration of the burst ($T_{90}$). Column (9) gives the $K$-correction factor used to convert observed quantities into the rest-frame 15–150 keV band. Columns (10) and (11) present the isotropic-equivalent gamma-ray energy ($E_{\gamma,iso}$) and the peak luminosity ($L_p$), respectively, both calculated in the rest-frame 15–150 keV band.

All relevant data in the first eight columns are taken from the Swift GRB table (https://swift.gsfc.nasa.gov/archive/grb_table/). This table is divided into two parts by a single horizontal line. The first part represents the GRBs with spectral parameters obtained by the PL fitting, and the second part represents the GRBs with spectral parameters obtained by the CPL fitting.

**Table 4**
Host Galaxy Properties of SN-less GRBs

| GRB (1) | $R_{\mathrm{off}}$ (kpc) (2) | SFR ($M_\odot$ yr$^{-1}$) (3) | $\log(Z/Z_\odot)$ (4) |
|---|---|---|---|
| 970508A | 0.091 | 1.14 | ··· |
| 970828A | 3.8 | 0.87 | ··· |
| 971214A | 1.06 | 11.4 | ··· |
| 980326A | 1.118 | ··· | ··· |
| 980329A | 0.28 | ··· | ··· |
| 980519A | 9.466 | ··· | ··· |
| 980613A | 0.74 | 4.7 | ··· |
| 980703A | 0.91 | 16.57 | −0.2 |
| 981226A | 6.3 | ··· | ··· |
| 990123A | 5.8 | 5.72 | ··· |
| 990308A | 8.959 | ··· | ··· |
| 990506A | 2.54 | 2.5 | ··· |
| 990510A | 0.57 | ··· | ··· |
| 990705A | 6.8 | 6.96 | ··· |
| 991216A | 2.94 | ··· | ··· |
| 000210A | 7.773 | 2.28 | ··· |
| 000301C | 0.59 | ··· | ··· |
| 000301D | 0.622 | ··· | ··· |
| 000418A | 0.192 | 10.35 | ··· |
| 000926A | 0.275 | 2.28 | ··· |
| 010222A | 0.38 | 0.34 | ··· |
| 010921A | 1.89 | 2.5 | −0.6 |
| 011030A | 2.69 | ··· | ··· |
| 011211A | 4.266 | 4.9 | ··· |
| 020305A | 0.567 | ··· | ··· |
| 020427A | 0.774 | ··· | ··· |
| 020813A | ··· | 6.76 | ··· |
| 020819B | 16.5 | ··· | ··· |
| 021004A | 2.357 | 29 | ··· |
| 030323A | 1.072 | ··· | ··· |
| 030328A | ··· | 3.2 | ··· |
| 030429A | 8.199 | ··· | ··· |
| 030528A | ··· | 15.07 | −0.56 |
| 040812 | 2.283 | ··· | ··· |
| 040912A | ··· | 6.3 | ··· |
| 050219A | 15.899 | 0.06 | ··· |
| 050223A | ··· | 1.44 | ··· |
| 050315A | 0.98 | ··· | ··· |
| 050401A | 0.66 | ··· | ··· |
| 050406 | 1.579 | ··· | ··· |
| 050408A | 1.29 | ··· | ··· |
| 050416 | ··· | 4.5 | −0.23 |
| 050525 | 0.185 | 0.07 | ··· |
| 050714B | ··· | 12.9 | ··· |
| 050730 | ··· | ··· | −1.96 |
| 050819A | ··· | 22 | ··· |
| 050820A | 3.6 | ··· | −0.76 |
| 050826A | 1.777 | 9.13 | 0.14 |
| 050904A | 0.75 | ··· | ··· |
| 050908A | 0.284 | ··· | ··· |
| 050915A | 6.8 | 196 | ··· |
| 050922C | ··· | ··· | −1.88 |
| 051001A | ··· | 110 | ··· |
| 051006A | ··· | 98 | ··· |
| 051008A | 5.5 | ··· | ··· |
| 051016B | 2.023 | 10.2 | −0.42 |
| 051022A | 1.283 | 60 | −0.2 |
| 051117B | ··· | 4.7 | ··· |
| 060115A | 2.11 | ··· | ··· |
| 060124A | 0.278 | ··· | ··· |
| 060202A | 11.6 | ··· | ··· |
| 060204B | ··· | 78 | ··· |

**Table 4**
(Continued)

| GRB (1) | $R_{\mathrm{off}}$ (kpc) (2) | SFR ($M_\odot$ yr$^{-1}$) (3) | $\log(Z/Z_\odot)$ (4) |
|---|---|---|---|
| 060206A | 2.09 | ··· | ··· |
| 060223A | 0.77 | ··· | ··· |
| 060306A | 6.9 | 17.6 | ··· |
| 060319A | 10.3 | ··· | ··· |
| 060418A | 3.5 | ··· | ··· |
| 060502A | 0.435 | ··· | ··· |
| 060505A | 7.16 | 0.43 | −0.22 |
| 060522A | 1.93 | ··· | ··· |
| 060602A | 1.21 | ··· | ··· |
| 060604A | ··· | 7.2 | −0.59 |
| 060605A | 0.18 | ··· | ··· |
| 060614A | 0.801 | 0.01 | ··· |
| 060707A | ··· | 19.9 | ··· |
| 060719A | 1.743 | 7.1 | −0.08 |
| 060805A | ··· | 9 | ··· |
| 060814A | 3.1 | 54 | ··· |
| 060912A | 5.155 | 5.1 | −0.08 |
| 060923A | 2.46 | ··· | ··· |
| 060923B | ··· | 3 | ··· |
| 060926A | ··· | 26 | ··· |
| 061007A | 0.184 | ··· | ··· |
| 061021A | ··· | 0.05 | ··· |
| 061110A | 0.992 | 0.23 | ··· |
| 061110B | 0.449 | ··· | ··· |
| 061121A | ··· | 27 | ··· |
| 061126A | 8.4 | 2.38 | ··· |
| 061202A | ··· | 43 | ··· |
| 061222A | 0.368 | ··· | ··· |
| 070103A | ··· | 43 | ··· |
| 070110A | ··· | 8.9 | ··· |
| 070125A | 3.2 | ··· | ··· |
| 070129A | ··· | 20 | ··· |
| 070208A | 0.83 | ··· | ··· |
| 070224A | ··· | 3.2 | ··· |
| 070306A | 0.78 | 101 | −0.15 |
| 070318A | 0.84 | 0.79 | ··· |
| 070328A | ··· | 8.4 | ··· |
| 070419B | ··· | 21 | ··· |
| 070508A | 3.3 | ··· | ··· |
| 070521A | 7.7 | 26 | ··· |
| 070612A | ··· | 81 | −0.4 |
| 070721B | 0.327 | ··· | ··· |
| 070802A | 1.1 | 24 | ··· |
| 071010A | 0.106 | ··· | ··· |
| 071010B | 0.86 | ··· | ··· |
| 071021A | 4 | 32 | ··· |
| 071031 | ··· | ··· | −1.85 |
| 071122A | 0.63 | ··· | ··· |
| 080207A | 6.6 | 77 | ··· |
| 080319C | 0.655 | ··· | ··· |
| 080325A | 6 | 9 | ··· |
| 080413B | ··· | 2.1 | −0.4 |
| 080430A | 0.81 | ··· | ··· |
| 080517A | 5 | 16 | ··· |
| 080520A | 4.1 | ··· | ··· |
| 080602A | ··· | 125 | ··· |
| 080603A | 0.753 | ··· | ··· |
| 080605A | 0.7 | 47 | −0.15 |
| 080607A | 5.4 | 19 | ··· |
| 080707A | 0.76 | ··· | ··· |
| 080804A | ··· | 15.2 | −0.75 |
| 080805A | 4 | 45 | −0.2 |

**Table 4**
(Continued)

| GRB (1) | $R_{off}$ (kpc) (2) | SFR ($M_{\odot}$ yr$^{-1}$) (3) | $\log(Z/Z_{\odot})$ (4) |
|---|---|---|---|
| 080916A | 0.075 | … | … |
| 080928A | 14.8 | … | … |
| 081008A | 14.3 | … | −0.52 |
| 081109A | 1.62 | 11.8 | … |
| 081121A | 2 | … | … |
| 081210A | … | 15.3 | … |
| 081221A | 3.3 | 35 | … |
| 090102A | 0.78 | … | … |
| 090113 | 1.573 | 17.9 | … |
| 090201A | … | 48 | … |
| 090205A | 2.7 | … | … |
| 090323A | … | 24 | 0.07 |
| 090328A | 1.035 | 3.6 | … |
| 090404 | 4.709 | … | … |
| 090407A | 1.49 | 13.8 | … |
| 090417B | 0.64 | … | … |
| 090418A | 0.812 | … | … |
| 090424A | 2.62 | 0.8 | −0.3 |
| 090709A | 2.393 | … | … |
| 090926 | … | … | −2.18 |
| 090926B | 5.1 | 26 | −0.35 |
| 091018A | … | 1.29 | 0.09 |
| 091208B | 0.79 | … | … |
| 100219A | 2.69 | … | −0.95 |
| 100424A | … | 21 | −0.76 |
| 100508A | … | 2.6 | −0.01 |
| 100526A | 1.645 | … | … |
| 100606A | … | 4.9 | 0.02 |
| 100615A | 2.073 | 8.6 | −0.29 |
| 100621A | 0.284 | 8.7 | −0.17 |
| 100724A | … | 3.2 | … |
| 100728A | … | 14.5 | … |
| 100814A | … | 3.2 | … |
| 110709B | 0.564 | … | … |
| 110731A | 1.423 | … | … |
| 110808A | … | 8.3 | −0.76 |
| 110818A | … | 44 | −0.44 |
| 110918A | 12 | 23 | 0.24 |
| 111005A | 0.951 | 0.16 | … |
| 111008A | … | … | −1.7 |
| 111123A | … | 77 | −0.68 |
| 111129A | … | 5.1 | … |
| 111215A | 2.844 | … | … |
| 120118B | … | 28 | −0.8 |
| 120119A | 0.099 | 43 | −0.09 |
| 120327A | … | … | −1.2 |
| 120624B | … | 30 | −0.26 |
| 120711A | 0.137 | … | … |
| 120722A | … | 22 | −0.21 |
| 120815A | … | 2.3 | −1.15 |
| 121024A | … | 37 | −0.28 |
| 121027A | … | 24 | … |
| 121201A | … | 30 | … |
| 130131B | … | 8 | … |
| 130606A | 0.36 | … | −1.1 |
| 130701A | … | 0.78 | … |
| 130925A | 0.765 | 2.9 | 0.04 |
| 131103A | … | 4.4 | −0.21 |
| 131105A | … | 31 | −0.08 |
| 131231A | … | 1.38 | −0.24 |
| 140213A | … | 0.72 | … |
| 140301A | … | 106 | … |
| 140331A | 11.55 | … | … |
| 140430A | … | 8.5 | −0.02 |
| 140506A | … | 0.35 | … |
| 140515A | 1.21 | … | … |

**Note.** Column (1) lists the names of the GRBs. Column (2) provides the projected offset ($R_{off}$) of each GRB. Column (3) reports the star formation rate (SFR). Column (4) gives the logarithmic metallicity ($\log(Z/Z_{\odot})$).

**References**. Swift Gamma-Ray Bursts Catalog (https://heasarc.gsfc.nasa.gov/W3Browse/swift/swiftgrb.html); https://www.grbhosts.org/; T. Krühler et al. (2015); Y. Li et al. (2016); J. S. Wang et al. (2016); P. K. Blanchard et al. (2016); F.-F. Wang et al. (2018a); A. A. Chrimes et al. (2019); B. O'Connor et al. (2022); X.-J. Li et al. (2024b).

association (i.e., the $S_w$ bursts) are consistent with the Amati relation and are within the $2\sigma$ confidence range. Only one $S_w$ burst clearly deviates from the $2\sigma$ range, i.e., GRB 090519. The optical afterglow of GRB 090519 is among the faintest ever observed (D. Turpin et al. 2016), which may be due to a combination of its high redshift and an intrinsic weakness of the afterglow. On the contrary, the distribution of the $S_a$ sample is somewhat different. We see that nine SN-GRBs deviate from the Amati relation, accounting for 19% of the $S_a$ sample.

GRB 980425 is the GRB confirmed to be associated with an SN. It has a low peak bolometric luminosity of about $5.8 \times 10^{46}$ erg s$^{-1}$. Such low-luminosity GRBs are considered to be a special category of GRBs, which may have a special origin (F. J. Virgili et al. 2009). Some low-luminosity bursts, such as GRBs 031203, 161219B, and 171205A (A. M. Soderberg et al. 2004; Z. Cano et al. 2017a; V. D'Elia et al. 2018), deviate from the Amati relation, as shown in Figure 1. However, other low-luminosity bursts including GRBs 020903, 060218, and GRB 201015A (H. Sun et al. 2015; M. Patel et al. 2023; C. M. Irwin & K. Hotokezaka 2025) follow the Amati relation. This point has already been noted by G. Finneran et al. (2025a), who found the inconsistency between SN-GRBs and GRBs without SNe association when the Amati relation is involved. They argued that GRBs not associated with an SN do not extend to the range of low-energy GRBs. It is not clear whether the deviation from the Amati relation is an intrinsic property of SN-GRBs or a more general intrinsic property of LLGRBs.

The deviation of LLGRBs from the Amati relation may be caused by a relatively large viewing angle, which makes the bursts follow a flatter track on the $E_p - E_{iso}$ plane (E. Ramirez-Ruiz et al. 2005; S. Dado & A. Dar 2012; R. Farinelli et al. 2021). Recently, F. Xu et al. (2023) considered the effect of the viewing angle. They found that the PL index should be between $1/4$ and $4/13$. In fact, we notice that all nine bursts deviating from the Amati relation, i.e., GRBs 980425, 031203, 100316D, 120422A, 140606B, 150818, 161219B, 171205A, and 190829A, are classified as off-axis events (Y. Sato et al. 2021; A. Kumar et al. 2022; F. Xu et al. 2023; X.-J. Li et al. 2024b). However, note that off-axis bursts are not necessarily

**Table 5**
Parameters of SNe

| SN (1) | Type (2) | Redshift (3) | $L_{\rm r,p}$ [a] (erg $s^{-1}$) (4) |
|---|---|---|---|
| 2002ap | Ic-BL | 0.002187 | $5.44 \times 10^{42}$ |
| 2003jd | Ic-BL | 0.018826 | $1.57 \times 10^{43}$ |
| 2005kz | Ic-BL | 0.027 | $5.26 \times 10^{42}$ |
| 2007D | Ic-BL | 0.023146 | $1.21 \times 10^{43}$ |
| 2007ru | Ic-BL | 0.01546 | $1.92 \times 10^{43}$ |
| 2009bb | Ic-BL | 0.009937 | $1.12 \times 10^{43}$ |
| 2010ah | Ic-BL | 0.0498 | $8.89 \times 10^{42}$ |
| 2010ay | Ic-BL | 0.067 | $4.39 \times 10^{43}$ |
| 2018bvw | Ic-BL | 0.054 | $1.13 \times 10^{43}$ |
| 2018ell | Ic-BL | 0.0638 | $8.60 \times 10^{42}$ |
| 2018gep | Ic-BL | 0.0442 | $2.18 \times 10^{43}$ |
| 2018hsf | Ic-BL | 0.1184 | $2.85 \times 10^{43}$ |
| 2018keq | Ic-BL | 0.0384 | $3.65 \times 10^{42}$ |
| 2019hsx | Ic-BL | 0.020652 | $2.22 \times 10^{42}$ |
| 2019gwc | Ic-BL | 0.038 | $8.06 \times 10^{42}$ |
| 2019lci | Ic-BL | 0.0292 | $5.53 \times 10^{42}$ |
| 2019moc | Ic-BL | 0.055 | $1.50 \times 10^{43}$ |
| 2019oqp | Ic-BL | 0.03082 | $2.64 \times 10^{42}$ |
| 2019pgo | Ic-BL | 0.05 | $1.37 \times 10^{43}$ |
| 2019qfi | Ic-BL | 0.028 | $5.23 \times 10^{42}$ |
| 2020zg | Ic-BL | 0.0557 | $1.97 \times 10^{43}$ |
| 2020ayz | Ic-BL | 0.025 | $1.81 \times 10^{42}$ |
| 2020bvc | Ic-BL | 0.0252 | $1.32 \times 10^{43}$ |
| 2020dgd | Ic-BL | 0.032 | $4.08 \times 10^{42}$ |
| 2020hes | Ic-BL | 0.07 | $1.86 \times 10^{43}$ |
| 2020hyj | Ic-BL | 0.055 | $6.41 \times 10^{42}$ |
| 2020jqm | Ic-BL | 0.03696 | $6.58 \times 10^{42}$ |
| 2020lao | Ic-BL | 0.030814 | $9.52 \times 10^{42}$ |
| 2020rfr | Ic-BL | 0.0725 | $1.18 \times 10^{43}$ |
| 2020rph | Ic-BL | 0.042 | $3.21 \times 10^{42}$ |
| 2020tkx | Ic-BL | 0.027 | $8.14 \times 10^{42}$ |
| 2020wgz | Ic-BL | 0.1785 | $7.42 \times 10^{43}$ |
| 2020abxl | Ic-BL | 0.0815 | $1.62 \times 10^{43}$ |
| 2020abxc | Ic-BL | 0.06 | $1.71 \times 10^{43}$ |
| 2020adow | Ic-BL | 0.0075 | $5.04 \times 10^{42}$ |
| 2021bmf | Ic-BL | 0.017 | $1.05 \times 10^{43}$ |
| 2021epp | Ic-BL | 0.0385 | $3.24 \times 10^{42}$ |
| 2021fop | Ic-BL | 0.077 | $8.14 \times 10^{42}$ |
| 2021hyz | Ic-BL | 0.046 | $1.11 \times 10^{43}$ |
| 2021ktv | Ic-BL | 0.07 | $1.59 \times 10^{43}$ |
| 2021ncn | Ic-BL | 0.02461 | $2.16 \times 10^{42}$ |
| 2021qjv | Ic-BL | 0.03803 | $5.79 \times 10^{42}$ |
| 2021too | Ic-BL | 0.035 | $2.39 \times 10^{43}$ |
| 2021ywf | Ic-BL | 0.028249 | $2.26 \times 10^{42}$ |
| PTF10bzf | Ic-BL | 0.0498 | $6.83 \times 10^{42}$ |
| PTF10ciw | Ic-BL | 0.115 | $8.45 \times 10^{42}$ |
| PTF10gvb | Ic-BL | 0.098 | $1.21 \times 10^{43}$ |
| PTF10qts | Ic-BL | 0.0907 | $1.46 \times 10^{43}$ |
| PTF10tqv | Ic-BL | 0.0795 | $6.17 \times 10^{42}$ |
| PTF10vgv | Ic-BL | 0.015 | $7.56 \times 10^{42}$ |
| PTF10xem | Ic-BL | 0.0566 | $6.71 \times 10^{42}$ |
| PTF10ysd | Ic-BL | 0.0963 | $1.73 \times 10^{43}$ |
| PTF10aavz | Ic-BL | 0.063 | $1.40 \times 10^{43}$ |
| PTF11cmh | Ic-BL | 0.1055 | $7.92 \times 10^{42}$ |
| PTF11img | Ic-BL | 0.158 | $1.68 \times 10^{43}$ |
| PTF11lbm | Ic-BL | 0.039 | $6.96 \times 10^{42}$ |
| PTF12as | Ic-BL | 0.0332 | $5.90 \times 10^{42}$ |
| PTF12eci | Ic-BL | 0.0874 | $8.60 \times 10^{42}$ |
| iPTF13u | Ic-BL | 0.0991 | $8.29 \times 10^{42}$ |
| iPTF13alq | Ic-BL | 0.054 | $1.05 \times 10^{43}$ |
| iPTF13bxl | Ic-BL | 0.145 | $1.23 \times 10^{43}$ |
| iPTF13dnt | Ic-BL | 0.137 | $1.30 \times 10^{43}$ |

**Table 5**
(Continued)

| SN (1) | Type (2) | Redshift (3) | $L_{\rm r,p}$ [a] (erg $s^{-1}$) (4) |
|---|---|---|---|
| iPTF13ebw | Ic-BL | 0.0686 | $5.33 \times 10^{42}$ |
| iPTF14dby | Ic-BL | 0.0736 | $4.19 \times 10^{42}$ |
| iPTF14gaq | Ic-BL | 0.0826 | $6.41 \times 10^{42}$ |
| iPTF15dqg | Ic-BL | 0.0577 | $1.90 \times 10^{43}$ |
| iPTF15eov | Ic-BL | 0.0535 | $6.96 \times 10^{43}$ |
| iPTF16asu | Ic-BL | 0.1874 | $2.69 \times 10^{43}$ |
| iPTF16gox | Ic-BL | 0.042 | $4.08 \times 10^{42}$ |
| iPTF16ilj | Ic-BL | 0.0397 | $1.30 \times 10^{43}$ |
| iPTF17cw | Ic-BL | 0.093 | $1.65 \times 10^{43}$ |

**Notes.** Column (1) lists the names of the SNe. Column (2) provides the type of each SN. Column (3) reports the redshift ($z$) of each event. Column (4) gives the peak luminosity in the $r$-band.
[a] The absolute magnitude in the $r$-band of SN 1998bw used in calculating the peak luminosity is −19.36 (T. J. Galama et al. 1998), and the absolute magnitude of the Sun is 4.83.
**References**. Z. Cano (2013); K. Takaki et al. (2013); F.-F. Wang et al. (2018a); H. Kuncarayakti et al. (2018); F. Taddia et al. (2019); D. Xiang et al. (2019); A. Gangopadhyay et al. (2020); C. Barbarino et al. (2021); S. Gomez et al. (2022); G. P. Srinivasaragavan et al. (2024). Parameters of Type I SLSNe are all taken from S. Gomez et al. (2024).

associated with SNe. For example, GRB 061021 is a likely long off-axis burst, but no SN was found to be connected to it (L. Nava et al. 2012; F. Xu et al. 2023).

### 3.2. $T_{90}$ versus Redshift

Figure 2 illustrates the two samples on the $T_{90}$–$z$ plane. We see that there is no correlation between the redshift and the duration for both $S_{\rm a}$ and $S_{\rm w}$ bursts. The redshift of GRBs without an SN association can be very large, i.e., up to ∼10. However, the redshift of SN-GRBs is generally less than 1.1, which is probably caused by the observational selection effect. SNe have a relatively low luminosity and are difficult to detect at high redshift. Additionally, even at low redshift, the afterglow emission in some bright GRBs may overshadow the supernova component and make the SN undetectable. This may happen in GRB 221009A, the brightest GRB ever observed to date at a redshift of $z = 0.151$ (D.-F. Kong et al. 2024). While the redshift distributions are not identical, the GRBs without SN associations exhibit a broader redshift range that substantially includes the redshift span of SN-GRBs.

The duration of LGRBs without an SN association is generally between 2 and 400 s, while the duration of some LGRBs accompanied by SNe can be less than 2 s or longer than $10^3$ s. For example, the duration of GRB 200826A is 0.65 s, which seems to suggest that it is a short burst. However, it was found to be associated with an SN (T. Ahumada et al. 2021), strongly pointing to a collapsar origin, although a special binary origin still could not be expelled yet (B. B. Zhang et al. 2021; A. Rossi et al. 2022c; Z.-k. Peng et al. 2024).

LGRBs longer than $10^3$ s are defined as ULGRBs (B. Gendre et al. 2013; A. J. Levan et al. 2014). GRB 111209A is the longest event ever observed. It is associated with a bright SN, SN 2011kl, which has an absolute magnitude of about −20 mag at the peak (J. Greiner et al. 2015;

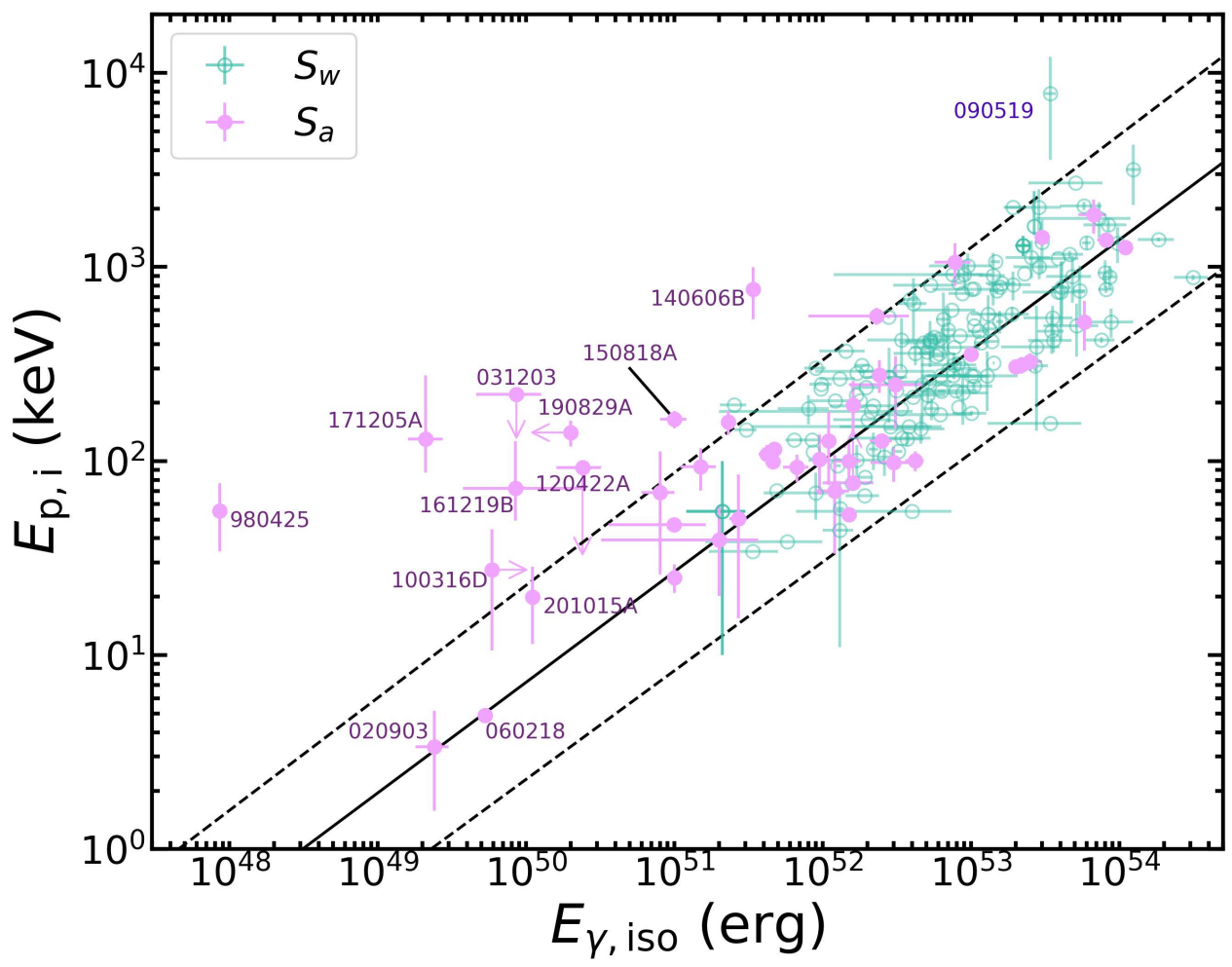

**Figure 1.** Comparison of the Amati relation for $S_a$ and $S_w$ GRBs. The open circles are GRBs without SN association, i.e., the $S_w$ sample. The corresponding data are taken from Tables 2, 3, and Y. Li et al. (2016). The filled circles are GRBs with SN association, i.e., the $S_a$ sample. The data are listed in Table 1. The solid and dashed lines represent the original Amati relation obtained by L. Amati et al. (2008; with a PL index of $m = 0.57 \pm 0.01$) and the $2\sigma$ uncertainty.

D. A. Kann et al. 2019). ULGRBs associated with a luminous SN may have a special origin (T. J. Moriya et al. 2020). They may originate from blue supergiant collapsars (K. Kashiyama et al. 2013), magnetars (J. Greiner et al. 2015), or white dwarf tidal disruption events (B. Gendre et al. 2013). While a magnetar origin is a plausible explanation for both luminous and long-duration GRBs, alternative interpretations cannot be ruled out (A. Fiore et al. 2025). SN-GRBs 100316D and 101225A, as shown in Figure 2, are also special ULGRBs (K. Ioka et al. 2016). On the contrary, although GRBs 060218 and 171205A last longer than ordinary long bursts, their spectra are soft, which differentiates them from other ULGRBs. Such a soft spectrum may be due to the contribution of the SN shock breakout (E. Waxman et al. 2007; D. Xiang et al. 2019; A. Kumar et al. 2022; A. K. Ror et al. 2024a). Note that not all ULGRBs are associated with an SN. For example, GRB 130925A has a $T_{90}$ of about 4500 s (A. Horesh et al. 2015), and GRB 141121A has a $T_{90}$ of about 1410 s (A. Cucchiara et al. 2015b), but they both are not associated with any SNe. Since these two events were detected by the Konus–Wind satellite, they are not shown in Figure 2 because our $S_w$ sample after 2004 only includes Swift bursts.

### 3.3. Host Galaxy Properties

#### 3.3.1. Projected Offset

Figure 3 shows the cumulative distribution function (CDF) of projected offsets for LGRBs with and without SN detections. Since many GRBs in the $S_a$ and $S_w$ samples do not have an offset measurement, only 25 SN-GRBs and 124 SN-less LGRBs are included in the plot. For SN-GRBs, the offset ranges from 0.079 to 8.17 kpc, with a mean value of 1.1 kpc and a median value of 1.53 kpc. Approximately 90% of these events occur within a 5 kpc range with respect to their host galaxy center. Two apparent gaps in the distribution, at 0.3–0.6 kpc and 0.7–1.3 kpc, may be fluctuations caused by the limited sample size.

The projected offset of SN-less LGRBs ranges from 0.075 to 16.5 kpc, with a mean value of 1.54 kpc and a median value of 1.57 kpc. About 90% of them are located within 8 kpc of their host center. There is an obvious rise in the distribution at 10 kpc; the curve is generally smooth. This pattern aligns with theoretical expectations that LGRBs typically trace the star-forming regions of galaxies (P. K. Blanchard et al. 2016; Y. Li et al. 2016; J. Japelj et al. 2018).

For comparison, the projected offset distributions of 66 SLSNe (B. Hsu et al. 2024) and 70 SGRBs (N. Gaspari et al. 2025) are also plotted in the figure. The SLSN offset ranges from 0.072 to 6.69 kpc, with a median value of 0.73 kpc and a mean value of 0.77 kpc. They are slightly closer to their host center than LGRBs. In contrast, SGRBs display a significantly larger offset (0.7–76.2 kpc), with a median value of 9.6 kpc and a mean value of 8.4 kpc. This distinct distribution supports the idea that SGRBs should be dominantly produced by the mergers of binary compact stars (N. Gaspari et al. 2025).

To quantify the difference of the distribution functions of the three samples, we have performed Kolmogorov–Smirnov (KS) tests on them. It is found that the distributions of SN-LGRBs and SN-less LGRBs are statistically indistinguishable at a significance level of $\alpha = 0.01$, with a probability parameter of $p = 0.45$, while the SN-LGRBs and SLSNe samples are consistent with each other ($p = 0.12$). However, the SN-less LGRB and SLSN samples differ significantly, with $p = 3.8 \times 10^{-3} < \alpha$. We see that while the circumburst environment of SN-LGRBs and SN-less LGRBs are similar, only a small fraction of LGRBs share similar properties with SLSNe in this aspect. These overlapping GRBs were predominantly found in dwarf galaxies or active star-forming (blue) galaxies (J. Greiner et al. 2015; R. Lunnan et al. 2015).

#### 3.3.2. Star Formation Rate

Figure 4 shows the SFR distributions of host galaxies for LGRBs with and without an SN association. 31 SN-GRBs and 113 SN-less LGRBs are included in the plot. The SFR is in a broad range of 0.01–200 $M_\odot\ \mathrm{yr}^{-1}$, as reported for typical LGRB host galaxies (D. A. Perley et al. 2016; J. T. Palmerio et al. 2019), but note that the SNe-GRBs exhibit a slightly narrower distribution.

The cumulative distribution of SN-less LGRBs is generally smooth, with no notable gaps. These bursts have a mean SFR of 7.87 $M_\odot\ \mathrm{yr}^{-1}$ and a median of 9.13 $M_\odot\ \mathrm{yr}^{-1}$. In contrast, SN-GRBs are mainly in moderately star-forming hosts, with a significantly lower mean and median SFRs of 0.77 and 0.96 $M_\odot\ \mathrm{yr}^{-1}$, respectively. It is consistent with the notion that they represent a distinct subset of the LGRB population (S. Savaglio et al. 2009; T. Krühler et al. 2015).

#### 3.3.3. Metallicity

LGRBs are known to preferentially occur in low-metallicity host galaxies compared to typical star-forming galaxies (S. C. Yoon et al. 2006; J. T. Palmerio et al. 2019; M. Modjaz et al. 2020; B. Schneider et al. 2022). Here, we further examine their host metallicity. Usually, the metallicity is measured in units of the solar metallicity as

$$\log(Z/Z_\odot) = 12 + \log(\mathrm{O/H}) - [12 + \log(\mathrm{O/H})_\odot], \quad (8)$$

where $Z_\odot$ denotes the solar metallicity. Figure 5 shows the metallicity distributions of the host galaxies of LGRBs. 22 SN-

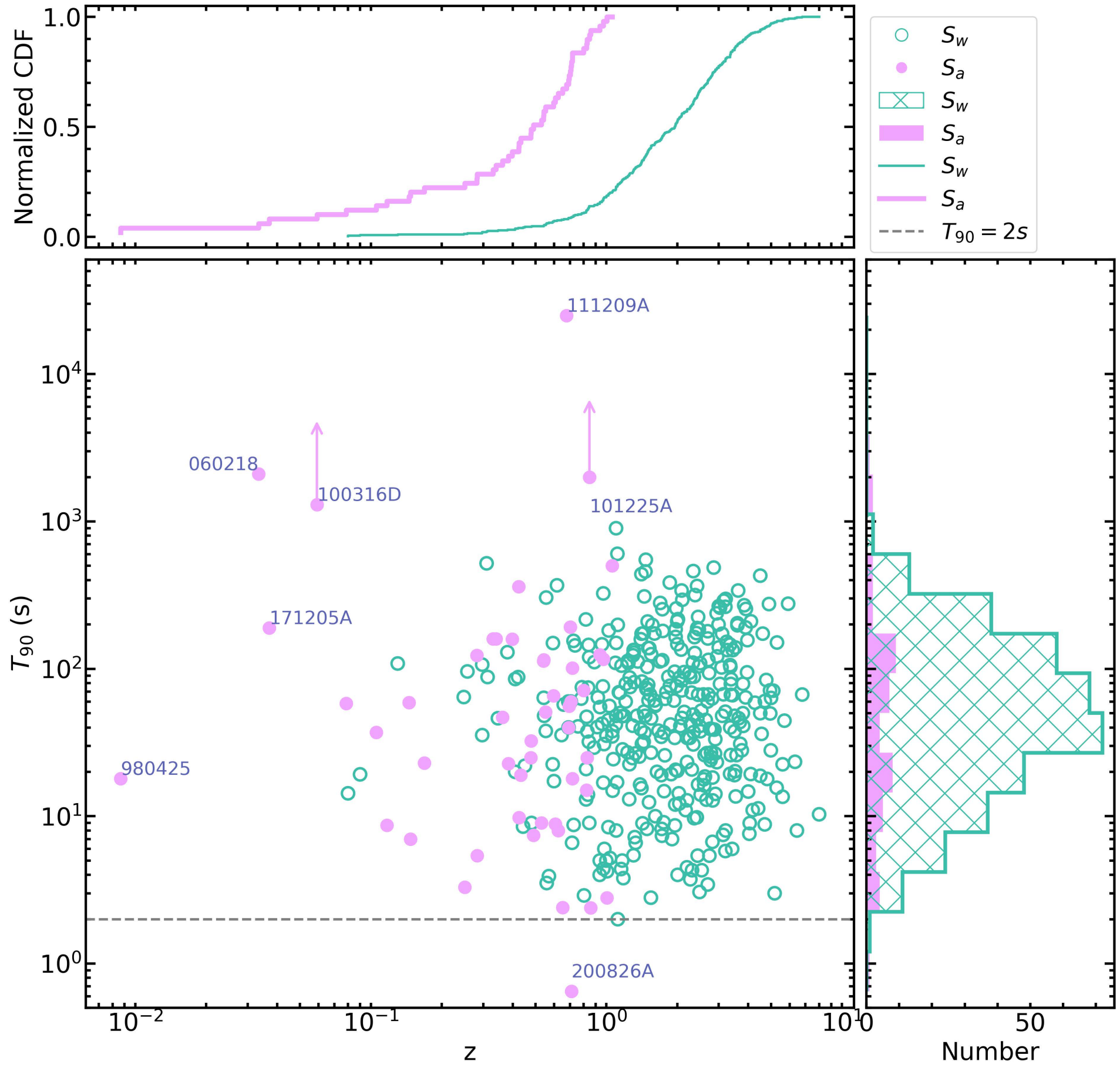

**Figure 2.** Redshift vs. $T_{90}$ for $S_a$ and $S_w$ GRBs. The solid circles represent the $S_a$ sample, and the hollow circles represent the $S_w$ sample. The top panel shows the normalized cumulative distribution function (CDF) of their redshifts, with the thick line representing $S_a$ and the thin line representing $S_w$. The right panel illustrates the distribution of their durations, where the filled histogram represents $S_a$ and the grid histogram represents $S_w$.

GRBs and 51 SN-less LGRBs in the $S_a$ and $S_w$ samples are included in the plot.

The SN-GRB hosts exhibit a narrow metallicity range from $-0.74$ to $-0.05$, with a mean of $-0.48$ and a median of $-0.43$. This limited range leads to a steep cumulative distribution, which is consistent with our previous knowledge of a low-metallicity environment for LGRBs (S. E. Woosley & A. Heger 2006a; E. M. Levesque et al. 2010; F. Mannucci et al. 2011). However, this range lies entirely within the broader metallicity distribution of typical LGRB hosts, which makes it difficult to distinguish SN-GRBs from the general population based solely on metallicity.

In contrast, the SN-less LGRBs span a much wider metallicity range, from $-2.0$ to 0.2, with a mean value of $-0.51$ and a median value of $-0.29$, consistent with previous results (A. Cucchiara et al. 2015a; Y. Li et al. 2016). Notably, all bursts with an extremely low metallicity ($\log(Z/Z_\odot) < -1.5$) are found at high redshifts ($z > 2$), consistent with simulations of early-Universe GRB host environments (R. Salvaterra et al. 2013).

Although LGRBs generally favor low-metallicity galaxies, several events occur in near-solar or even supersolar metallicity environments. For instance, GRB 110918A resides in a nearly solar-metallicity host (J. Elliott et al. 2013), while GRB 050825A has a metallicity of $\log(\mathrm{O/H}) + 12 = 8.83 \pm 0.1$ (E. M. Levesque et al. 2010). This wide distribution underscores the diversity of LGRB formation environments, suggesting that metallicity itself is not a strict criterion for their progenitors.

### 3.4. Event Rate

In this section, we present the evolution of the comoving event rate density with respect to redshift for various GRB samples. For the $S_w$ sample, we use the peak luminosity ($L_p$) listed in Tables 2 and 3 to calculate the event rate, which is the usual practice for the purpose (H. Yu et al. 2015; A. Pescalli et al. 2016; G. Q. Zhang & F. Y. Wang 2018; G.-X. Lan et al. 2019; X. F. Dong et al. 2022; Y. Liu et al. 2025). However, $L_p$ values are not available for most GRBs in the $S_a$ sample; we therefore adopt the mean isotropic luminosity $L_{\gamma,\mathrm{iso}} = E_{\gamma,\mathrm{iso}}(1 + z)/T_{90}$ (Z. Cano et al. 2017b) as a substitute.

Since the GRB samples consist of truncated observational data, it is necessary to correct for selection effects. We adopt the Lynden-Bell C-method (D. Lynden-Bell 1971) to estimate the GRB event rates, accounting for the observational truncation introduced by instrumental detection thresholds. The C-estimator is a nonparametric maximum likelihood method developed for randomly truncated univariate data and has been rigorously established as the unique and optimal solution under the assumption of random truncation (M. Woodroofe 1985; M.-C. Wang et al. 1986). Analogous to the Kaplan–Meier estimator for censored data (E. L. Kaplan & P. Meier 1958), the Lynden-Bell estimator provides a robust framework for inferring intrinsic distributions from truncated samples. This methodology also underpins more advanced statistical inferences such as the two-sample test, correlation, linear regression, and Bayesian analysis (B. Efron & V. Petrosian 1992; E. D. Feigelson & G. J. Babu 2012; A. Dörre & T. Emura 2019).

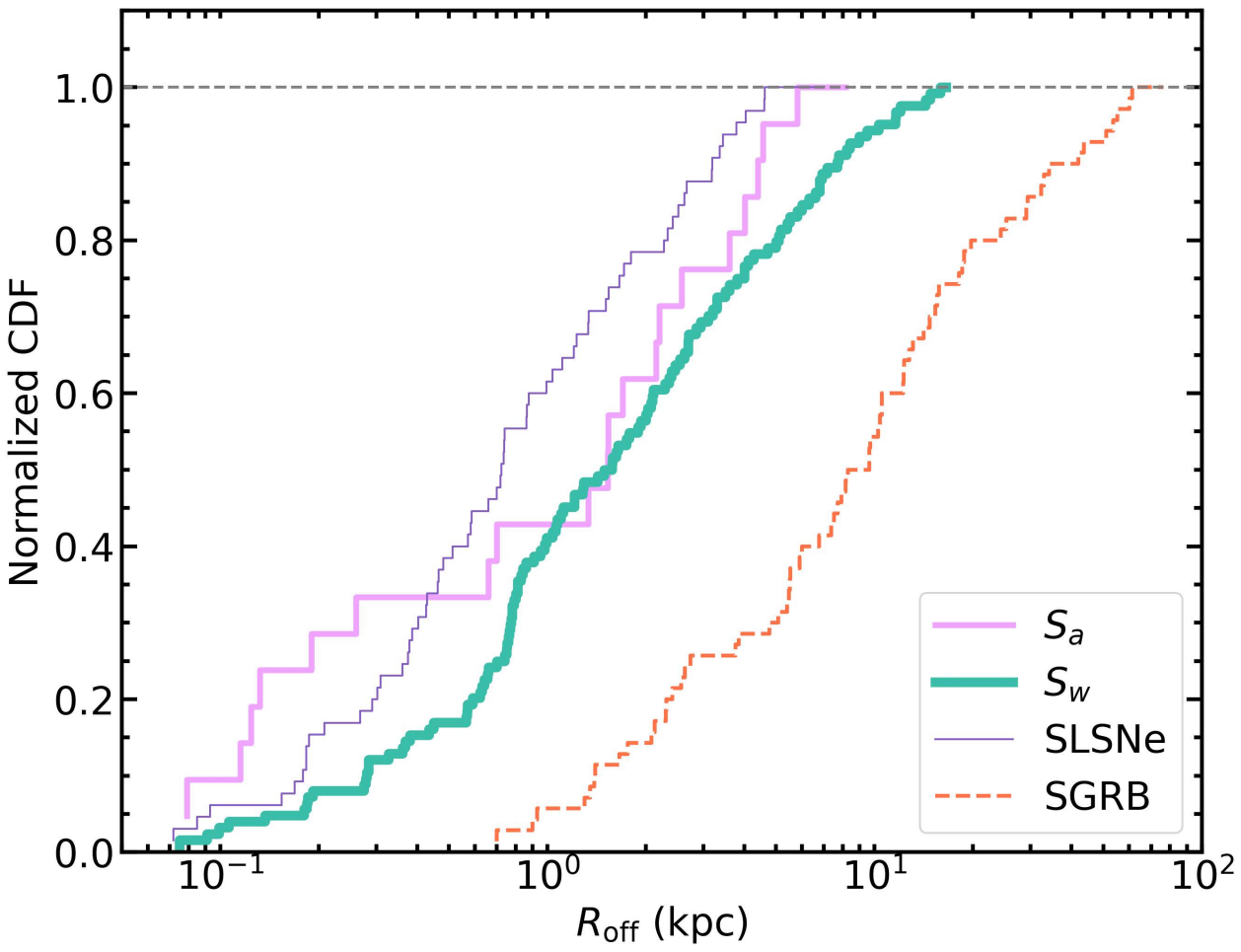

**Figure 3.** Normalized CDF of the projected offset for $S_{\rm a}$, $S_{\rm w}$, and SLSNe samples. The thickest line represents the GRB sample without an SN association, i.e., $S_{\rm w}$. The medium-thick line represents the sample with SN association, i.e., $S_{\rm a}$. The thinnest line represents SLSNe (B. Hsu et al. 2024). The dashed line represents SGRBs (N. Gaspari et al. 2025).

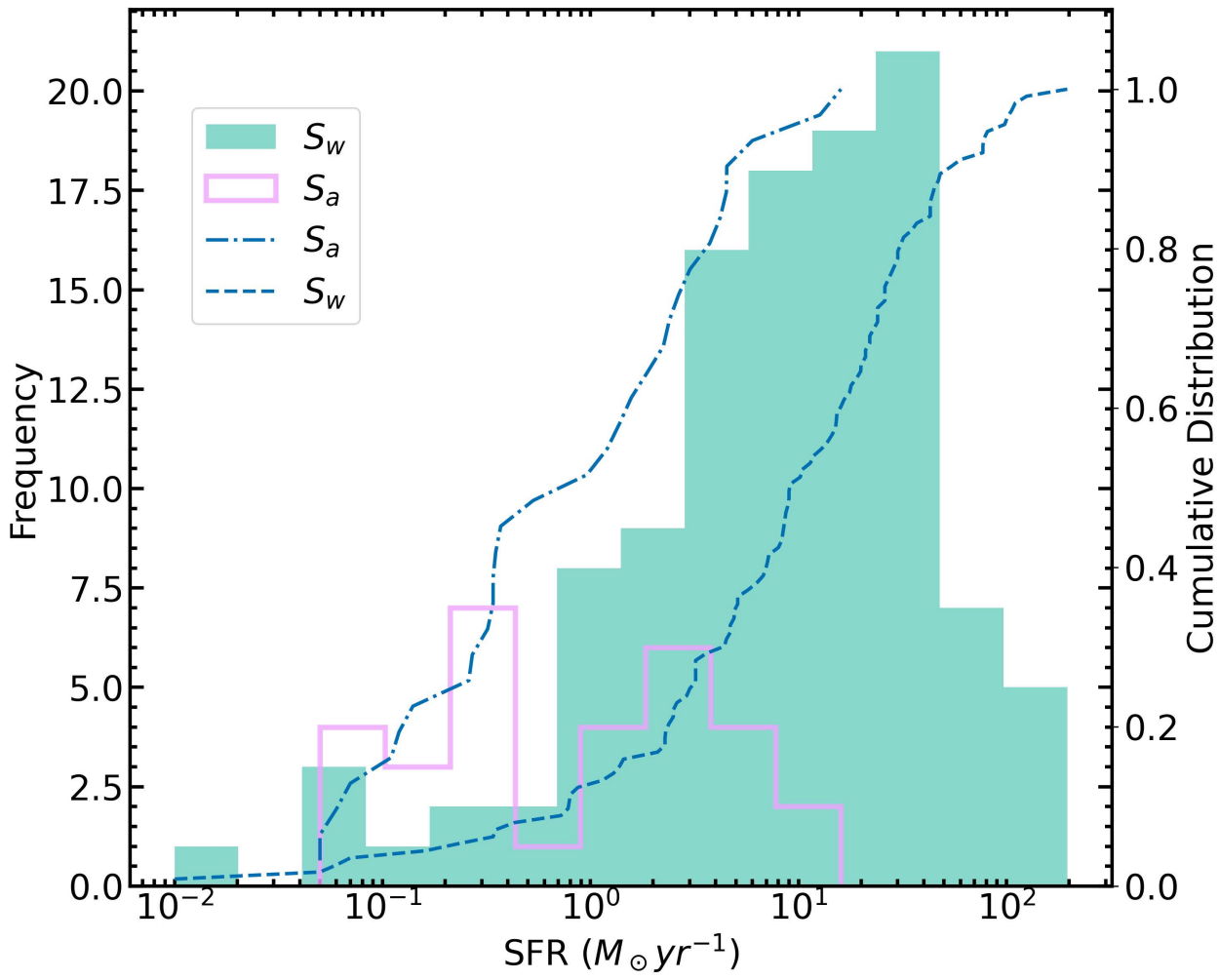

**Figure 4.** Distribution of SFR of host galaxy for the $S_{\rm a}$ and $S_{\rm w}$ GRBs. The hollow histogram and the solid, filled histogram represent $S_{\rm a}$ and $S_{\rm w}$ samples, respectively. The dotted–dashed line and the dashed line represent the normalized cumulative distribution curve of $S_{\rm a}$ and $S_{\rm w}$, respectively.

To apply the C-method, the truncation boundary should be well understood. It is actually determined by the instrument sensitivity. The C-method then enables an unbiased reconstruction of the underlying GRB rate distribution. We follow this nonparametric approach based on the redshift and luminosity distributions as described by X. F. Dong et al. (2022).

To ensure sample completeness, we adopt the limiting flux ($F_{\rm lim}$) following X. F. Dong et al. (2023). Rather than selecting the lowest flux in the differential distribution as the threshold, we adopt a higher flux value, such that GRBs above this threshold are approximately uniformly distributed in space. Although this approach excludes some GRBs, it significantly reduces selection bias and enhances the completeness of the events (A. Pescalli et al. 2016).

Figure 6 presents the evolution of GRB event rate with respect to redshift. The step lines correspond to the derived event rates, while the scatter points represent the observed SFR data. We see that the event rates derived from the $S_{\rm a}$ and $S_{\rm w}$ samples are obviously different. The event rate of SN-less GRBs is nearly constant in the redshift range of $z \sim$ 1–5, which is largely consistent with the cosmic SFR history. In contrast, the event rate of SN-GRBs declines monotonically as redshift increases in the range of $z < 1$. This behavior deviates significantly from both the event rate of SN-less GRBs and the local SFR. It indicates that the $S_{\rm w}$ and $S_{\rm a}$ samples should be of different origin.

## 4. Discussion

### 4.1. SN-GRBs and the Associated SNe

The detailed analysis in Section 3 reveals that while $S_{\rm a}$ and $S_{\rm w}$ GRBs are similar in their prompt emission and host galaxy properties, their event rate differs markedly. In this section, we concentrate on the SN-GRBs to try to reveal the intrinsic connection between the GRB parameters and the properties of the accompanying SNe.

Among the 61 SN-GRBs, 29 events have spectroscopically confirmed SNe, of which 26 GRBs have well-sampled optical observations, allowing us to extract their bolometric peak luminosities. Figure 7 plots the optical peak luminosity of the accompanying SN ($L_{\rm p,SN}$) versus the redshift ($z$), duration ($T_{90}$), peak energy ($E_{\rm p}$), and isotropic energy ($E_{\rm iso}$) of the corresponding GRBs. We see that the peak luminosity of SN distributes in a narrow range of $4 \times 10^{42}$–$3 \times 10^{43}$ erg, with an average value of $L_{\rm p,SN,avg} = 9.42 \times 10^{42}$ erg s$^{-1}$. It is consistent with the result of Y. Aimuratov et al. (2023a). Similarly, F. Taddia et al. (2019) also reported an average peak $r$-band luminosity of $9.07^{+5.3}_{-3.3} \times 10^{42}$ erg s$^{-1}$ for 34 Ic-BL SNe.

We use the Kendall's $\tau$ rank correlation test to assess potential correlation between $L_{\rm p,SN}$ and the GRB parameters shown in Figure 7. The $\tau$ values for peak luminosity versus $z$, $T_{90}$, $E_{\rm p}$, and $E_{\rm iso}$ are 0.058, −0.062, −0.058, and −0.037, with corresponding $p$-values of 0.69, 0.66, 0.69, and 0.79. None of the correlations are statistically significant. It indicates that the peak optical luminosity of the SNe associated with GRBs is largely independent of the redshift, duration, spectral peak, or isotropic energy of the GRBs.

Y. Aimuratov et al. (2023a) also explored the correlations between the peak luminosity and peak time of SNe, as well as the redshift and energy of their associated GRBs. They also found no clear correlations. The independence between SN parameters and GRB parameters indicates that the SN properties are primarily governed by intrinsic processes such as the decay of radioactive nickel, rather than by the GRB emission itself. The results are somewhat consistent with the binary-driven hypernova model, in which the GRB and SN properties are largely independent.

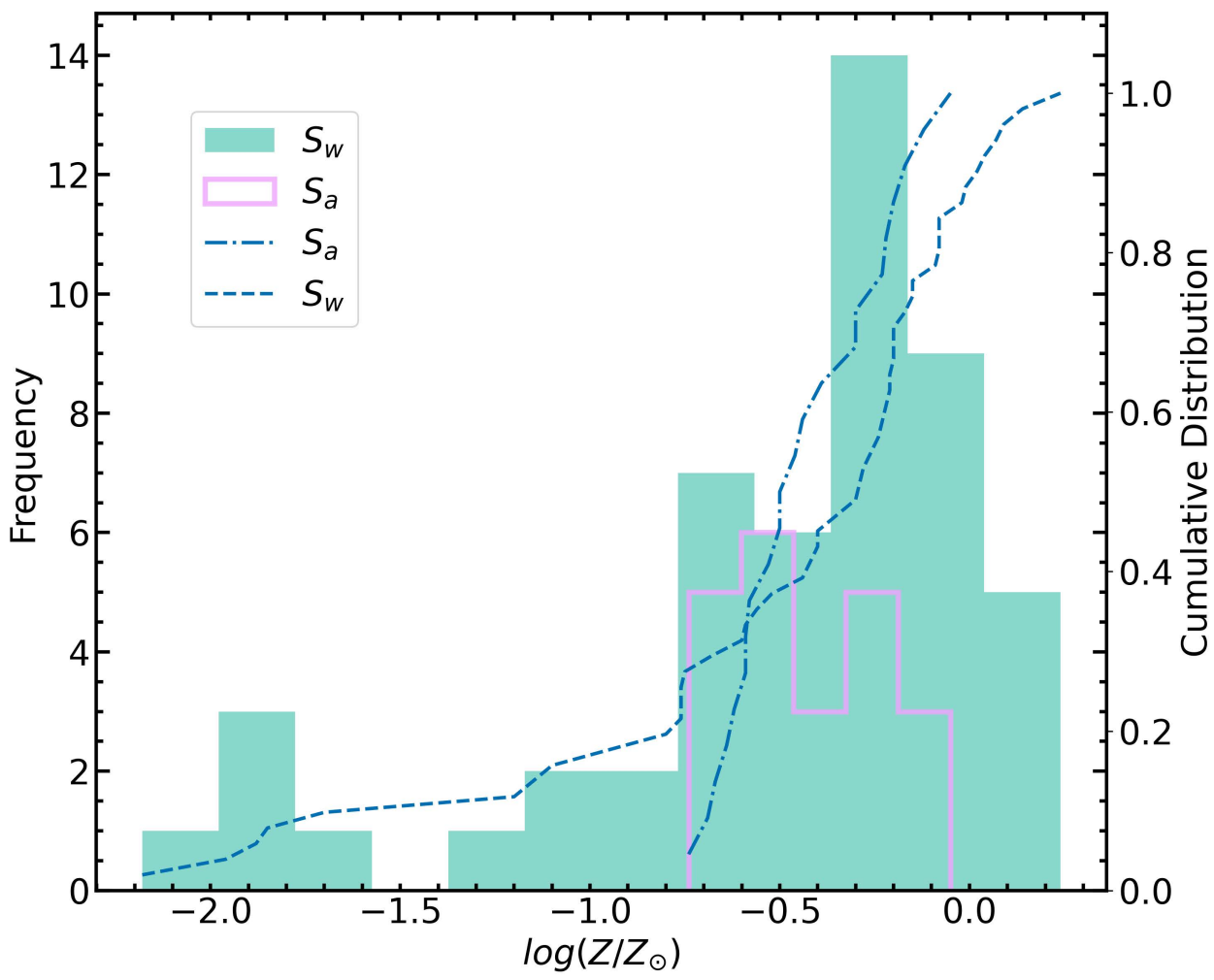

**Figure 5.** Distribution of host metallicity ($\log(Z/Z_\odot)$) for the $S_a$ and $S_w$ GRBs. The hollow histogram and the solid-filled histogram represent $S_a$ and $S_w$ samples, respectively. The dotted–dashed line and the dashed line represent the normalized cumulative distribution curve of $S_a$ and $S_w$ GRBs, respectively.

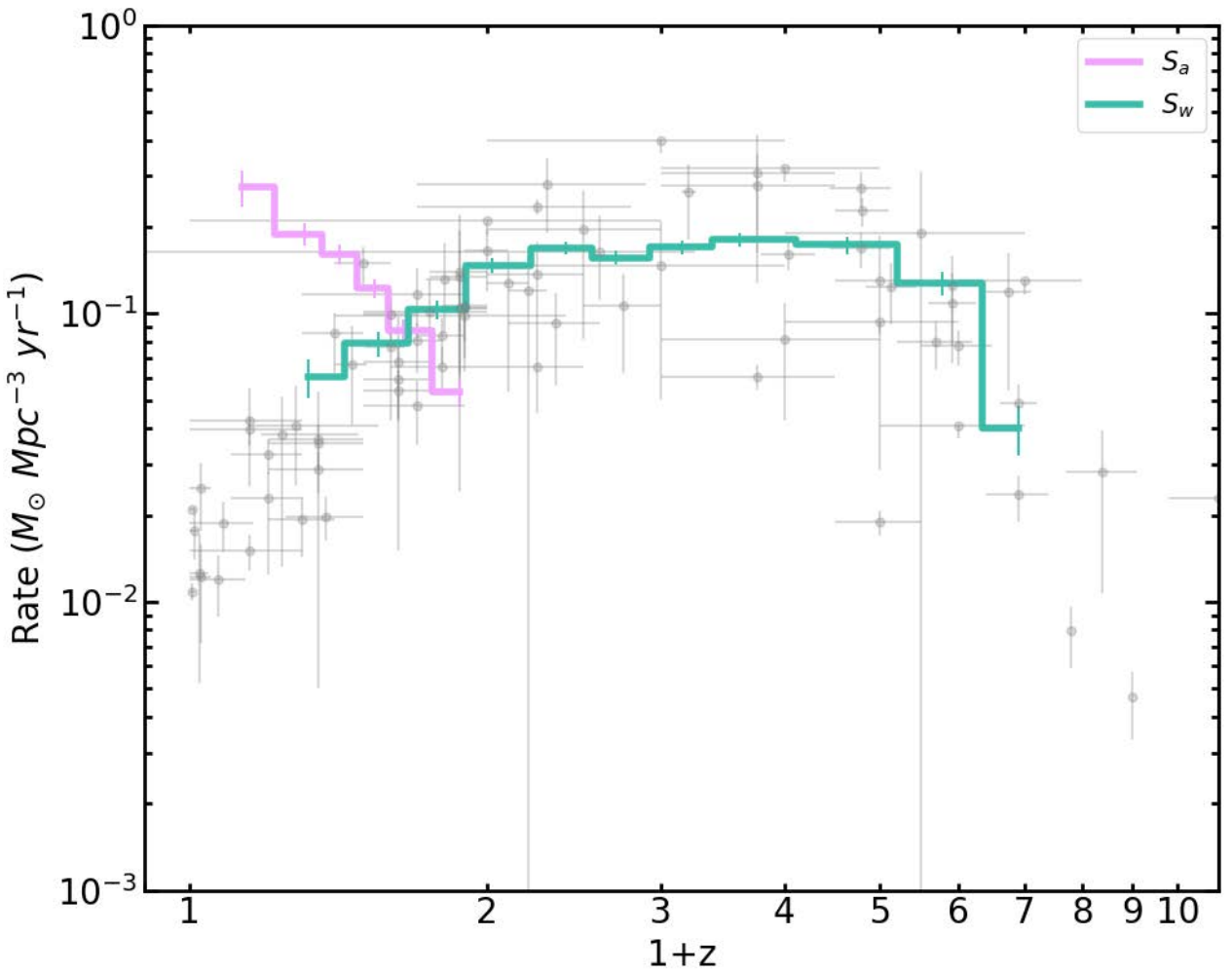

**Figure 6.** Event rates of GRBs derived from the $S_a$ and $S_w$ samples. The scattered points represent the observational data of SFR (X. F. Dong et al. 2022). Note that the GRB event rates were scaled by an arbitrary factor.

### 4.2. Event Rate Disparity between SN-GRBs, Ic-BL SNe, SLSNe, and Star Formation

It is found that the event rate of SN-GRBs does not follow the cosmic SFR in the $z < 1$ range. It strongly challenges the link between LGRBs and massive star collapse (J. Hjorth et al. 2003; D. Malesani et al. 2004). While the limited sample size of SN-GRBs may be a potential reason for the deviation, some intrinsic mechanisms still could not be expelled.

Interestingly, almost all SNe associated with GRBs are Type Ic-BL SNe (M. Modjaz et al. 2016), although GRB 111209A stands out as a possible exception, being likely associated with an SLSN (J. Greiner et al. 2015). We thus have computed the event rates of Ic-BL SNe and SLSNe separately based on two samples of Ic-BL SNe and SLSNe that are not associated with GRBs ($SN_{\rm IcBL}$ and $SN_{\rm SL}$). The same nonparametric method is used in our analysis, adopting an apparent limiting magnitude of 20 as the detection threshold. The event rates derived based on the $S_a$, $SN_{\rm IcBL}$, and $SN_{\rm SL}$ samples are shown in Figure 8 and are compared with SFR.

In Figure 8, we see that the derived event rates of SN-GRBs, Ic-BL SNe, and SLSNe all decrease with the increasing redshift. Among them, the event rate of $\rm SN_{IcBL}$ shows the steepest decline, followed by $\rm SN_{SL}$, while the SN-GRB event rate exhibits the shallowest redshift dependence. The different behaviors of $\rm SN_{IcBL}$ and $\rm SN_{SL}$ may reflect the intrinsic difference in their host galaxy environments (R. Lunnan et al. 2014; S. Schulze et al. 2018; M. Modjaz et al. 2020; A. Aamer et al. 2025; B. Zhang et al. 2025). Additionally, no SLSN is confirmed to be associated with a GRB. It seems to indicate that the circumburst environment of SLSNe may inhibit the occurrence of a successful GRB (J. Greiner et al. 2015). A. Kumar et al. (2024) leveraged machine learning algorithms to explore the origin of GRB-SNe within the magnetar framework. Their results show that these SNe occupy distinct regions in the parameter space compared to SLSNe, normal LGRBs, and SGRBs, particularly in terms of the median values of ejecta mass, the magnetar's initial spin period, and the magnetic field strength.

The difference between the event rates of SN-GRBs and $\rm SN_{IcBL}$ is also reasonable. Although all GRB-associated SNe identified to date are Type Ic-BL events, only a specific subset (∼1%–10%) of Ic-BL SNe is actually linked to GRBs (A. M. Soderberg et al. 2006b; D. A. Perley et al. 2020; G. Schroeder et al. 2025). This suggests that GRB-SNe may represent a distinct subgroup of the Ic-BL population, potentially differing in spectral features (M. Modjaz et al. 2016), host galaxy environments (J. Japelj et al. 2018), and other properties. G. Finneran et al. (2025b) investigated the expansion velocity evolution of GRB-SNe and ordinary Ic-BL SNe to assess potential jet-related signatures. However, they found no significant difference between the two types. The result further highlights the complexity of identifying GRB progenitors within the Ic-BL population.

Notably, none of these events follow the redshift evolution of the cosmic SFR. Host galaxy metallicity plays a crucial role in shaping stellar evolution. In a high-metallicity environment, the stellar wind will be strong, leading to a large loss rate of angular momentum during earlier evolutionary phases (A. Levan et al. 2016). This procedure will prevent the formation of a rapidly rotating compact core required for producing a GRB, such as an accreting magnetar or black hole. Finally, these stars, especially those medium-mass stars, are more likely to produce Type II SNe (S. C. Yoon & N. Langer 2005; S. E. Woosley & A. Heger 2006b; L. Dessart et al. 2013), which may further account for the mismatch between the $SN_{\rm IcBL}$ event rate and the cosmic SFR. Additionally, binary interactions in a high-metallicity environment, such as tidal spin-up or mass transfer, may help sustain the core's rapid rotation despite angular momentum loss from stellar winds (M. Cantiello et al. 2007). These channels can introduce additional delays between star formation and GRB production, potentially leading to a GRB event rate evolution that deviates from the cosmic SFR (E. Zapartas et al. 2017).

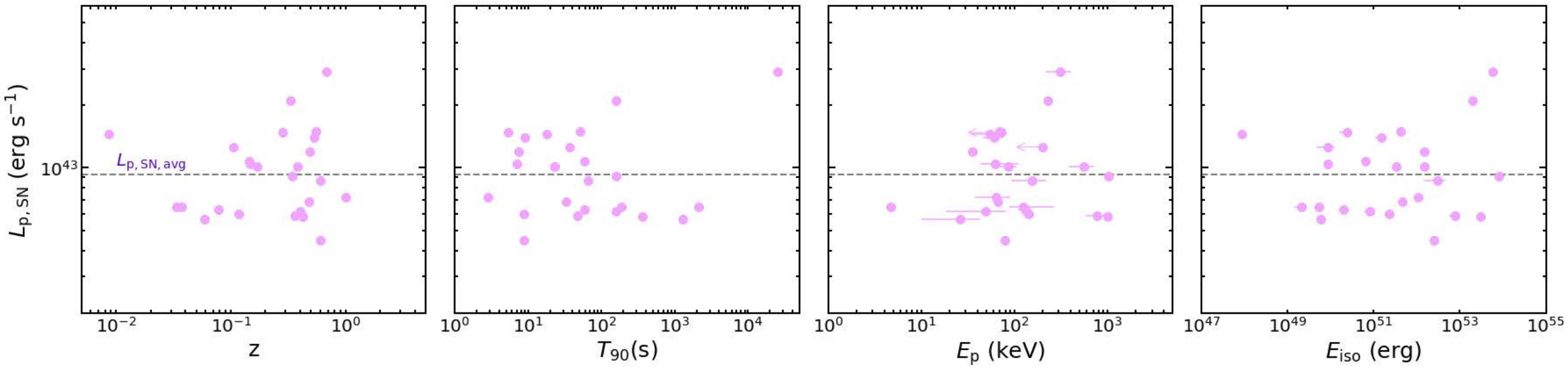

**Figure 7.** Redshift $z$, duration $T_{90}$, peak energy $E_{\rm p}$, and isotropic energy $E_{\rm iso}$ plotted versus the corresponding bolometric peak luminosity $L_{\rm p,SN}$ of the accompanying SN for the SN-GRBs (i.e., the $S_{\rm a}$ sample). The bolometric peak luminosity of SNe is taken from S. Klose et al. (2019) and Y. Aimuratov et al. (2023a). The dashed line represents the average value of SN peak luminosity, i.e., $L_{\rm p,SN,avg} = 9.42 \times 10^{42}$erg s$^{-1}$.

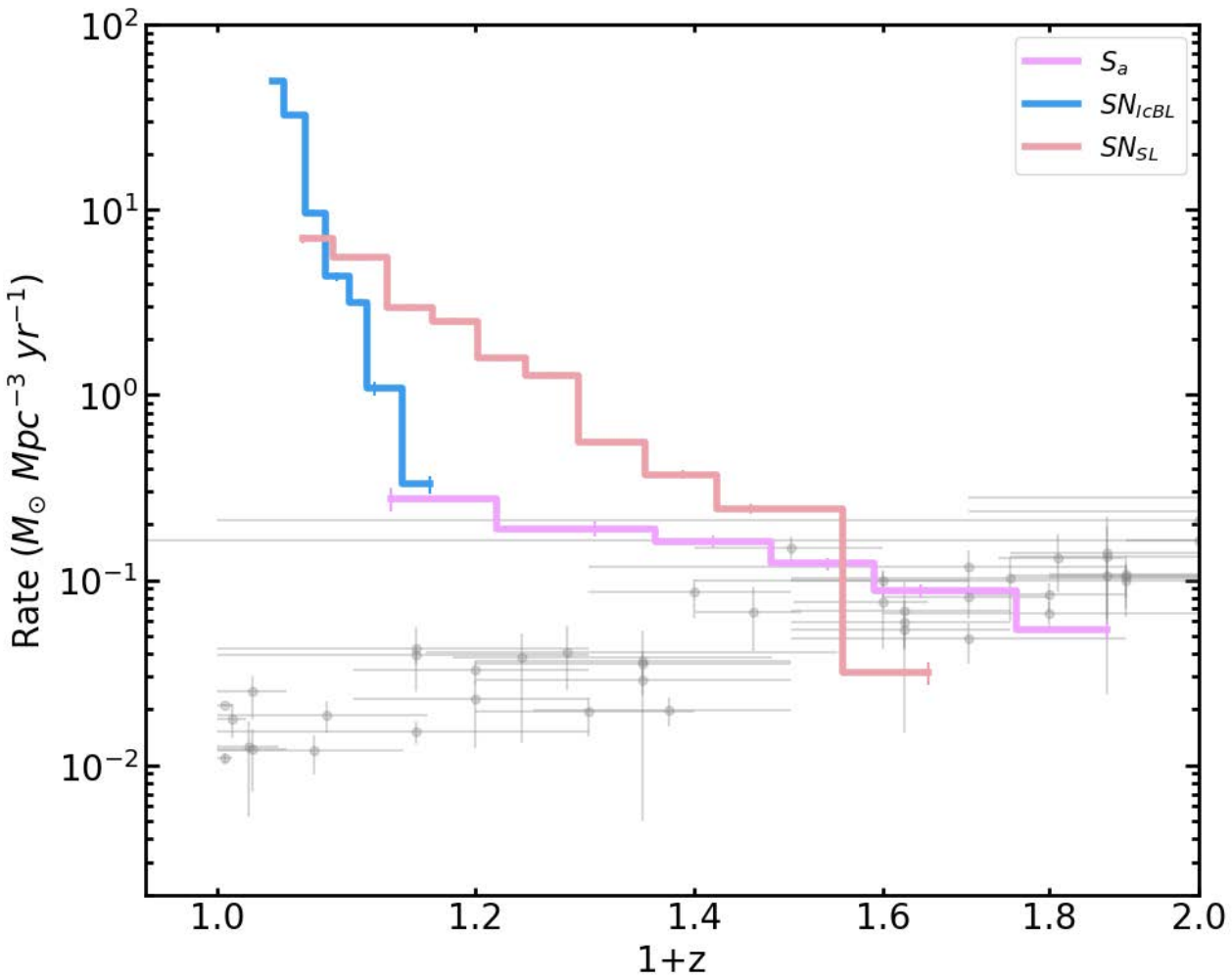

**Figure 8.** Event rates derived from the $S_{\rm a}$, $\rm SN_{IcBL}$, and $\rm SN_{SL}$ samples. The SLSN data are taken from S. Gomez et al. (2024). The Ic-BL SNe data are presented in Table 5. The scattered points represent the observational data of SFR (X. F. Dong et al. 2022). The $y$-axis is SFR. Note that the event rates derived for GRBs and SNe are scaled by an arbitrary factor.

## 5. Conclusions

In this study, a sample of 53 SN-associated LGRBs ($S_{\rm a}$) is built. The parameters involving the prompt emission, host galaxy, and accompanying SNe are collected. For comparison, another sample including 373 long GRBs without an SN association ($S_{\rm w}$) is also compiled. The majority (337 GRBs) of these SN-less events were observed by Swift, while the remaining 36 bursts were detected by various other instruments. The energetics of all the SN-less LGRBs are corrected to the rest-frame 15–150 keV band. The properties of these two samples are comprehensively analyzed and compared, aiming at elucidating the peculiarities of the SN-GRBs. Our main conclusions are summarized as follows.

*Similarity between the two samples*. The $S_{\rm a}$ and $S_{\rm w}$ samples are quite similar in their prompt emission and host galaxy properties. No single parameter in these aspects can be reliably used to distinguish SN-GRBs and SN-less GRBs. Although some SN-GRBs have an unusually long duration or deviate from the usual Amati relation, these traits are not exclusive or decisive.

*Difference in event rate*. The evolution of the derived event rate with respect to redshift differs substantially between $S_{\rm a}$ and $S_{\rm w}$ samples. While the event rate of SN-less GRBs aligns well with SFR, the rate of SN-GRBs deviates significantly. This discrepancy suggests that SN-less LGRBs constitute an intrinsically distinct subclass with a different kind of progenitor.

These conclusions would not be significantly affected by the properties of associated SNe, as no clear correlation is found between GRB parameters and the peak luminosity of their associated SNe. Moreover, the event rates of SN-GRBs, Ic-BL SNe, and SLSNe exhibit distinct deviations from the cosmic SFR. Improved SN samples in the future will allow for more accurate constraints on their cosmic evolution.

## Acknowledgments

We acknowledge useful comments by the anonymous referee. We are grateful to Yi-Han Iris Yin for helpful discussions. This study was supported by the National Natural Science Foundation of China (grant Nos. 12233002, U2031118, 12447179, 12273113), by the National SKA Program of China No. 2020SKA0120300, by the National Key R&D Program of China (2021YFA0718500), by the Postgraduate Research & Practice Innovation Program of Jiangsu Province (No. KYCX25_0197). Y.F.H. acknowledges the support from the Xinjiang Tianchi Program. J.J.G. acknowledges support from the Youth Innovation Promotion Association (2023331).

## ORCID iDs

Xiao-Fei Dong https://orcid.org/0009-0000-0467-0050
Yong-Feng Huang https://orcid.org/0000-0001-7199-2906
Zhi-Bin Zhang https://orcid.org/0000-0003-4859-682X
Jin-Jun Geng https://orcid.org/0000-0001-9648-7295
Chen Deng https://orcid.org/0000-0002-2191-7286
Ze-Cheng Zou https://orcid.org/0000-0002-6189-8307
Chen-Ran Hu https://orcid.org/0000-0002-5238-8997
Orkash Amat https://orcid.org/0000-0003-3230-7587